\documentclass{jfm}
\usepackage{graphicx}
\usepackage{natbib}
\usepackage{hyperref}
\usepackage{amsmath}
\usepackage{ar}
\usepackage{tikz}
\usepackage{pgfplots}
\usepackage{tikz-3dplot}
\usepackage{bm}
\usepackage{float}
\usepackage{epstopdf, epsfig}

\newcommand\ie{{i.e.}\ }

\title{Wake transition and aerodynamics of a dragonfly-inspired airfoil}

\author[A. Chiarini and G. Nastro]{Alessandro Chiarini\aff{1}\corresp{Present address: Dipartimento di Scienze e Tecnologie Aerospaziali, Politecnico di Milano, via la Masa 34, 20156 Milano, Italy.}\corresp{\email{alessandro.chiarini@polimi.it}} and Gabriele Nastro\aff{2}\corresp{\email{gabriele.nastro@isae-supaero.fr}}}
\affiliation{
\aff{1} Complex Fluids and Flows Unit, Okinawa Institute of Science and Technology Graduate University, 1919-1 Tancha, Onna-son, Okinawa 904-0495, Japan
\aff{2} ISAE-SUPAERO, Université de Toulouse, France
}

\begin{document}
\maketitle

%%%%%%%%%%%%%%%%%%%%%%%%%%%%%%%%%%%%%%%%%%%%%%%%%%
\begin{abstract}
We investigate the dynamics and the stability of the incompressible flow past a corrugated dragonfly-inspired airfoil in the two-dimensional (2D) $\alpha-Re$ parameter space, where $\alpha$ is the angle of attack and $Re$ is the Reynolds number. The angle of attack is varied between $-5^\circ \le \alpha \le 10^\circ$, and $Re$ (based on the free-stream velocity and the airfoil chord) is increased up to $Re=6000$. The study relies on linear stability analyses and three-dimensional (3D) nonlinear direct numerical simulations. For all $\alpha$ the primary instability consists of a Hopf bifurcation towards a periodic regime. The linear stability analysis reveals that two distinct modes drive the flow bifurcation for positive and negative $\alpha$, being characterised by a different frequency and a distinct triggering mechanism. The critical $Re$ decreases as $|\alpha|$ increases, and scales as a power law for large positive/negative $\alpha$. At intermediate $Re$, different limit cycles arise depending on $\alpha$, each one characterised by a distinctive vortex interaction, leading thus to secondary instabilities of different nature. For intermediate positive/negative $\alpha$ vortices are shed from both the top/bottom leading- and trailing-edge shear layers, and the two phenomena are frequency locked. By means of Floquet stability analysis, we show that the secondary instability consists of a 2D subharmonic bifurcation for large negative $\alpha$, of a 2D Neimark--Sacker bifurcation for small negative $\alpha$, of a 3D pitchfork bifurcation for small positive $\alpha$, and of a 3D subharmonic bifurcation for large positive $\alpha$. The aerodynamic performance of the dragonfly-inspired airfoil is discussed in relation to the different flow regimes emerging in the $\alpha-Re$ space of parameters.
\end{abstract}

\begin{keywords}
\end{keywords}

\section{Introduction}

Over the years, the structure of the wings of animals like insects, birds, and bat flies has attracted the attention of many scholars, because of its relevance that goes beyond a fundamental interest to encompass emerging engineering applications such as micro air vehicles (MAV) and unmanned aerial vehicles (UAV)~\citep{liu-etal-2016,kang-shyy-2013}.
These vehicles typically operate over a wide range of Reynolds numbers ($Re$), $\Rey \approx 10^2-10^5$, and face distinctive challenges in generating sufficient aerodynamic forces to sustain flight and perform complex manoeuvres. 
Many bio-inspired vehicle designs have been proposed, including fixed-wing, rotary- and flapping-wing configurations \citep{phan2019}.
Mimetics in bio-inspired flight systems can indeed offer new insights and breakthrough technologies in the field of low-speed aerodynamics and flight control. 

\subsection{Dragonfly-inspired airfoil}

\begin{figure}
\centering
\includegraphics[width=\textwidth]{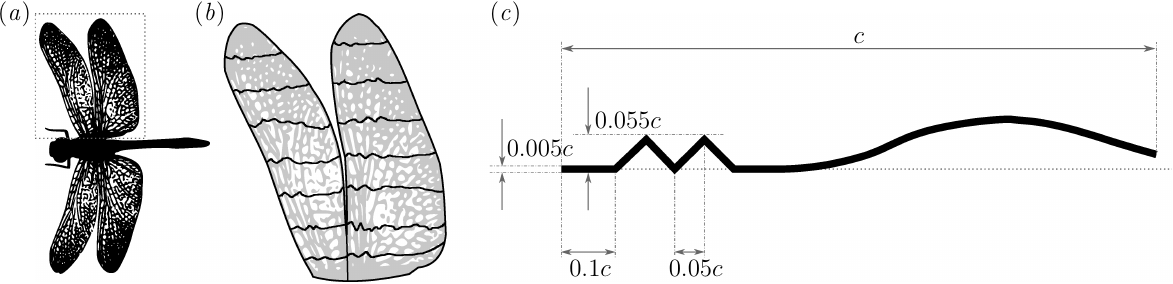}
\caption{Drawing of (\textit{a}) a dragonfly and (\textit{b}) close-up inset of its fore- and hind wings whose profile cross sections are qualitatively depicted in black~\citep[see also][]{kesel-2000,bomphrey-etal-2016}. Geometry of (\textit{c}) a simplified dragonfly airfoil based on the work of \cite{newman-etal-1977} about gliders.}
\label{fig:dragonfly_sketch}
\end{figure}

In this context, dragonflies (see figure \ref{fig:dragonfly_sketch}\textit{a}) have always drawn a great deal of interest, inspiring many biomimetic flying robots.
Dragonfly wings are not smooth or simple cambered surfaces~(see for instance the numerical reconstruction of a dragonfly wing in figure 1\textit{d} of \citealp{bomphrey-etal-2016}).
As illustrated in figure \ref{fig:dragonfly_sketch}(\textit{b}), the wing cross-section is highly irregular and low-cambered and varies significantly along the spanwise direction~\citep{okamoto-etal-1996}. Dragonfly wings present high stiffness and strength, and exhibit excellent aerodynamic performance due to their extremely \emph{corrugated configuration}~\citep{jongerius-lentink-2010}. 
Moreover, in contrast to other four-winged insects, the fore and hind wings of dragonflies work independently.
This double flight-power system allows large dragonflies to perform complex manoeuvres~\citep{ruppell-1989}, and gliding flight~\citep{wakeling-ellington-1997}. Thanks to their unique aerodynamic characteristics, dragonfly-inspired wings are widely spread and studied~\citep{newman-etal-1977,ruppell-1989,wakeling-ellington-1997,kesel-2000,levy-seifert-2009,bomphrey-etal-2016,bauerheim-chapin-2020}, especially in the context of MAV/UAV's research area. They indeed represent a relevant configuration of study in the perspective of tandem-wings and flapping flight~\citep[see \S 3 and \S 4 in][]{bomphrey-etal-2016}. However, despite the interest, even the gliding-flight dynamics of an isolated cross-section profile is not entirely clear, and the link between the flow dynamics and the enhanced aerodynamic performance in the complete $\alpha-Re$ parameter space requires further investigation.

Several profile models that replicate the general dragonfly-wing cross section have been introduced over the years. In this work, we consider the cross-section model introduced by \cite{newman-etal-1977}, and later used by other authors \citep[see for example][]{levy-seifert-2009,bauerheim-chapin-2020}.
As illustrated in figures \ref{fig:dragonfly_sketch}(\textit{c}), this simplified dragonfly-inspired configuration exhibits two sharp-corner corrugations with a rear arc that combines the mean geometrical characteristics of the profile cross sections of dragonfly fore and hind wings. The coordinates of some points along the lower surface are presented in table \ref{tab:points} and the thickness $t$ is $0.5 \%$ of the chord $c$.

\cite{levy-seifert-2009} showed that for $\Rey \leq 8000$ and $-2^\circ \leq \alpha \leq 10^\circ$ the aerodynamic performance of this two-dimensional (2D) dragonfly-inspired airfoil is superior to that of similar, but smooth airfoils (e.g. the traditional low-$Re$ airfoil Eppler-E61). 
In particular, they found that dragonfly-inspired airfoils exhibit a lower drag and a higher aerodynamic efficiency. \citet{bauerheim-chapin-2020} investigated the dynamics of the flow past this specific corrugated airfoil at $\Rey=6000$, varying the angle of attack ($\alpha$) in the $6^\circ \le \alpha \le 10^\circ$ range. Interestingly, they found that a small variation of $\alpha$ may lead to a sudden change of regime and, therefore, to a large variation of the aerodynamic performance. 
For $7.7^{\circ} \leq \alpha \leq 8.5^{\circ}$ they showed a sudden transition to chaos, which has been explained with the complex interaction between the main unstable shear layer produced at the leading-edge and the reverse flow in the cavity between the second corrugation and the rear arc on the suction side. In contrast, for higher $\alpha$ ($\alpha > 8.5^{\circ}$), they observed a periodic nonlinear regime. They found that the maximum lift-to-drag ratio is obtained just before the transition to chaos ($\alpha < 7.7^{\circ}$), whereas the highest lift coefficient is reached in the non-linear periodic regime ($\alpha > 8.5^{\circ}$). More recently, \cite{fujita-iima-2023}, considered the corrugated airfoil introduced by \cite{kesel-2000}, and investigated by 2D direct numerical simulations its aerodynamic performance (in an impulsively started configuration) at $\Rey=4000$ and $20^\circ \le \alpha \le 40^\circ$. They observed that the corrugated geometry enhances the aerodynamic performance when compared to a flat plate, but only for $\alpha>30^\circ$. They suggest that the mechanisms that are responsible for the lift increase at these large $\alpha$ are (i) the pressure reduction over the top side due to the complex vortex interaction, and (ii) the generation of a low-pressure region within the grooves near the leading-edge.

A comprehensive characterisation of the flow around dragonfly-inspired airfoils in a gliding flight configuration in the complete $\alpha-\Rey$ parameter space is missing, and the understanding of the underlying physics still remains elusive.
In this work, we do a step forward in this direction. We use the cross-section geometry introduced by \cite{newman-etal-1977} to investigate the sequence of bifurcations the flow undergoes up $\Rey=6000$ for both positive and negative $\alpha$, and relate the different flow regimes with the (enhanced) aerodynamic performance.

\subsection{Two- and three-dimensional wake transition}

The flow past smooth airfoils has been largely investigated over the years, with a particular focus on the transition from the steady 2D wake flow at low $\Rey$, to the three-dimensional (3D) and fully turbulent regime at large $\Rey$. The primary instability leads from the low-$Re$ steady state to a periodic vortrex street \citep{vonKarman-1954}, and consists of a 2D von K\'arm\'an mode that arises from a supercritical Hopf bifurcation \citep{sreenivasan-etal-1987}; see \cite{he-etal-2017,nastro-etal-2023,gupta-etal-2023} for symmetric and non-symmetric airfoils. Several authors found that the critical $Re$ and $\alpha$, i.e. the values of $\Rey$ and $\alpha$ corresponding to the first onset of the primary bifurcation, scale as a power law of $\alpha$ and $\Rey$ respectively~\citep{nastro-etal-2023,gupta-etal-2023}. At larger $\Rey$ and/or $\alpha$ the flow past smooth airfoils undergoes a secondary instability and becomes 3D. Recently, \cite{gupta-etal-2023} investigated the three-dimensionalisation of the wake past NACA airfoils at different $\alpha$.  
They found that the transition from 2D to 3D flow follows a power law $\alpha_{\text{3D}} \sim \Rey^{-0.5}$, and occurs through a subharmonic (period-doubling) mode, called mode $C$, having a characteristic spanwise wavelength of $0.1 \lessapprox \lambda_z/c \lessapprox 0.4$ that increases with $\alpha$. 
The same mode has been observed to be responsible of the three-dimensionalisation of the flow past stalled or post-stalled NACA0012 and NACA0015 airfoils \citep{meneghini-etal-2011,deng-eta-2017}, and of the flow past inclined flat plates at $20^{\circ} \leq \alpha \leq 30^{\circ}$ \citep{yang-etal-2013}. 
Further increasing $Re$ beyond criticality, \cite{deng-eta-2017} found via Floquet analysis that further subdominant modes become unstable, i.e. the synchronous mode $A$, the quasi-periodic mode $QP$, and the subharmonic modes $SL$ and $SS$. 

The bifurcation scenario valid for smooth NACA airfoils is expected to not hold for non-smooth airfoils. In fact, the shape and the aspect ratio $\AR$ (the chord-to-thickness ratio) of the body are known to influence the type and the sequence of bifurcations the flow undergoes. In the simple case of symmetric 2D bluff bodies in free stream, for example, the nature of the secondary bifurcation changes with the geometry and $\AR$, despite the primary bifurcation consists of a Hopf bifurcation towards a 2D and time-periodic state for all cases \citep{jackson-1987-finiteelementstudy,chiarini-quadrio-auteri-2022}. For the circular \citep{williamson-1988,barkley-henderson-1996} and square \citep{robichaux-balachandar-vanka-1999} cylinders, the flow three-dimensionalisation is driven by the synchronous and long-wavelength mode $A$, while the synchronous mode $B$, characterised by a shorter wavelength and a different spatio-temporal symmetry, becomes amplified at larger $\Rey$ \citep{barkley-etal-2000}. Further increasing $\Rey$, an additional quasi-periodic $QP$ mode with frequency that is not commensurate with that of the periodic base flow arises \citep{blackburn-lopez-2003,blackburn-etal-2005}. Mode $A$ drives the 3D wake transition also for elliptic cylinders with different $\AR$ \citep{leontini-etal-2015}. For rectangular cylinders with aerodynamic leading edge, instead, mode $A$ is the leading mode for $\AR <7.5$ only \citep{ryan-thompson-hourigan-2005}. For larger $\AR$ the flow three-dimensionalisation is instead driven by the so-called mode $B'$, which possess the same spatio-temporal symmetries as mode $B$, but has a much larger wavelength. For rectangular cylinders with sharp corners and $\AR \le 1$, mode $A$ drives the flow three-dimensionalisation for $\AR \ge 0.1$ only \citep{choi-yang-2014}. For smaller $\AR$, the flow three-dimensionalisation is due to the so-called mode $QP2$, that differs from mode $QP$ due to the larger wavelength. For elongated cylinders, i.e. $\AR \sim 5$, \cite{chiarini-quadrio-auteri-2022d} found that the scenario further complicates, and that the flow becomes 3D due to a quasi-subharmonic mode $QS$ with a characteristic wavelength similar to that of mode $A$. This bifurcation is triggered by the mutual inviscid interaction between the recirculating regions that coexist over the cylinder side at a given time.

\subsection{Focus of the present work}

In this work we consider a dragonfly-inspired airfoil and characterise the sequence of bifurcations the flow undergoes in the 2D space of parameters of $\alpha$ and $\Rey$. The aim of the present study is to extend the work of \cite{bauerheim-chapin-2020} to a wider range of parameters. In particular, we aim to (i) characterise the sequence of bifurcations the flow undergoes for different values of $\alpha$ as $Re$ increases, (ii) to provide a comprehensive description of the flow regimes in the $\alpha-\Rey$ parameter space, and (iii) to relate them with the enhanced aerodynamic performance of the airfoil, with a look at the MAV's and UAV's research field. 
The angle of attack is varied between $-5^{\circ} \leq \alpha \leq 10^{\circ}$, while $Re$ is increased up to $\Rey = 6000$. The flow behaviour largely changes with $\Rey$ and $\alpha$, ranging from a 2D and steady flow for small $\Rey$ and/or $|\alpha|$ to a 3D and non-periodic state at larger $\Rey$ and/or $|\alpha|$. Based on 2D and 3D high-fidelity direct numerical simulations (DNS) and stability analyses, we provide a complete picture of the dynamics around this geometry in a gliding flight configuration, and clarify the different bifurcations experienced by the flow and how they are related with the distinctive aerodynamic performance of the airfoil.

The remainder of the work is organised as follows.
The mathematical framework and the numerical methods are presented in \S \ref{sec:methods}.
In \S\ref{sec:flow-reg} the flow regimes are introduced. The primary bifurcation is then addressed in \S\ref{sec:prim-bif}, while the dynamics of the time-periodic flow at intermediate $Re$ and the secondary flow bifurcations are discussed in \S\ref{sec:inter-re} and \S\ref{sec:sec-bif}, respectively. In \S\ref{sec:aero-perf} we discuss the aerodynamic performance of the corrugated airfoil in relation with the flow regimes and bifurcations. Eventually, a concluding discussion and perspectives are provided in \S\ref{sec:conc_pers}.

\section{Methods} 
\label{sec:methods}

\subsection{Flow configuration and governing equations}
\label{subsec:2.1}

\begin{figure}
\centering
\includegraphics[width=\textwidth]{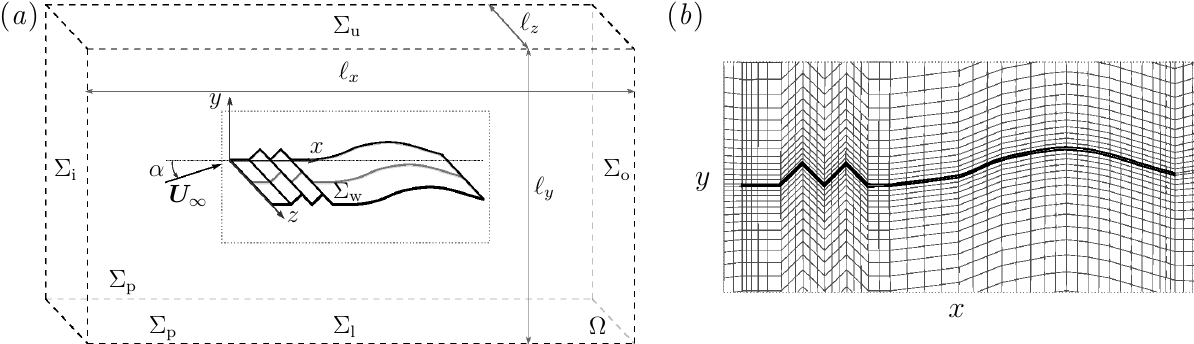}
\caption{Sketches of (\textit{a}) the extruded geometry and the flow computational domain for three-dimensional DNS and (\textit{b}) the grid details in a cross section (evidenced by the dotted-line rectangle on the left) around the dragonfly-inspired airfoil.}
\label{fig:sketch}
\end{figure}
The incompressible flow over a dragonfly-inspired airfoil (see figure \ref{fig:sketch}) is investigated via nonlinear 2D and 3D simulations and stability analyses. 
The geometry considered in this work is modelled by extruding the two-dimensional profile introduced by \citet{newman-etal-1977}, which has been largely used to study the flow past dragonfly-inspired wings in different regimes~\citep{levy-seifert-2009,bauerheim-chapin-2020}.
The geometry is discretised after a linear interpolation of the $32$ points shown in table \ref{tab:points}, $16$ for each airfoil side; the geometry has thickness $t=0.005c$ ($c$ is the chord) and sharp corners.
It has been verified that a spline-type interpolation between the $32$ points gives the same results in the considered range of $\alpha$ and $Re$.
\begin{table}
\centering
\begin{tabular}{ccccccccccccccccc}
& $P_1$ & $P_2$ & $P_3$ & $P_4$ & $P_5$ & $P_6$ & $P_7$ & $P_8$ & $P_9$ & $P_{10}$ & $P_{11}$ & $P_{12}$ & $P_{13}$ & $P_{14}$ & $P_{15}$ & $P_{16}$ \\
%\hline
$x_b/c$ & $0$ & $0.09$ & $0.14$ & $0.19$ & $0.24$ & $0.29$ & $0.34$ & $0.5$ & $0.6$ & $0.7$ & $0.75$ & $0.8$ & $0.85$ & $0.9$ & $0.95$ & $1$ \\
$y_b/c$ & $0$ & $0$ & $0.055$ & $0$ & $0.055$ & $0$ & $0$ & $0.02$ & $0.06$ & $0.08$ & $0.085$ & $0.08$ & $0.07$ & $0.055$ & $0.04$ & $0.025$ \\
\end{tabular}
\caption{Coordinates $(x_b,y_b)$ of the bottom side of the dragonfly-inspired airfoil, made dimensionless with the chord $c$. The coordinates of the top side $(x_t,y_t)$ are $(x_t,y_t)=(x_b,y_b+t)$, were $t=0.005c$ is the body thickness.}
\label{tab:points}
\end{table}
A Cartesian reference system is placed at the leading edge of the airfoil, with the $x$ axis being aligned with the airfoil chord and the $y$ and $z$ axes denoting the vertical and spanwise directions, respectively.

The Newtonian fluid flow is governed by the incompressible Navier--Stokes (NS) equations for the velocity $\bm{u}=(u,v,w)$ and pressure $p$ fields:
\begin{equation}
  \frac{\partial \bm{u}}{\partial t} + \bm{u} \cdot \bm{\nabla} \bm{u} = -\bm{\nabla} p + \frac{1}{\Rey} \bm{\nabla}^2 \bm{u}, \ \bm{\nabla} \cdot \bm{u} = 0.
\label{eq:NS}
\end{equation}
The NS equations \eqref{eq:NS} are completed with the following boundary conditions: an undisturbed uniform velocity is assumed at the inlet $\Sigma_\text{i}$ and on the lower and upper boundaries $\Sigma_\text{l}$ and $\Sigma_\text{u}$, a stress-free boundary condition $p \bm{n} - \Rey^{-1} \nabla \bm{u} \cdot \bm{n} = 0$ at the outlet $\Sigma_\text{o}$, periodic conditions on the lateral boundaries $\Sigma_\text{p}$ and no-slip and no-penetration conditions on the solid walls $\Sigma_\text{w}$.

The angle of attack $\alpha$, i.e. the angle between the incoming flow $\bm{U}_\infty=(U_\infty \cos(\alpha), U_\infty \sin(\alpha), 0)$ and the $x$ axis, is varied between $-5^{\circ} \le \alpha \le 10^{\circ}$. 
The Reynolds number based on the chord $c$ of the airfoil and the free stream velocity magnitude $U_\infty$ is increased up to $\Rey = U_\infty c / \nu=6000$ ($\nu$ is the fluid kinematic viscosity).
The Strouhal number describing the oscillating flow mechanisms is defined as $St = f c/U_{\infty}$ with $f$ being the frequency.
The drag and lift forces are reported in their non-dimensional form through
\begin{equation}
    C_d = \frac{D}{\tfrac{1}{2} \rho U^2 _{\infty} c}\quad \text{and} \quad C_\ell = \frac{L}{\tfrac{1}{2} \rho U^2 _{\infty} c},
\end{equation}
where $D$ and $L$ are the aerodynamic forces per unit width in directions parallel and perpendicular to the free stream, respectively, and $\rho$ is the constant fluid density. 
To describe the aerodynamic performance, it is useful to introduce the aerodynamic efficiency, i.e. the lift-to-drag ratio $E=L/D$. 

\subsection{Primary instability}

The onset of the primary two-dimensional instability is studied using linear theory, which is based on a normal-mode analysis~\citep{Theofilis2003,Theofilis2011}. 
The field $\{\bm{u},p\}$ of velocity and pressure is divided into a time-independent base flow $\{\bm{U}_1,P_1\}$, and an unsteady contribution $\{\bm{u}_1,p_1\}$ with small amplitude $\epsilon$,
\begin{equation}
  \bm{u}(\bm{x},t) = \bm{U}_1(\bm{x}) + \epsilon \bm{u}_1(\bm{x},t) \ \text{and} \
  p(\bm{x},t) = P_1(\bm{x}) + \epsilon p_1 (\bm{x},t).
\end{equation}
Using this decomposition in the NS equations \eqref{eq:NS}, the steady two-dimensional NS equations for $\{\bm{U}_1,P_1\}$ are obtained at order $\epsilon^0$. 
The governing equations for the small perturbations $\{\bm{u}_1,p_1\}$, instead, are obtained at order $\epsilon^1$, and are the linearised Navier--Stokes (LNS) equations. 
Using the normal-mode ansatz $\{\bm{u}_1,p_1\} = \{\hat{\bm{u}}_1,\hat{p}_1\}e^{\gamma_1 t} + c.c.$ ($c.c.$ stands for complex conjugate) the LNS equations yield an eigenvalue problem for the complex eigenvalue $\gamma_1 = \sigma_1 + i \omega_1$ --- with $\sigma_1$ being the temporal growth rate and $\omega_1$ the angular frequency --- and the complex eigenvector $\{\hat{\bm{u}}_1,\hat{p}_1\}$,
\begin{equation}
  \gamma_1 \hat{\bm{u}}_1 + \bm{U}_1 \cdot \bm{\nabla} \hat{\bm{u}}_1 + \hat{\bm{u}}_1 \cdot \bm{\nabla} \bm{U}_1 = -\bm{\nabla} \hat{p}_1 + \frac{1}{\Rey} \bm{\nabla}^2 \hat{\bm{u}}_1, \ \ \bm{\nabla} \cdot \hat{\bm{u}}_1=0.
  \label{eq:sLNSE}
\end{equation}
The linear stability of the system is determined by the sign of the real part of $\gamma_1$. 
If all $\sigma_1<0$ the perturbations decay and the flow is linearly stable. If at least one $\sigma_1>0$ exists, the perturbations grow exponentially and the flow is linearly unstable. 
The imaginary part $\omega_1$ determines whether the time-independent base flow $\{\bm{U}_1,P_1\}$ experiences a pitchfork or transcritical ($\omega_1=0$) or Hopf-type ($\omega_1 \neq 0$) bifurcation.
Like for low-$Re$ smooth airfoils \citep{he-etal-2017,nastro-etal-2023}, for all $\alpha$ considered in this work the primary instability results to be due to a 2D oscillatory mode which arises from a Hopf bifurcation and leads to a time-periodic flow, \ie a limit cycle solution.

\subsection{Secondary instability}

When the Reynolds number (or the angle of attack) is above the critical value $\Rey_{c1}$ (or $\alpha_{c1}$, respectively), the flow approaches a 2D periodic limit cycle. 
The Floquet theory is then used to study the linear stability of the 2D and time-periodic base flows with respect to 2D and 3D perturbations. 
The flow is written as the sum of a 2D base flow $\{\bm{U}_2,P_2\}$, which is periodic with period $T$, and an unsteady 3D perturbation with small amplitude $\epsilon$, i.e.
\begin{equation}
   \{\bm{u},p\}(x,y,z,t) = \{\bm{U}_2,P_2\}(x,y,t) + \frac{\epsilon}{\sqrt{2 \pi}} \int_{-\infty}^\infty \{\bm{u}_2,p_2\}(x,y,\beta,t) e^{i\beta z} \text{d} \beta,
\label{eq:ans}
\end{equation}
where $i$ is the imaginary unit, $\bm{u}_2$ and $p_2$ indicate the Fourier transform of the velocity and pressure disturbances in the homogeneous spanwise direction $z$, and $\beta = 2 \pi/\lambda_z$ is the corresponding wavenumber, with $\lambda_z$ being the wavelength.

Introducing the decomposition \eqref{eq:ans} into the NS equations \eqref{eq:NS}, the 2D periodic base flow is obtained at order $\epsilon^0$, while the eigenproblem describing the linear evolution of the perturbations is obtained at order $\epsilon^1$. 
By applying the Fourier transform in the $z$ direction, the unsteady linearised Navier--Stokes equations for each $\beta$ read
\begin{equation}
\frac{\partial \bm{u}_2}{\partial t} + \bm{U}_2 \cdot \bm{\nabla}_\beta \bm{u}_2 + \bm{u}_2 \cdot \bm{\nabla}_\beta \bm{U}_2 - \frac{1}{\Rey} \bm{\nabla}_\beta^2 \bm{u}_2
+\bm{\nabla}_\beta p_2 = 0, \ \bm{\nabla}_\beta \cdot \bm{u}_2 = 0.
\label{eq:ULNS}
\end{equation}
Here $\bm{\nabla}_\beta \equiv (\partial/\partial x, \partial/ \partial y, i \beta)$ is the Fourier-transformed gradient operator. 
Following the Floquet theory, we assume the functional form for the perturbation field $\{\bm{u}_2,p_2\}$ given by
\begin{equation}
\{\bm{u}_2,p_2\}(x,y,\beta,t) = \{\hat{\bm{u}}_2,\hat{p}_2\}(x,y,\beta,t) e^{\gamma_2 t},
\label{eq:decay}
\end{equation}
where $\gamma_2 = \sigma_2 + i \omega_2$ is the Floquet exponent, and $\{\hat{\bm{u}}_2,\hat{p}_2\}$ is the Floquet mode associated with $\gamma_2$ and $\beta$, and possesses the same periodicity of the base flow
\begin{equation}
  \{\hat{\bm{u}}_2,\hat{p}_2\}(x,y,\beta,t+T) = \{\hat{\bm{u}}_2,\hat{p}_2\}(x,y,\beta,t).
\end{equation}
The stability of the system is determined by the sign of $\sigma_2$ or, equivalently, by the modulus of the Floquet multiplier $\mu = e^{\gamma_2 T}$. 
If all $\sigma_2 < 0$, or $|\mu|<1$, the perturbations decay and the flow remains two-dimensional and periodic. 
Otherwise, if at least one exponent exists with $\sigma_2 > 0$ or $|\mu|>1$, the perturbations grow exponentially, and the flow bifurcates becoming 3D if $\beta \neq 0$ or remaining 2D if $\beta = 0$.
Bifurcations of the limit cycle solution depend on how the corresponding Floquet multiplier crosses the unit circle in the complex plane.
If $\mu = +1$, the periodic orbit presents a (secondary) pitchfork or saddle-node bifurcation yielding a new stable limit cycle with the same period as the starting one.
If $\mu \neq +1$, \ie $\omega_2 \neq 0$, a new frequency emerges and the resulting flow attractor draws a torus in the phase space.
This bifurcation is called secondary Hopf or Neimark--Sacker bifurcation. 
As Neimark--Sacker bifurcations introduce a new temporal frequency, in general incommensurate with that of the original orbit, the resulting solutions are referred to as temporally quasi-periodic. 
Within this case (\ie $\omega_2 \neq 0$), the particular condition $\mu = -1$, \ie $\omega_2 = \omega_1/2$ with $\omega_1$ the base flow angular frequency, is called period-doubling, flip, or subharmonic bifurcation.
In this case the bifurcation leads to a new limit cycle that presents twice the periodicity of the starting one.

\subsection{Numerical methods}

Two different numerical methods are used. The 3D direct numerical simulations (DNS) and the primary instability analysis are carried out using the open source high-performance solver Nek5000~\citep{nek5000-web-page}, which is based on spectral elements.
The secondary stability analysis is instead based on finite elements and the simulations have been implemented in the non-commercial software FreeFem++~\citep{hecht-2012}. 

The spatial discretisation for the nonlinear 3D simulations relies on a square-grid mesh (see figure \ref{fig:sketch}). The streamwise, cross-stream and spanwise extents are set to $\ell_x=20c$, $\ell_y=20c$ and $\ell_z=c$, respectively. The choice of $\ell_z = c$ is justified by the fact that the three-dimensional modes that emerge and drive the dynamics in the considered range of $Re$ are characterised by $\beta \geq 2 \pi$ (see \S\ref{sec:sec-bif}).
The grid consists of approximately $1.2 \times 10^6$ spectral elements with polynomial order $P = 6$.
The total number of degrees of freedom $N_{\text{dof}}$ is thus approximately $2.6 \times 10^8$. 
A semi-implicit third-order accurate temporal scheme is used to integrate forward in time the NS equations with a target CFL of $0.5$~\citep[see][for more detail]{nek5000-web-page}.
Numerical simulations were performed for a final time greater than or equal to $200$ convective time scales in order to ensure the complete settlement of the flow regime.

Primary instability results are also obtained with Nek5000. 
The low-$\Rey$ steady two-dimensional base flow $\bm{U}_1$ is obtained using the selective frequency damping (SFD) method proposed by \citet{aakervik2006} wherein the NS steady solution is obtained by damping the unstable frequency via the addition of a dissipative relaxation term proportional to the frequency content of the velocity oscillations. 
The generalised eigenvalue problem \eqref{eq:sLNSE} for the onset of the primary instability is solved via Krylov techniques described extensively in \S 3 of \citet{frantz-etal-2023}. Consistently with the 3D DNS, the computational domain extends for $(\ell_x,\ell_y)=(20c,20c)$. The grid consists of $2.24 \times 10^5$ spectral elements with polynomial order $P=6$, yielding a total number of degrees of freedom of $N_{\text{dof}} \approx 8.06 \times 10^6$. The convergence of the results is discussed in \S\ref{sec:sens_grid_res_pb}.

The results of the secondary stability analysis are obtained with the numerical code used and validated by \cite{chiarini-quadrio-auteri-2022d}, which is based on finite elements. The finite-element formulation employs quadratic elements (P2) for the velocity and linear elements (P1) for the pressure. 
The 2D periodic base flow $\bm{U}_2$ is computed integrating in time the 2D version of the unsteady NS equations. 
The time integration employs an explicit third-order low-storage Runge-Kutta method for the nonlinear term, combined with an implicit second-order Crank--Nicolson scheme for the linear terms \citep{rai-moin-1991}. 
The BoostConv algorithm is used to accelerate the convergence of the nonlinear simulations to the periodic limit cycle. 
It is an iterative algorithm inspired by the Krylov subspace methods introduced by \cite{citro-etal-2017}.
Convergence is reached when the absolute value of the largest difference over the computational domain between variables at $t$ and $t+T$ is below $10^{-10}$. 
The numerical method used for the Floquet analysis is similar to that used in the works by \cite{barkley-henderson-1996} and \cite{jallas-marquet-fabre-2017}. The evolution of the velocity perturbation with wavenumber $\beta$, i.e. $\bm{u}_\beta$, from time $t_0$ to time $t_0+T$ can be written using the linearised Poincar\'{e} map $\mathcal{P}_\beta$ as
\begin{equation}
  \bm{u}_\beta(t_0+T) = \mathcal{P}_\beta \bm{u}_\beta(t_0).
\end{equation}
The Floquet multipliers $\mu_\beta$ and the Floquet modes $\hat{\bm{u}}_{2,\beta}$ at the time $t_0$ correspond to the eigenvalues and eigenvector of $\mathcal{P}_\beta$. The eigenvalues of $\mathcal{P}_\beta$ with largest modulus and the corresponding eigenvectors are computed using the Arnoldi method \citep{saad-2011}, which implies the iterative action of the $\mathcal{P}_\beta$ operator. The modified Gram--Schmidt algorithm is used for the orthogonalisation of the eigenvectors. All the computed Floquet modes are normalised using their kinetic energy. After solving the eigenvalue problem, we evaluate the evolution of the Floquet mode over the period $T$ by integrating equation \ref{eq:ULNS} using the computed $\{\hat{\bm{u}}_{2,\beta},p_{2,\beta}\}(t_0)$ as initial condition. To account for the time decay/growth (see equation \ref{eq:decay}) the result of the integration is corrected by $e^{\gamma_2 t}$. For consistency, the time integration of the linearised Navier--Stokes equations is carried out with the same numerical scheme used for the computation of $\{\bm{U}_2,P_2\}$. When integrating the linearised Navier--Stokes equations, we evaluate the base flow at each time step by Fourier interpolation of $100$ instantaneous $\{\bm{U}_2,P_2\}$ fields stored uniformly along the period $T$. The adjoint modes needed to compute the structural sensitivities are obtained in a similar way \citep[see][for additional details]{chiarini-quadrio-auteri-2022d}.
For the secondary stability analysis two different computational domains have been used, after a preliminary convergence study (see appendix \S\ref{sec:grid-sec}). For $8^\circ \le \alpha \le 10^\circ$ the computational domain extends for $(\ell_x,\ell_y) =( 20c,30c)$ with a number of triangles of $N_{\text{el}} \approx 15.4 \times 10^4$. For $-5^\circ \le \alpha \le 7^\circ$, instead, the computational domain extends for $(\ell_x,\ell_y)=( 17.5c,14c)$ with $N_{\text{el}} \approx 13.8 \times 10^4$. The larger $\ell_y$ used for the larger $\alpha$ is needed to properly capture the growth rate of the ensuing leading mode (see appendix \S\ref{sec:grid-sec}). For all cases, the distribution and size of the triangles have been chosen to properly refine the region close to the body, paying particular attention to the near-corner regions and to the wake.

Hereinafter, unless otherwise indicated, all quantities are made dimensionless with $U_\infty$ and $c$.

\section{The flow regimes} 
\label{sec:flow-reg}

A variation of $\alpha$ modifies the low-$Re$ 2D flow and, therefore, changes the sequence of bifurcations the flow undergoes at larger $Re$. In this section we start by using the 3D nonlinear simulations to identify the different flow regimes in the considered portion of the $\alpha-Re$ space of parameters. The different flow bifurcations and regimes are then characterised in detail in \S\ref{subsec:pri-bif} and \S \ref{sec:sec-bif}, by means of stability analyses. In the following, we distinguish between steady, periodic and non-periodic flow regimes. A regime is said to be non-periodic when it is characterised by multiple, distinct, non-commensurate frequencies. 

\begin{figure}
    \centering
    \includegraphics[trim={0 0 0 0},clip,width=0.95\textwidth]{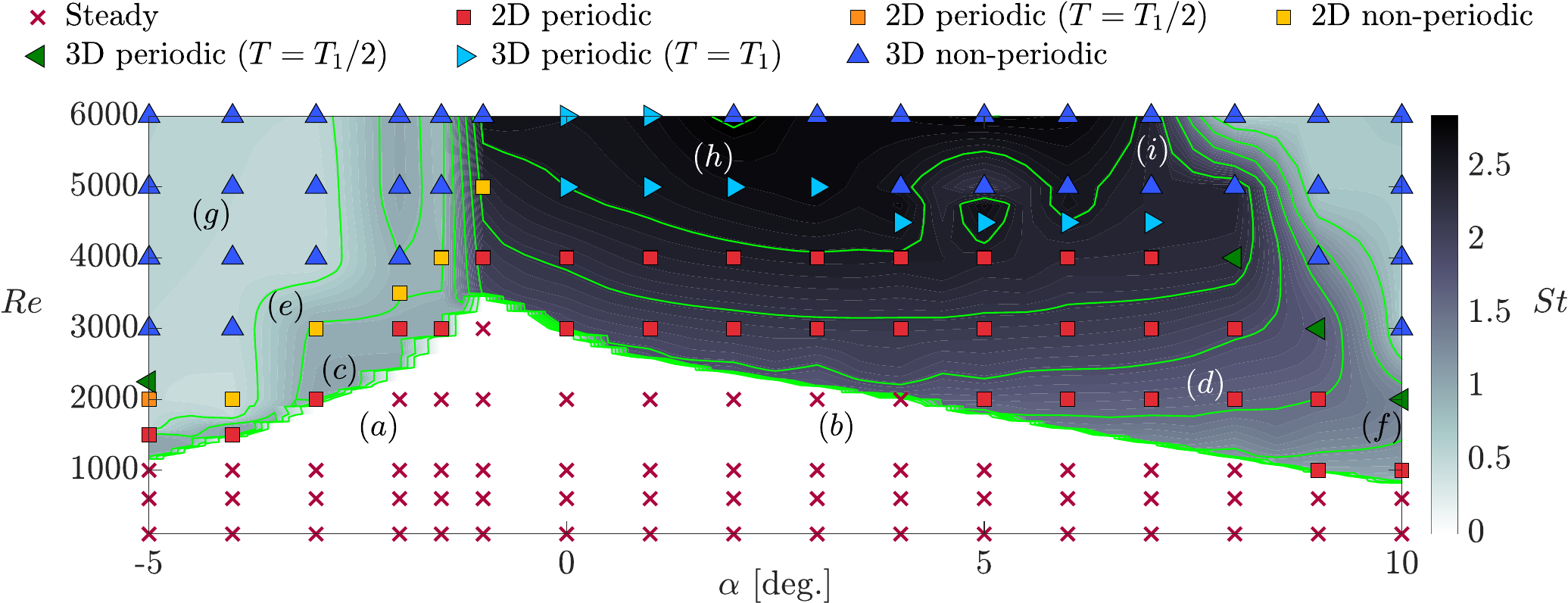}
    \includegraphics[trim={0 13 0 0},clip,width=0.95\textwidth]{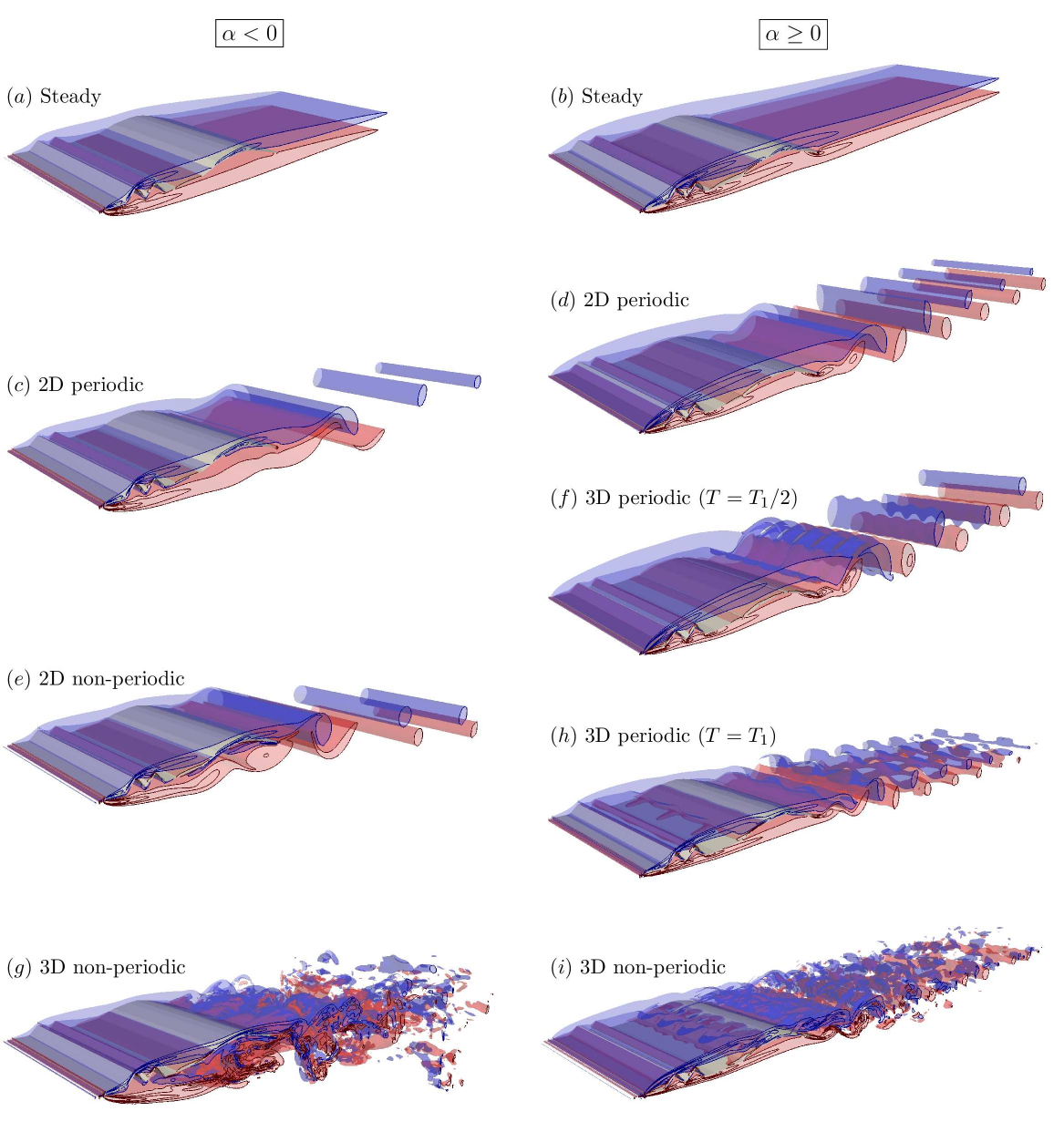}
    \caption{Dominant Strouhal number (measured from the frequency spectrum of the longitudinal velocity probed at $(x,y,z)=(1.3,0.17,0.5)$) map from DNS in the $\alpha-\Rey$ parameter space. Ten equally spaced green-coloured contour levels are used. Symbols indicate the wake regime observed from the results of numerical simulations whose spanwise vorticity is depicted at the bottom. The green left-oriented (blue right-oriented) triangle refers to a periodic 3D state that arises after a subharmonic (pitchfork) bifurcation of the 2D periodic regime at lower $Re$; see \S \ref{sec:sec-bif}.}
    \label{fig:St_Re-alpha_plane}
\end{figure}
Figure \ref{fig:St_Re-alpha_plane} shows the dominant Strouhal number, measured via inspection of the frequency spectrum of the longitudinal velocity probed at $(x,\,y,\,z) = (1.3,\,0.17,\,0.5)$, in the $\alpha-Re$ parameter space. From a qualitative viewpoint, the results are independent of the choice of the probe location.
Numerical simulations were run long enough to ensure that the transient was exceeded, and that the flow regime was completely settled;
considering that the simulations are advanced for at least $200 c/U_\infty$ after the initial transient, the frequency resolution results to be lower than or equal to $ \Delta f \sim 6 \times 10^{-3}$, which has been verified to ensure the appropriate depiction of the complete spectrum for each flow regime.
The flow regimes are indicated by different symbols in the $\alpha-Re$ map, and are illustrated with representative snapshots in the bottom panels.

The lowest contour level in figure \ref{fig:St_Re-alpha_plane}(a) delimits the threshold for unsteadiness.
Below this threshold (\ie in the white area of the map) the flow is steady and 2D.
Here the Reynolds number is indeed lower than the critical Reynolds number $Re_{c1}(\alpha)$ which denotes $Re$ corresponding to the first onset of the primary bifurcation at a given $\alpha$. $Re_{c1}$ decreases as $|\alpha|$ increases, with the maximum value $Re_{c1} \approx 3400$ found for $\alpha=-1^\circ$. 
Nevertheless, for all $\alpha$ the primary bifurcation consists of a Hopf bifurcation towards a 2D periodic limit cycle, characterised by an alternating vortex shedding in the wake. As detailed in \S \ref{sec:inter-re}, different limit cycles arise depending on $\alpha$, each one characterised by different vortex interactions. This is visualised in figure \ref{fig:St_Re-alpha_plane} by the large dependence of the dominant flow frequency on $\alpha$ at intermediate $Re$; for $\alpha \lesssim -1.25^\circ$ $St$ is smaller compared to the $\alpha>-1.25^\circ$ cases. 

\begin{figure}
    \centerline{
    \begin{tikzpicture}
    \node at (0.0,3.5) {\includegraphics[trim={0 0 0 0},clip,width=0.95\textwidth]{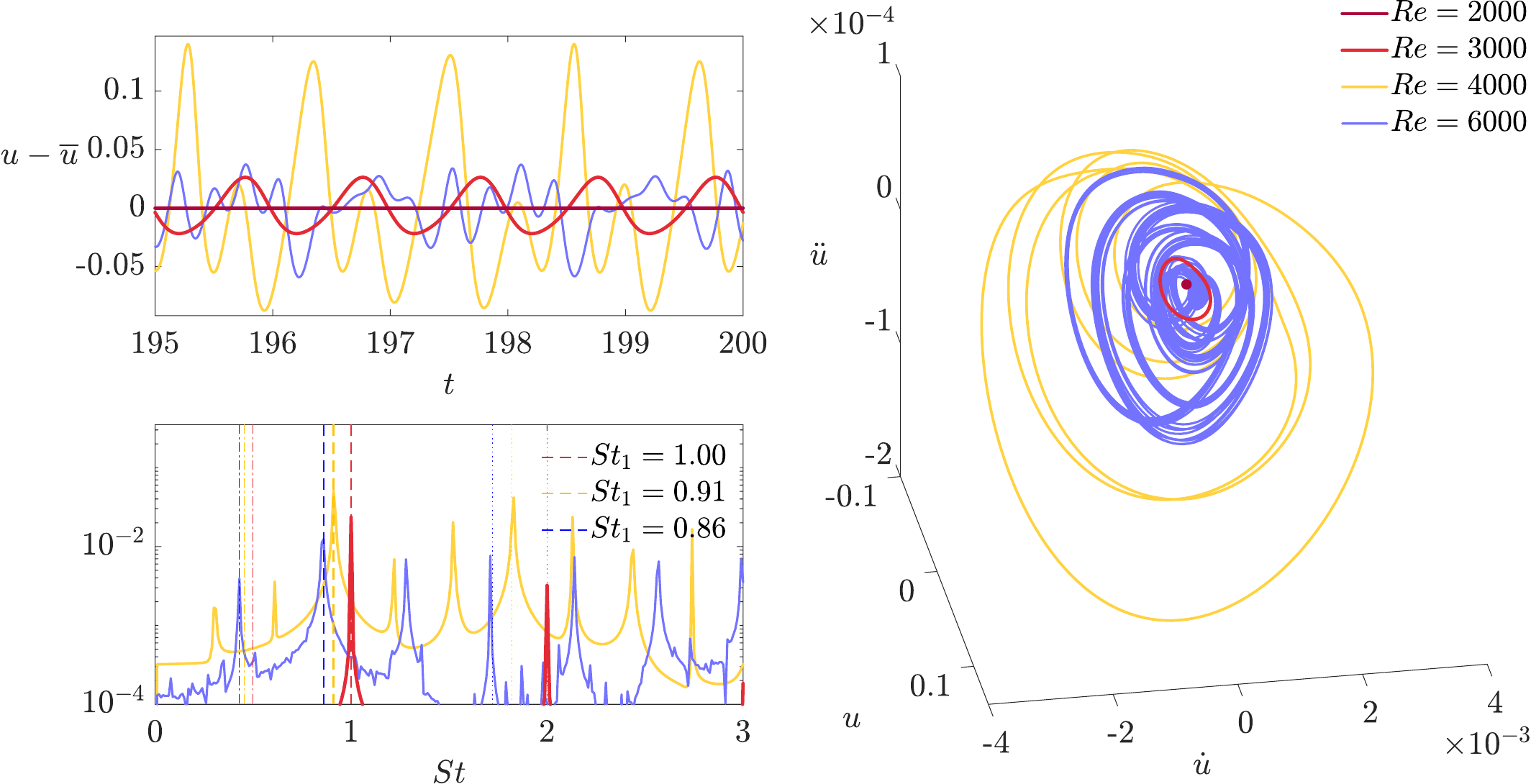}};
    \node at (-5.3,6.75) {$(a)$};
    \node at (-5.3,3.45) {$(b)$};
    \node at (1.5,6.75)  {$(c)$};
    \end{tikzpicture}
    }
    \caption{Longitudinal velocity signal at $\alpha=-1.5 ^\circ$ for different $\Rey$. Panel $(a)$: temporal fluctuations of the longitudinal velocity $u$ at $(x,\,y,\,z) = (1.3,\,0.17,\,0.5)$. Panel $(b)$: corresponding spectrum. Panel $(c)$: flow attractor. The $\overline{\cdot}$ operator indicates time average.}
    \label{fig:flow_attractors_AoA-1.5}
\end{figure}

The nature of the secondary bifurcation changes with $\alpha$. For small negative $\alpha$ ($-5^\circ < \alpha <-1.25^\circ$), the limit cycle becomes unstable via a 2D Neimark--Sacker bifurcation. At $Re = Re_{c2}$ the wake loses the temporal symmetry due to the emergence of an incommensurate frequency, but retains its 2D character; see the light orange squares in the $\alpha-Re$ map, and the 2D non-periodic state depicted in figure \ref{fig:St_Re-alpha_plane}(e). When increasing $Re$, the wake becomes first 3D and then approaches a more chaotic regime.
As an example, figure \ref{fig:flow_attractors_AoA-1.5} considers $\alpha = -1.5^{\circ}$, and shows the dependence of the frequency spectrum and of the flow attractor on $Re$.
At $\Rey = 2000$ the wake is steady, corresponding to a fixed point in the phase space.
At $Re = 3000$, instead, the flow is periodic with a vortex-shedding frequency of $St \approx 1$, and a limit cycle arises in the phase space.
At $\Rey = 4000$ several peaks emerge in the frequency spectra, and the wake loses its periodic behaviour.
A further increase in $Re$ ($\Rey = 6000$) leads to the flow three-dimensionalisation; see the blue triangles.
The redistribution of energy as a result of the three-dimensionalisation promotes a reduction in the amplitude of the longitudinal velocity oscillations in favour of the spanwise velocity oscillations (not shown). 

For large negative $\alpha$ ($\alpha \le -5^\circ$), instead, the secondary bifurcation is 2D and of subharmonic nature. In this case, a new limit cycle arises for $Re>Re_{c2}$, with twice the periodicity of the original one; see the dark orange square. Before the flow becomes non-periodic for $Re \geq 3000$, this second 2D limit cycle becomes unstable for $Re > Re_{c3}$ (with $Re_{c3} \approx 2125 \text{ at } \alpha -5^\circ$) and three-dimensionalises through a 3D subharmonic bifurcation; see the green left-oriented triangle.

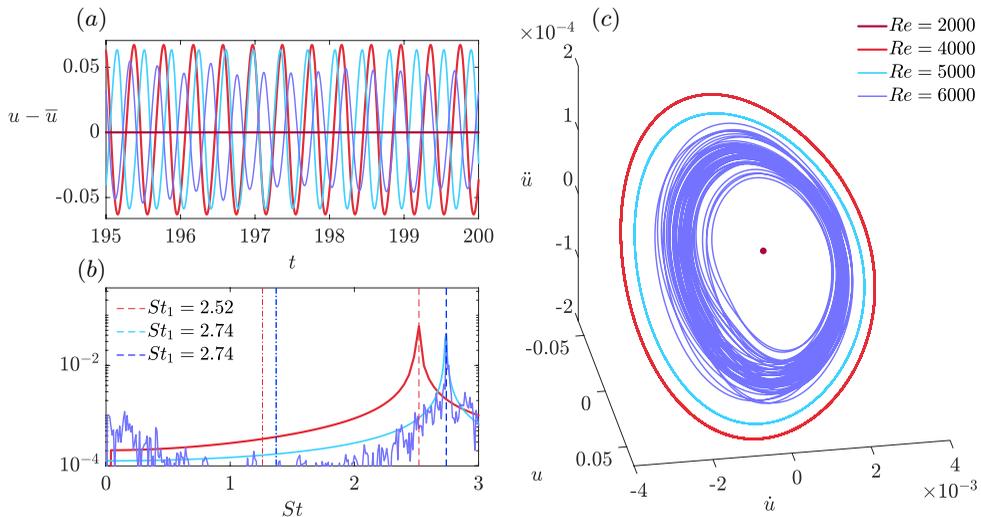
\begin{figure}
    \centerline{
    \begin{tikzpicture}
    \node at (0.0,3.5) {\includegraphics[trim={0 0 0 0},clip,width=0.95\textwidth]{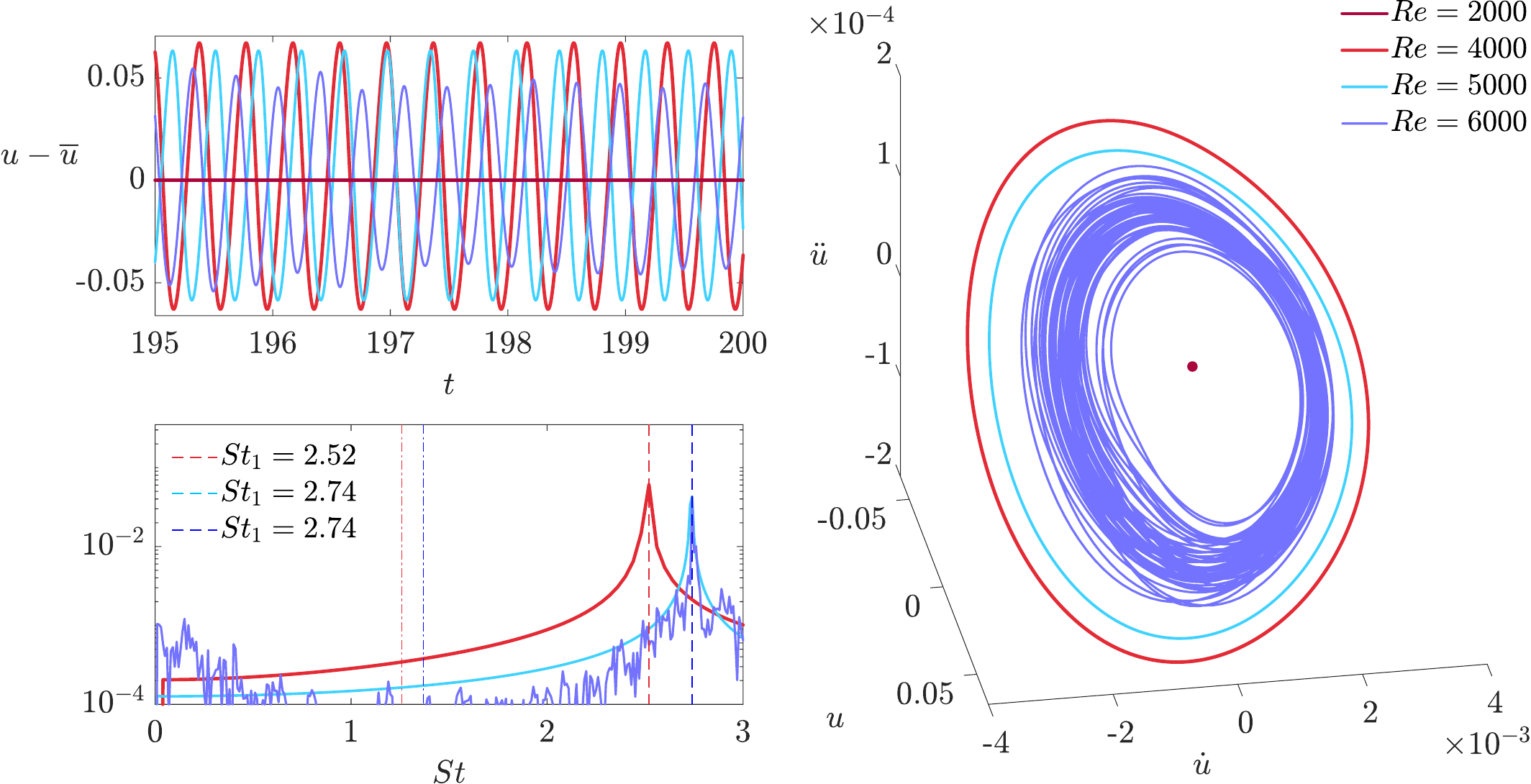}};
    \node at (-5.3,6.75) {$(a)$};
    \node at (-5.3,3.45) {$(b)$};
    \node at (1.5,6.75)  {$(c)$};
    \end{tikzpicture}
    }    
    \caption{As figure \ref{fig:flow_attractors_AoA-1.5} for $\alpha=3 ^\circ$.}    
    \label{fig:flow_attractors_AoA3}
\end{figure}

For $0^\circ \leq \alpha \leq 7^\circ$ the secondary instability consists of a 3D pitchfork bifurcation of the 2D periodic limit cycle; see the light blue right-oriented triangles. At $Re = Re_{c2}$ the wake becomes 3D, but retains the same temporal periodicity as at smaller $Re$.
Figure \ref{fig:flow_attractors_AoA3} considers $\alpha = 3^\circ$ and illustrates the velocity signal for different $Re$.
At $\Rey = 4000$ the 2D periodic limit cycle is characterised by a Strouhal number of $St \approx 2.52$.
At $\Rey = 5000$ the spectrum still presents a single peak at $St \approx 2.74$, and the wake becomes 3D.
While retaining the same (slightly larger) dominant frequency $St \approx 2.74$, an increase of $Re$ to $\Rey = 6000$ yields the emergence of low-frequency beating and the loss of the temporal periodicity of the 3D wake.

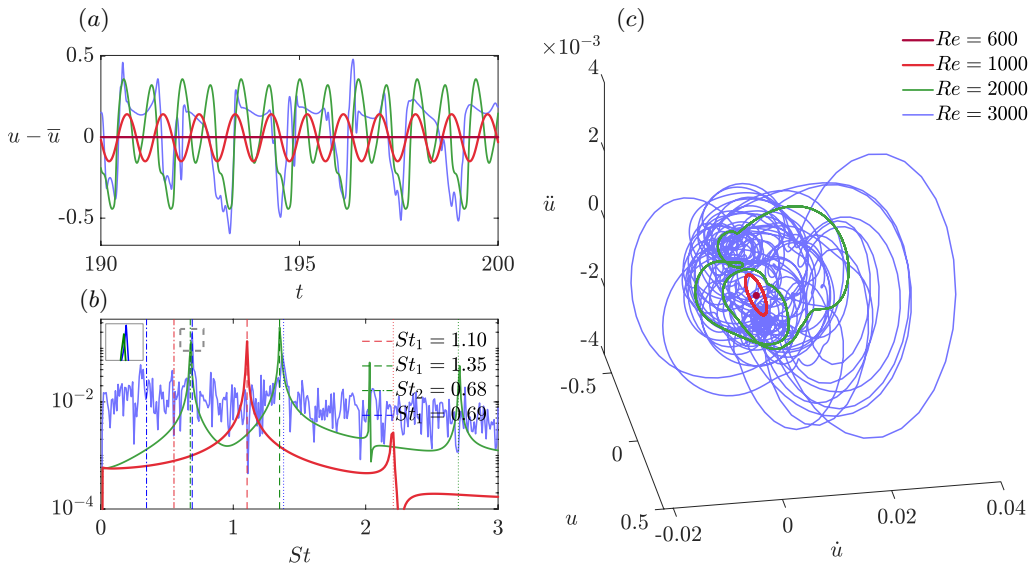
\begin{figure}
    \centerline{
    \begin{tikzpicture}
    \node at (0.0,3.5) {\includegraphics[trim={0 0 0 0},clip,width=\textwidth]{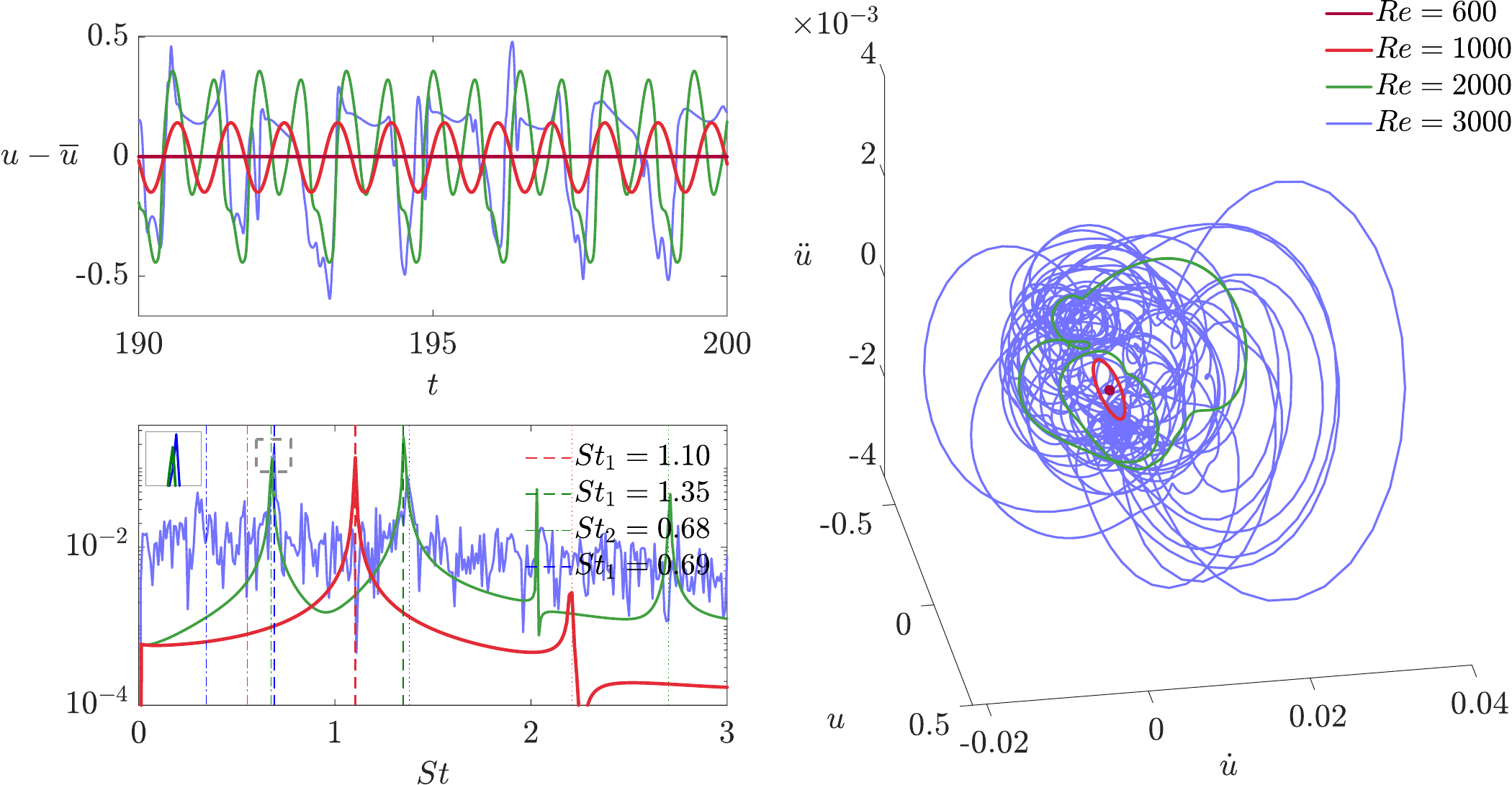}};
    \node at (-5.6,7.15) {$(a)$};
    \node at (-5.6,3.45) {$(b)$};
    \node at (1.5,7.15)  {$(c)$};
    \end{tikzpicture}
    }      
    \caption{As figure \ref{fig:flow_attractors_AoA-1.5} for $\alpha=10 ^\circ$.}
    \label{fig:flow_attractors_AoA10}
\end{figure}

For large positive $\alpha$ (i.e. $\alpha \geq 8^\circ$) the scenario is different; see figure \ref{fig:flow_attractors_AoA10} that illustrates the velocity signal at $\alpha = 10^\circ$ at different $Re$.
In this case the flow remains 2D and periodic for $Re \lesssim 2000$ with a frequency of $St \approx 1$. At $Re = 2000$ the flow becomes 3D and a new peak emerges in the spectrum at half of the vortex-shedding frequency, \ie $St_2 = St_1/2 = 0.68$. For $\alpha \ge 8^\circ$, indeed, the secondary instability consists of subharmonic 3D bifurcation of the 2D periodic limit cycle; see the green left-oriented triangles. Then, when $Re$ is increased up to $Re = 3000$ the flow loses the periodicity, and the dominant frequency becomes the subharmonic one, i.e. $St \approx 0.69$ (as evidenced by the close-up view in figure \ref{fig:flow_attractors_AoA10}b). 

%%%%%%%%%%%%%%%%%%%%%%%%%%%%%%%%%%%%%%%%%%%%%%%%%%%%%%%%%%%%%%%%%%%%%%%%%%%%%%%%%%%%%%%

\section{Low $\Rey$: the primary bifurcation}
\label{sec:prim-bif}

In this section we detail the primary bifurcation of the low-$\Rey$ steady flow that leads to the 2D and periodic regime. Both negative and positive $\alpha$ are considered, i.e. $-5^\circ \le \alpha \le 10^\circ$.

\subsection{The steady base flow}
\label{subsec:steady_bf}

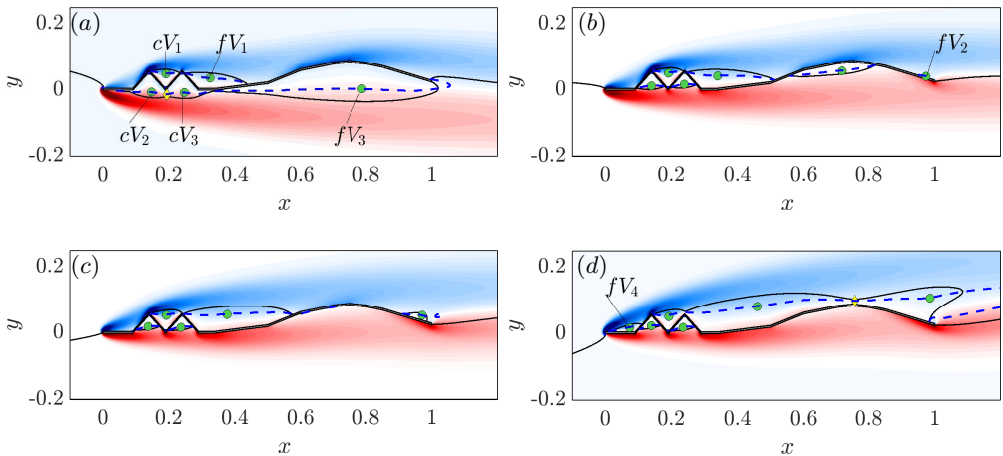
\begin{figure}
\centerline{
\begin{tikzpicture}
\node at (-0.15,8.4) {\includegraphics[trim={0 0 60 30},clip,width=0.49\textwidth]{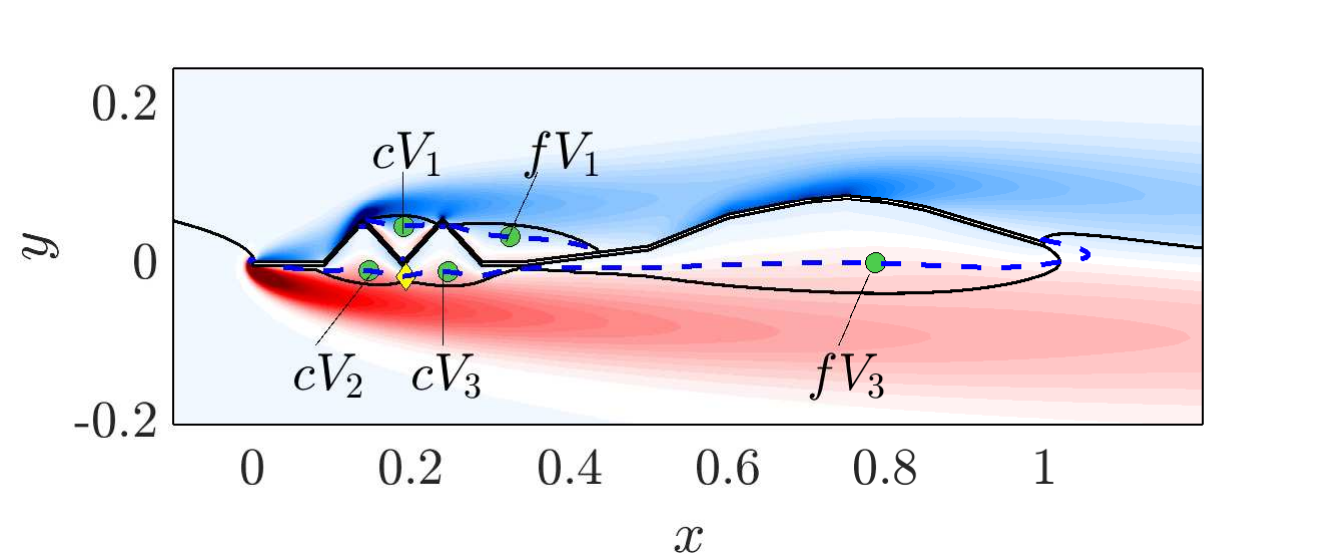}};
\node at (6.5,8.4) {\includegraphics[trim={0 0 60 30},clip,width=0.49\textwidth]{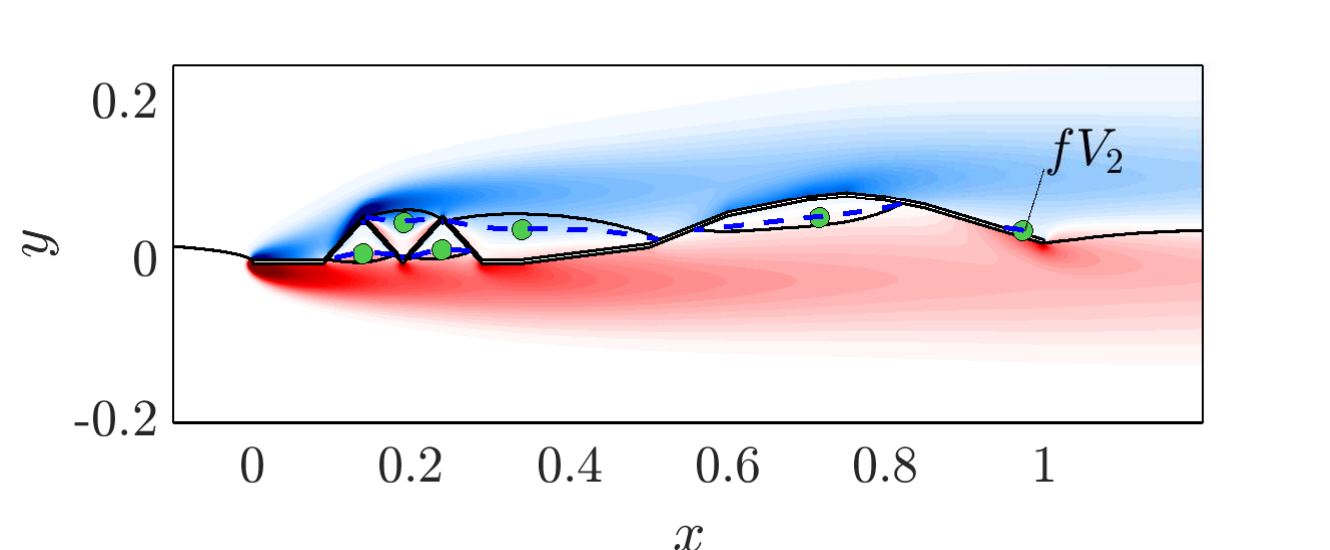}};
\node at (-0.15,5.2) {\includegraphics[trim={0 0 60 30},clip,width=0.49\textwidth]{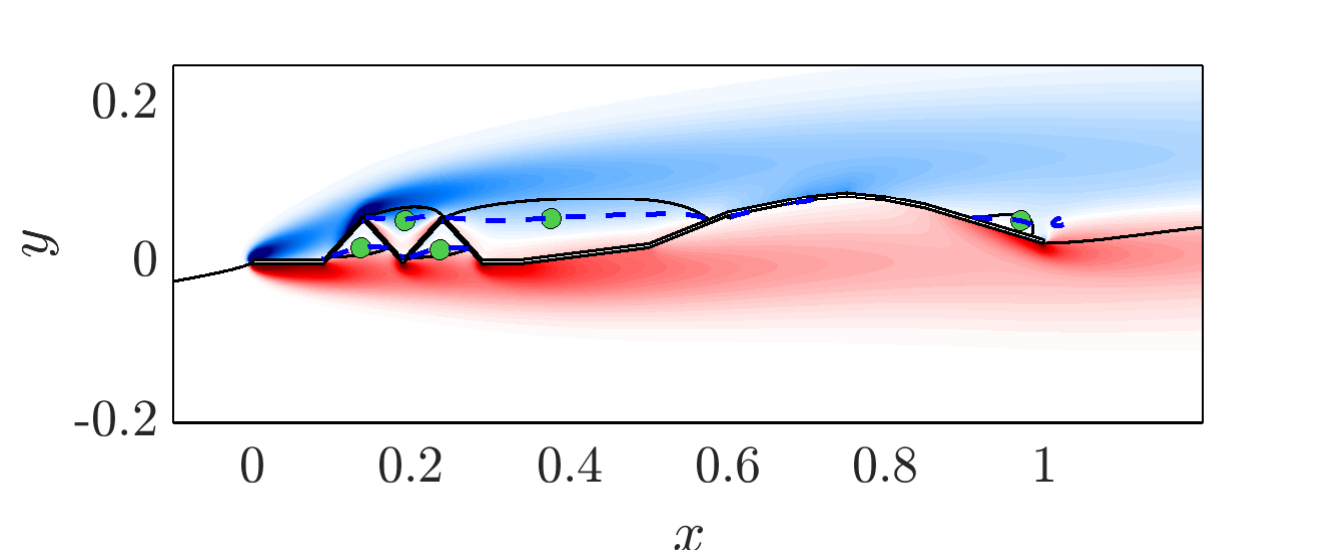}};
\node at (6.5,5.2) {\includegraphics[trim={0 0 60 30},clip,width=0.49\textwidth]{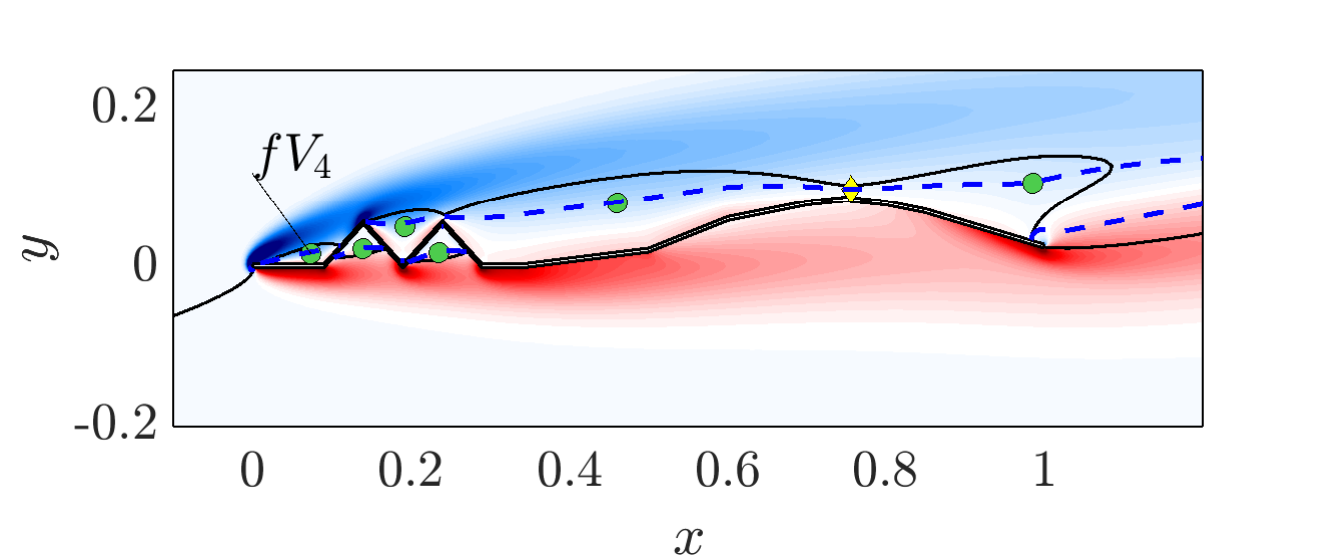}};
\node at (-2.3,9.5) {$(a)$};
\node at (4.4,9.5)  {$(b)$};
\node at (-2.3,6.3) {$(c)$};
\node at (4.4,6.3)  {$(d)$};
\end{tikzpicture}
}
\caption{Base flow near the first bifurcation at $\Rey=800$. Streamlines are superimposed on the map of the spanwise vorticity $\Omega_{z,1} = \partial V_1/\partial x - \partial U_1/\partial y$, with the blue-to-red colour map in the $-50 \le \Omega_{z,1} \le 50$ range. Panel $(a)$: $\alpha=-5^\circ$. Panel $(b)$: $\alpha=0^\circ$. Panel $(c)$: $\alpha=5 ^\circ$. Panel $(d)$: $\alpha=10^\circ$. Green circles are for the elliptic stagnation points. Yellow diamonds are for the hyperbolic stagnation points. The blue dashed line is for $U_1=0$. The tags of the recirculating regions (see text) are introduced only once, in the first panel the recirculating regions appear.}
\label{fig:BF1}
\end{figure}

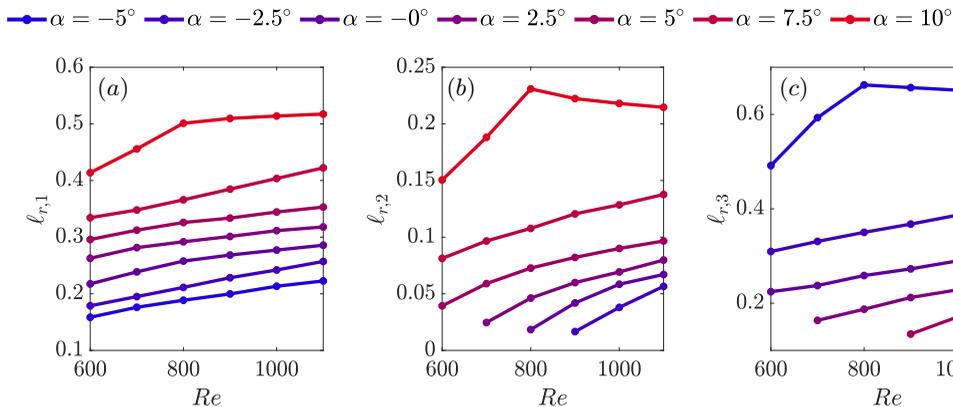
\begin{figure}
\centerline{
\begin{tikzpicture}
\node at (3.65,4) {\includegraphics[trim={0 10 70 0},clip,width=0.99\textwidth]{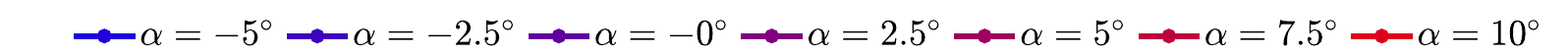}};
\node at (0,1.2) {\includegraphics[width=0.325\textwidth]{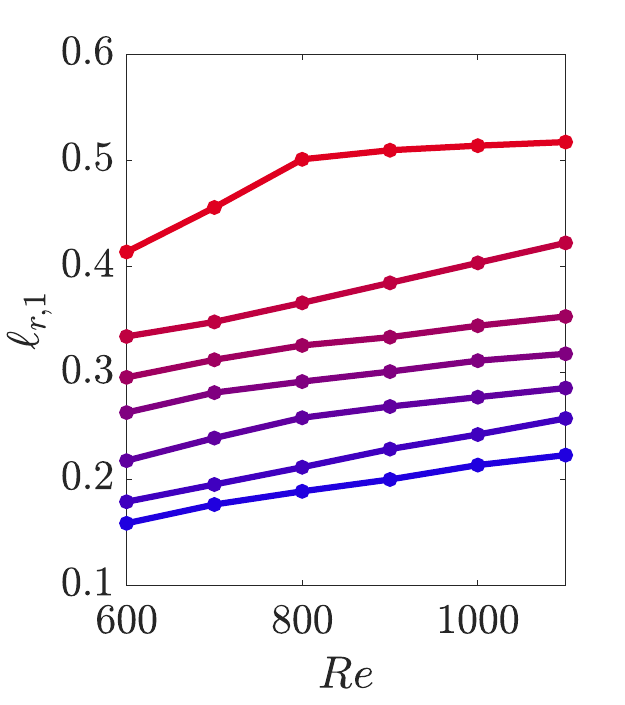}};
\node at (4.5,1.2) {\includegraphics[width=0.325\textwidth]{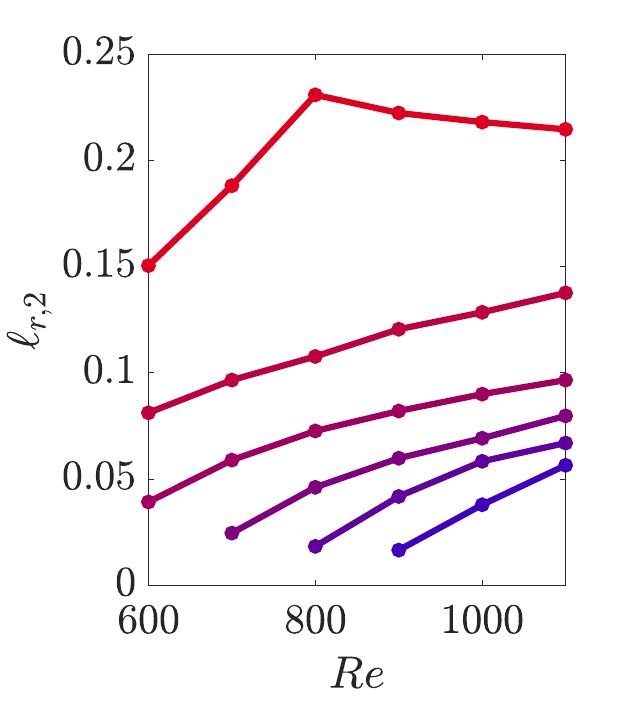}};
\node at (9.0,1.2) {\includegraphics[width=0.325\textwidth]{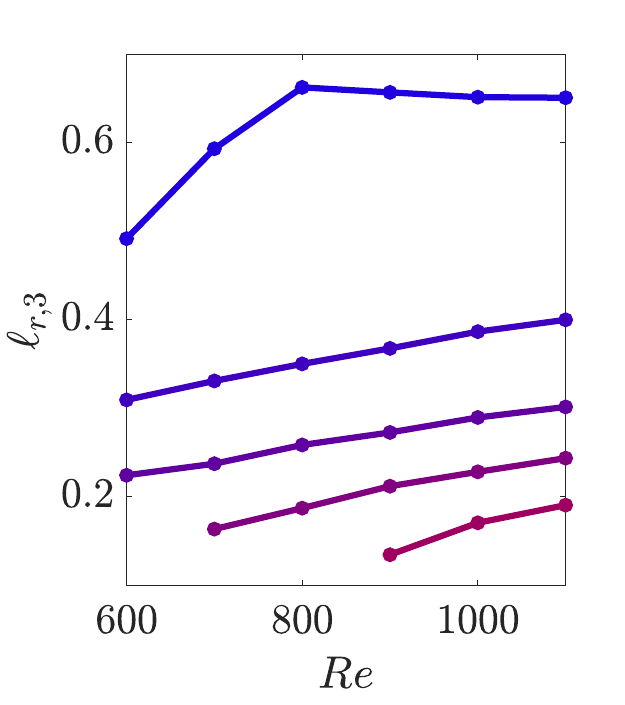}};
\node at (-1,3)  {$(a)$};
\node at (3.6,3) {$(b)$};
\node at (8.0,3) {$(c)$};
\end{tikzpicture}
}
\caption{Dependence of the length of the main base flow recirculating regions on $\Rey$. See text for the definition of $\ell_{r,1}$, $\ell_{r,2}$ and $\ell_{r,3}$. The size of the recirculating regions are measured using the delimiting $\Psi=0$ line; when the $\Psi=0$ line delimits two recirculating regions (see for example the bottom right panel of figure \ref{fig:BF1}), the position of the ensuing hyperbolic stagnation point is used as delimiting point.}
\label{fig:BF1_rec}
\end{figure}

Figure \ref{fig:BF1} shows the streamlines and vorticity $\Omega_{z,1}=\partial V_1/\partial x- \partial U_1/\partial y$ colour maps of the low-$Re$ steady base flow at $\Rey=800$ for $\alpha=-5^\circ,0^\circ,5^\circ,10^\circ$. The stagnation points are marked with symbols. Green circles are used for the elliptical stagnation points corresponding to a local maximum or minimum of the stream function $\Psi_1$, defined as $\bm{\nabla}^2 \Psi_1 = - \Omega_{z,1}$, and are used to detect the centre of rotation of the recirculating regions. Yellow diamonds, instead, refer to hyperbolic stagnation points corresponding to saddle points of $\Psi_1$. Quantitative information is provided in figure \ref{fig:BF1_rec}. 

Due to the complex geometry, the flow separates in several points, giving origin to many recirculating regions. Some recirculating regions are present at all $\alpha$, and their size only slightly changes with $Re$. Others instead, originate only for some values of $\alpha$ and $\Rey$, as they are related to pressure gradients that may be stronger/milder and adverse/favourable depending on the flow configuration. We refer to the recirculating regions with size that depends only on the geometry as \textit{constrained} recirculating regions ($cV$). The recirculating regions whose size and occurrence depend on $\alpha$ and $\Rey$, instead, are referred to as \textit{free} recirculating regions ($fV$).

Along the top side of the body, a constrained recirculating region ($cV_1$) is detected close to the leading edge (LE), within the cavity delimited by the two corrugations. Moving downstream, the flow separates from the corner of the downstream groove, and reattaches along the body: a large free recirculating region $fV_1$ arises, with size $\ell_{r1}$ that monotonically increases with $\Rey$ and $\alpha$ (see figure \ref{fig:BF1_rec}). 
Further moving downstream, the flow faces an adverse pressure gradient and, for some values of $\alpha$ and $\Rey$, it separates and gives rise to a second free recirculating region $fV_2$ close to the trailing edge (TE). The separating point of $fV_2$ moves downstream as $\alpha$ and/or $\Rey$ increase, and its size $\ell_{r,2}$ monotonically increases with $\alpha$ and $\Rey$ (see figure \ref{fig:BF1_rec}). Moving to the bottom side, two constrained recirculating regions ($cV_2$ and $cV_3$) arise close to the LE within the two grooves. Moving downstream, a further large recirculating region ($fV_3$) arises depending on $\Rey$ and $\alpha$. Its size $\ell_{r,3}$ increases with $\Rey$ but decreases with $\alpha$ (see figure \ref{fig:BF1_rec}): for large positive $\alpha$ and small enough Reynolds numbers it is not detected. Note that over the top side of the body, a further free recirculating ($fV_4$) region arises for large $\alpha$ and $\Rey$, in the area upstream the first corrugation.

When fixing $\alpha$, the increase of the size of the free recirculating regions with $Re$ is consistent with what observed for 2D and 3D bluff bodies \citep[see for example][]{giannetti-luchini-2007,MarquetLarsson2015}. 

\subsection{The critical Reynolds number and the global modes}
\label{subsec:pri-bif}

\begin{figure}
    \centerline{
    \begin{tikzpicture}
    \node at (0,0) {\includegraphics[trim={0 0 0 0},clip,width=\textwidth]{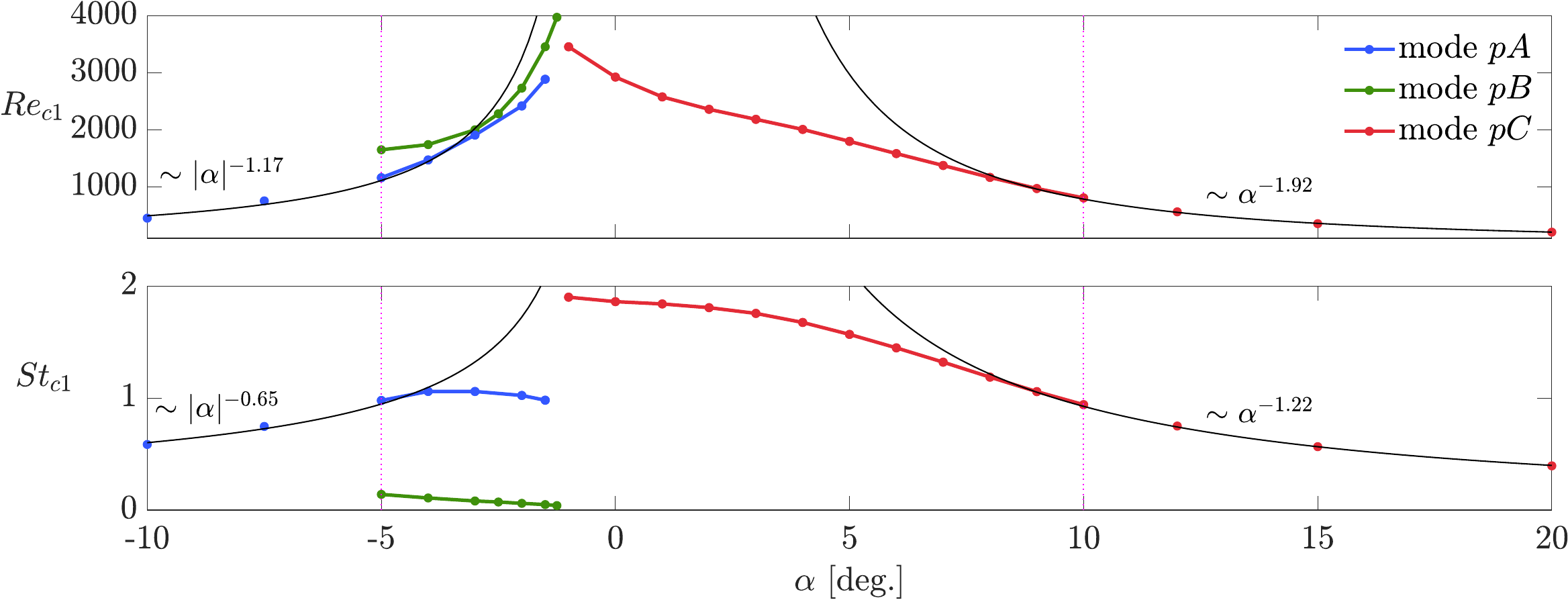}};
    \node at (-5,2.2) {$(a)$};
    \node at (-5,-0.17) {$(b)$};
    \end{tikzpicture}
    }
    \caption{Panel $(a)$: critical Reynolds number for $-10^\circ \le \alpha \le 20^\circ$. Panel $(b)$: critical frequency of the primary bifurcations.}
    \label{fig:marginal_curves}
\end{figure}

\begin{figure}
\centerline{
\begin{tikzpicture}
\node at (-0.15,8.4) {\includegraphics[width=0.49\textwidth]{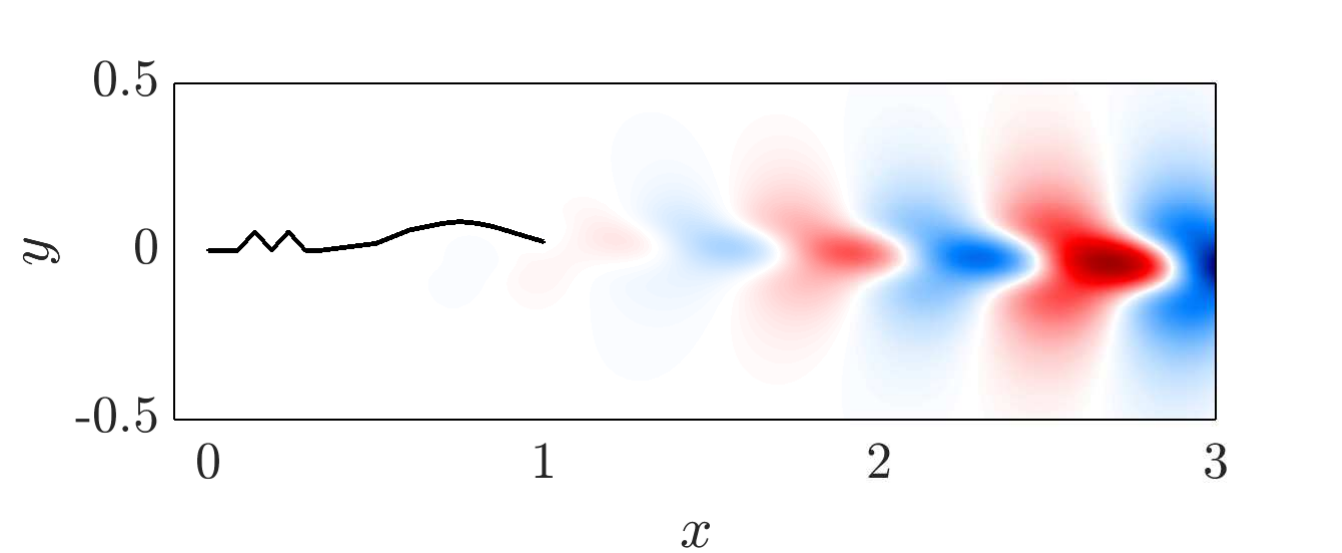}};
\node at (6.5,  8.4) {\includegraphics[width=0.49\textwidth]{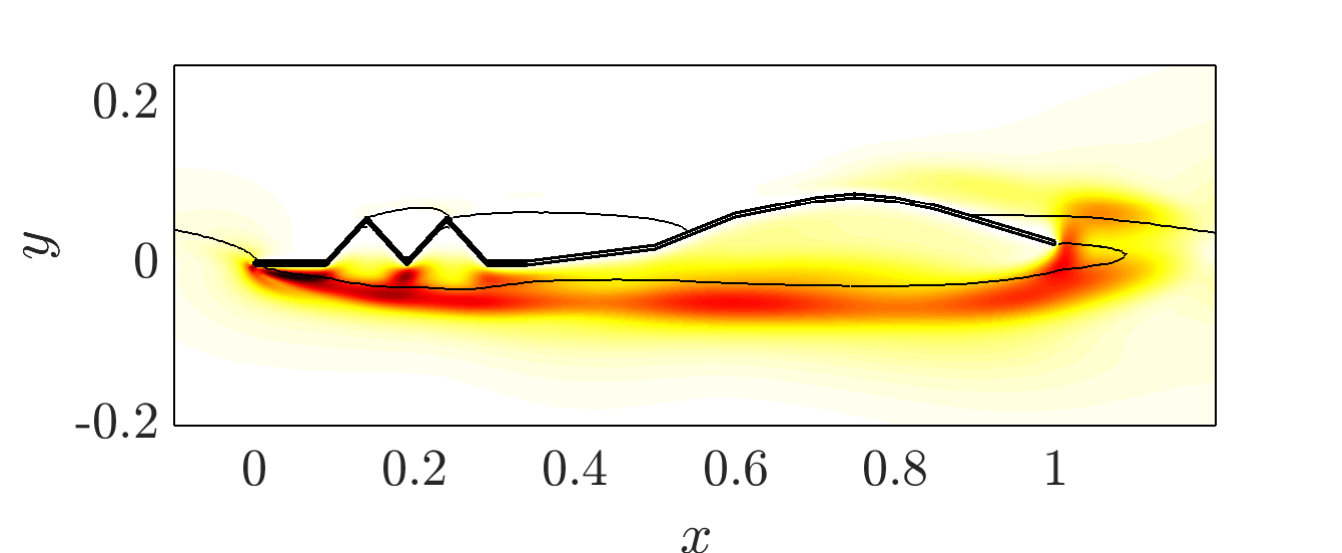}};
\node at (-0.15,5.5) {\includegraphics[width=0.49\textwidth]{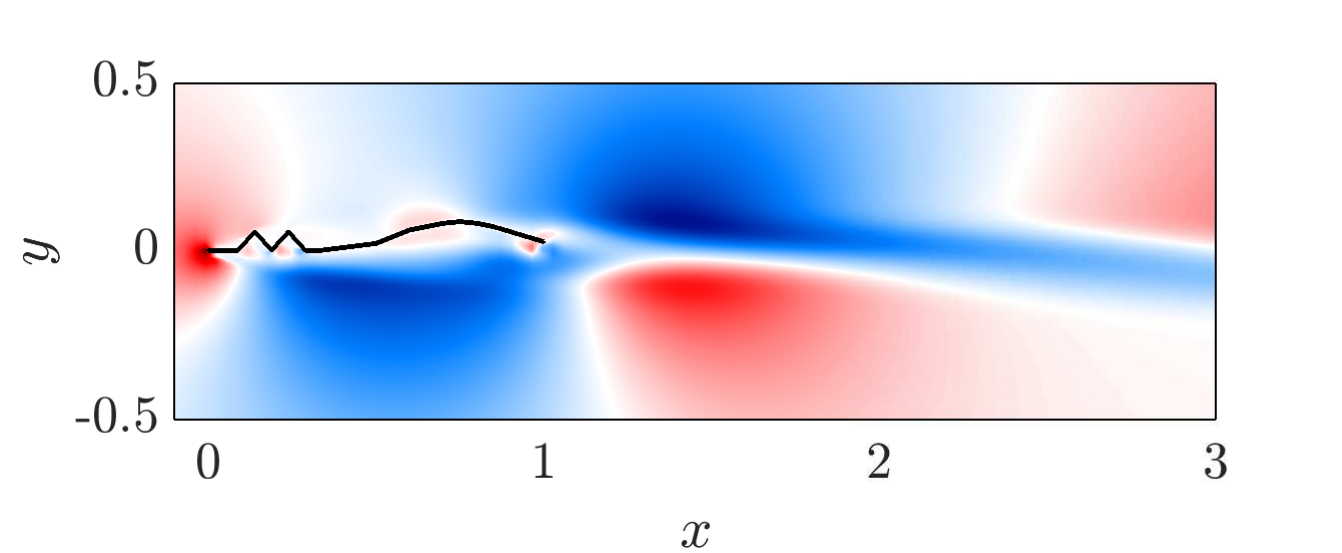}};
\node at (6.5,  5.5) {\includegraphics[width=0.49\textwidth]{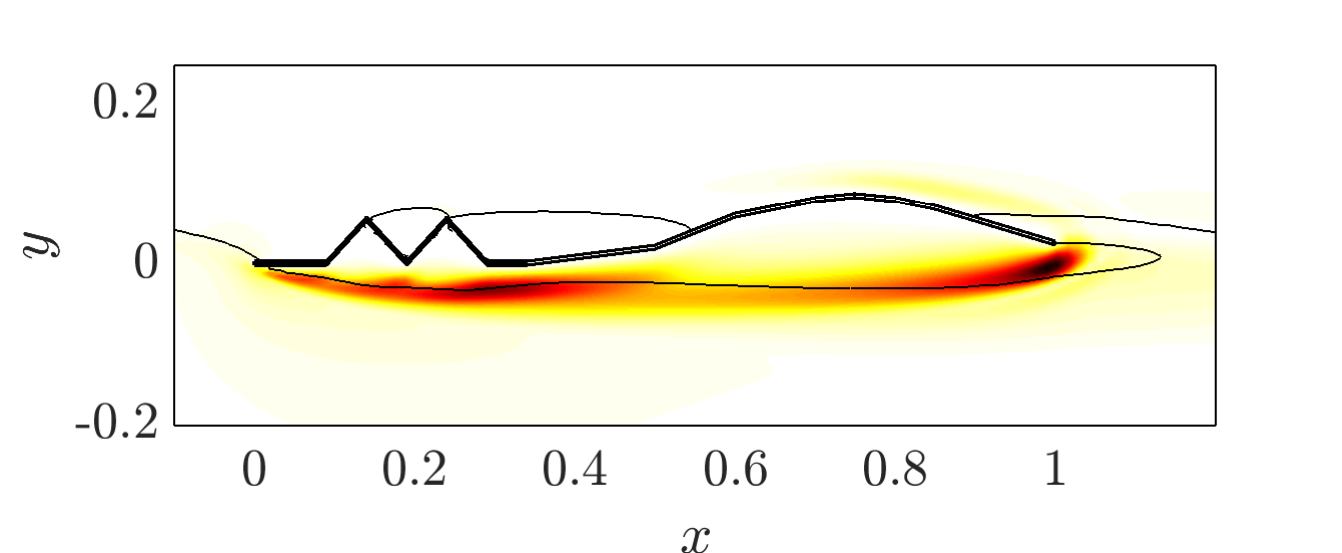}};
\node at (-0.15,2.6) {\includegraphics[width=0.49\textwidth]{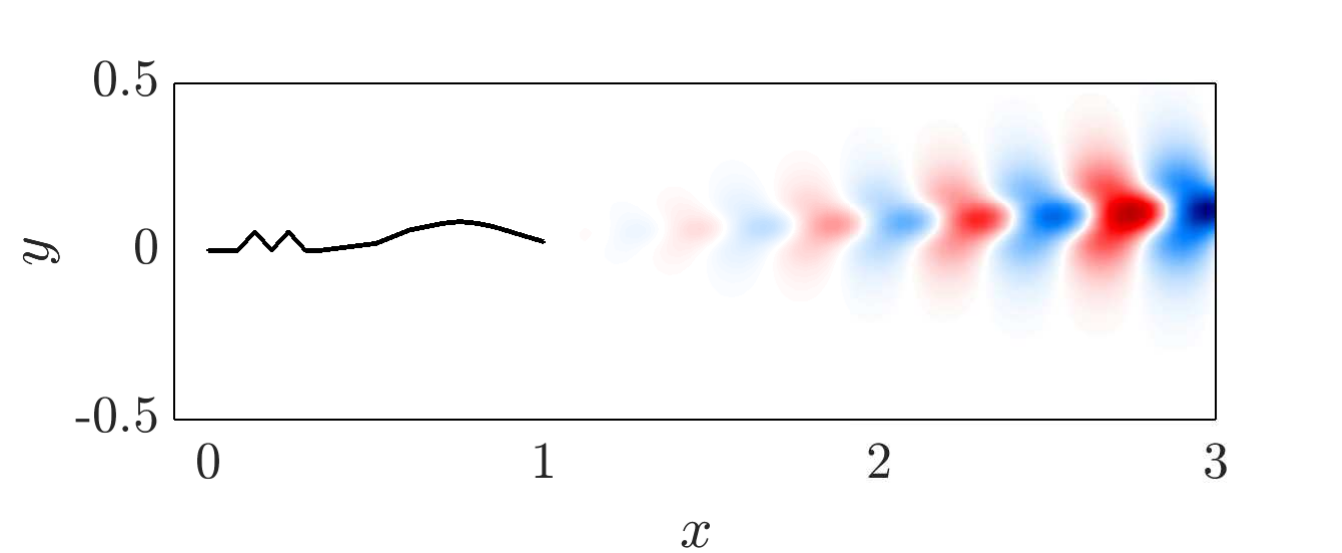}};
\node at (6.5,  2.6) {\includegraphics[width=0.49\textwidth]{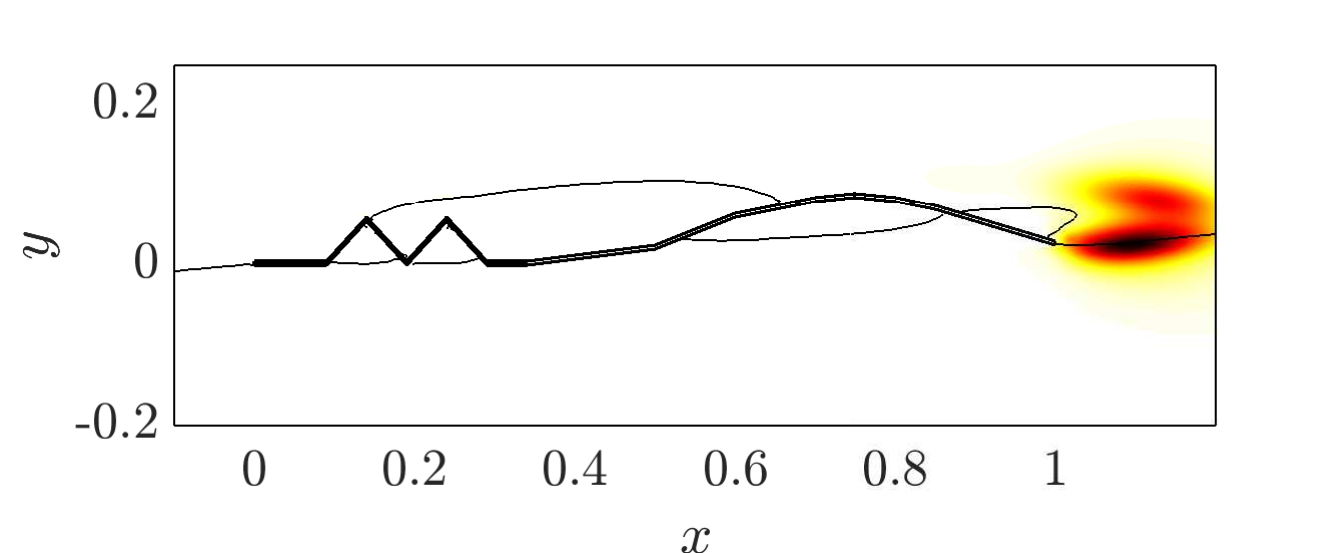}};
\node at (-2.3,9.1) {$(a)$};
\node at (4.35,9.1) {$(b)$};
\node at (-2.3,6.2) {$(c)$};
\node at (4.35,6.2) {$(d)$};
\node at (-2.3,3.3) {$(e)$};
\node at (4.35,3.3) {$(f)$};
\end{tikzpicture}
}
\caption{Spatial structure of the leading eigenmodes of the low-$\Rey$ steady base flow at $\Rey=\Rey_{c1}$. Left panels $(a,c,e)$: $\hat{v}_1$. Right panels $(b,d,f)$: structural sensitivity. Top panels $(a,b)$: mode $pA$ for $\alpha = -3^\circ$. Central panels $(c,d)$: mode $pB$ for $\alpha=-3^\circ$. Bottom panels $(e,f)$: mode $pC$ for $\alpha = 3^\circ$. The spatial structure of the $pA$ and $pB$ modes ($pC$ mode) does not change qualitatively with $\alpha$ for $-5^\circ \le \alpha \le -1.25^\circ$ ($-1.25 < \alpha \le 10^\circ$).}
\label{fig:eigenmode}
\end{figure}
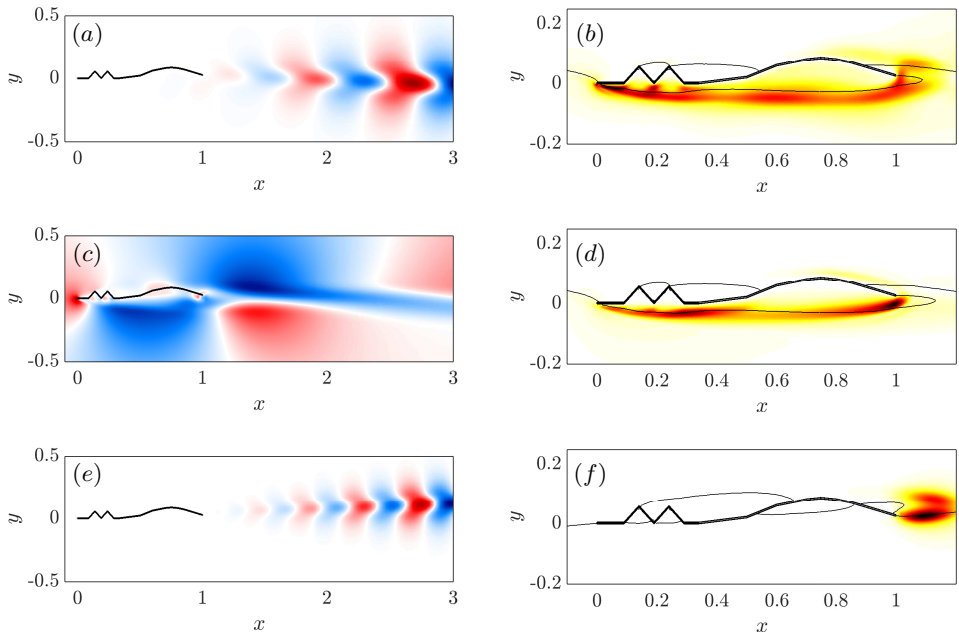

We now move to the results of the linear stability analysis. For each $\alpha$, we consider the steady base flow and look for the growing eigenmodes as $\Rey$ increases. 
Figure \ref{fig:marginal_curves} summarises the effect of $\alpha$ on the first bifurcation; figure \ref{fig:marginal_curves}$(a)$ shows the neutral curves, while figure \ref{fig:marginal_curves}$(b)$ depicts the dependence of the critical frequency of the unstable modes on $\alpha$. As mentioned in \S \ref{sec:flow-reg}, for all $\alpha$ the first instability consists of a Hopf bifurcation towards a 2D and oscillatory state, characterised by an alternating vortex-shedding in the wake. Three unstable modes are detected, i.e. modes $pA$, $pB$ and $pC$, with $pA$ and $pC$ being the leading modes for $\alpha \le -1.25^{\circ}$ and $\alpha > -1.25^\circ$, respectively. The spatial structure of the modes is shown in figure \ref{fig:eigenmode}$(a,c,e)$.

For $\alpha \le -1.25^{\circ}$, two bifurcations of the steady base flow are found to occur in short sequence. They correspond to the oscillatory modes $pA$ and $pB$, which are characterised by a different frequency, i.e. $f_{{c1}_{pA}} \approx 1$ and $f_{{c1}_{pB}} \approx 0.07$. For all negative $\alpha$, mode $pA$ exhibits a positive growth rate $\sigma_1$ for smaller Reynolds numbers with respect to mode $pB$, i.e. it emerges first as primary instability. For more negative $\alpha$, the critical Reynolds number of both modes $pA$ and $pB$ decreases and the flow becomes less stable. This is consistent with the increase of the size of the $fV_3$ recirculating region (where the mechanism triggering the instability is placed; see the following discussion) and with the stronger reverse flow that arises in it. In fact, the size of the recirculating region dictates the spatial extent of the absolute instability pocket \citep{chomaz-2005}, and the intensity of the reverse flow within it impacts the local amplification of the growing wave packets \citep{hammond-redekopp-1997}. The spatial structure of the mode $pB$ is largely different from that of the K\'{a}rm\'{a}n-like mode $pA$ (see figure \ref{fig:eigenmode}).
Unlike modes $pA$, and what observed for the flow past other 2D bluff bodies \citep{giannetti-luchini-2007,chiarini-quadrio-auteri-2021a}, the magnitude of mode $pB$ is large also close to the body sides, particularly within the $fV_3$ recirculating region. 
We anticipate that, besides $pB$ is an amplified mode of the low-$\Rey$ steady flow and becomes amplified for $\Rey>\Rey_{c1}$ only, it actually emerges also in the nonlinear unsteady simulations at Reynolds numbers close to those predicted by the linear stability analysis (see figure \ref{fig:St_Re-alpha_plane} and \S\ref{sec:sec-bif}). 

For $\alpha > -1.25 ^\circ$ the scenario changes and the leading mode is mode $pC$. This agrees with a substantial change of the base flow (see figure \ref{fig:BF1_rec}).
 Like mode $pA$ for negative $\alpha$, also mode $pC$ is oscillatory and leads to a vortex shedding in the wake. In this case, the frequency is larger and at criticality, i.e. for $Re=Re_{c1}(\alpha)$, decreases with $\alpha$, being $f_{{c1}_{pC}} \approx 2$ for small $\alpha$ and $f_{{c1}_{pC}} \approx 1$ for $\alpha \approx 10^\circ$. The critical Reynolds number decreases when increasing $\alpha$, in agreement with the enlargement of the $fV_2$ recirculating region. In the wake, the spatial structure of the mode resembles that of mode $pA$, but the streamwise wavelength of the wake is smaller in agreement with the larger shedding frequency (see figure \ref{fig:eigenmode}).

For large positive and negative $\alpha$ the critical Reynolds number $\Rey_{c1}$ for modes $pA$ and $pC$ scales as a power law of $\alpha$, similarly to what found by other authors for smooth NACA airfoils \citep{nastro-etal-2023,gupta-etal-2023}. For mode $pA$ we find $\Rey_{c1} \sim |\alpha|^{-1.17}$, while for mode $pC$ $\Rey_{c1} \sim \alpha^{-1.92}$. A power law has been observed also for the associated frequencies at criticality, being $f_{c1} \sim |\alpha|^{-0.65}$ for mode $pA$ and $f_{c1} \sim \alpha^{-1.22}$ for mode $pC$.

In passing, note that the stabilisation of the flow for small values of $|\alpha|$ is consistent with previous results in literature, which report a monotonic increase of the critical Reynolds number with the streamwise extent of the bodies; see for example \cite{chiarini-quadrio-auteri-2022} and \cite{jackson-1987-finiteelementstudy} for 2D symmetric bodies, and \cite{zampogna-boujo-2023} and \cite{chiarini-boujo-2024} for 3D rectangular prisms. In fact, for small $|\alpha|$ the characteristic size of the body, once projected in the streamwise direction, increases.

We now look at the structural sensitivity of the modes $pA$, $pB$ and $pC$. Figure \ref{fig:eigenmode}$(b,d,f)$ shows the map of
\begin{equation}
S(\bm{x}) = \frac{ \Vert \hat{\bm{u}}_1 (\bm{x}) \Vert  \ \Vert \hat{\bm{u}}_1^{\dag} (\bm{x}) \Vert  }{ \int_\Omega ( \hat{\bm{u}}_1^{\dag*} \cdot \hat{\bm{u}}_1 )\text{d}\Omega},
\end{equation}
where $\hat{\bm{u}}_1^{\dag}$ is the adjoint mode and the $\cdot^*$ superscript indicates the complex conjugate. The structural sensitivity was introduced by \cite{giannetti-luchini-2007} as an upper bound for the eigenvalue variation $|\delta \gamma_1|$ induced by a specific perturbation of the LNS operator. It is a ‘‘force-velocity coupling’’ representing a feedback from a localised velocity sensor to a localised force actuator at the same location. The structural sensitivity is thus an indicator of the eigenvalues sensitivity and identifies the wavemaker \citep{monkewitz-huerre-chomaz-1993}. The most sensitive regions, where the direct and adjoint modes overlap, are found close to the body for all three modes, which is typical of 2D and 3D bluff body wakes; see \cite{giannetti-luchini-2007}, \cite{zampogna-boujo-2023} and \cite{chiarini-boujo-2024}. 

Modes $pA$ and $pB$ are most sensitive along the streamline that separates from the bottom LE corner. In both cases the structural sensitivity does not show a localised peak, but it is large along a relatively wide region suggesting the existence of a non local feedback in the ensuing triggering mechanism. For mode $pA$ the structural sensitivity peaks close to the LE, while for mode $pB$ the maximum value is found close to the TE. Also, downstream the TE, the structural sensitivity of mode $pA$ resembles what found for the flow past 2D bluff bodies \citep{giannetti-luchini-2007}, with the classical structure with two symmetric lobes placed in correspondence of the $\Psi_1=0$ line. Mode $pC$, instead, is most sensitive in the region just downstream the TE, with the classical double-lobed conformation. For this mode, the intensity of the structural sensitivity is not symmetric being larger within the bottom lobe. This is typical for non symmetric configurations \citep[see also][]{chiarini-auteri-2023,nastro-etal-2023}, and indicates that the instability is mainly triggered by the bottom shear layer where the flow faces a larger acceleration. 
For mode $pC$ the structural sensitivity is negligible over the lateral sides of the body, indicating that the $fV_1$ and $fV_3$ recirculating regions do not play a significant role in the triggering mechanism, and that it is a leading mode of the wake. 

The different structural sensitivities found for modes $pA$ and $pC$ indicate that the physical mechanism that drives these two instabilities is not the same: it appears that for mode $pA$ the mechanism is not confined in the near wake, but involves also the shear layer that separates from the bottom LE corner. For negative $\alpha$, indeed, the flow separates at the LE and reattaches at the TE, resulting in a ``long-type" bubble that encompasses the entire bottom side of the airfoil. On the other hand, for mode $pC$ the absolute instability pocket is embedded in the $fV_2$ recirculating region in the near wake; for $\alpha \ge -1.25$ the flow does not separate at the bottom LE corner. Accordingly, the characteristic temporal frequencies of the two modes are distinct and exhibit a different dependence on $\alpha$. The frequency of mode $pA$ is lower ($f \approx 1$) and only marginally varies with $\alpha$; for $\alpha<0$ the extent of the ``long-type" bubble does not depend on $\alpha$. The frequency of mode $pC$, instead, is larger ($1 \lessapprox f \lessapprox 2$) and decreases for larger $\alpha$; the $fV_2$ recirculating region enlarges as $\alpha$ increases (see figure \ref{fig:BF1_rec}). 

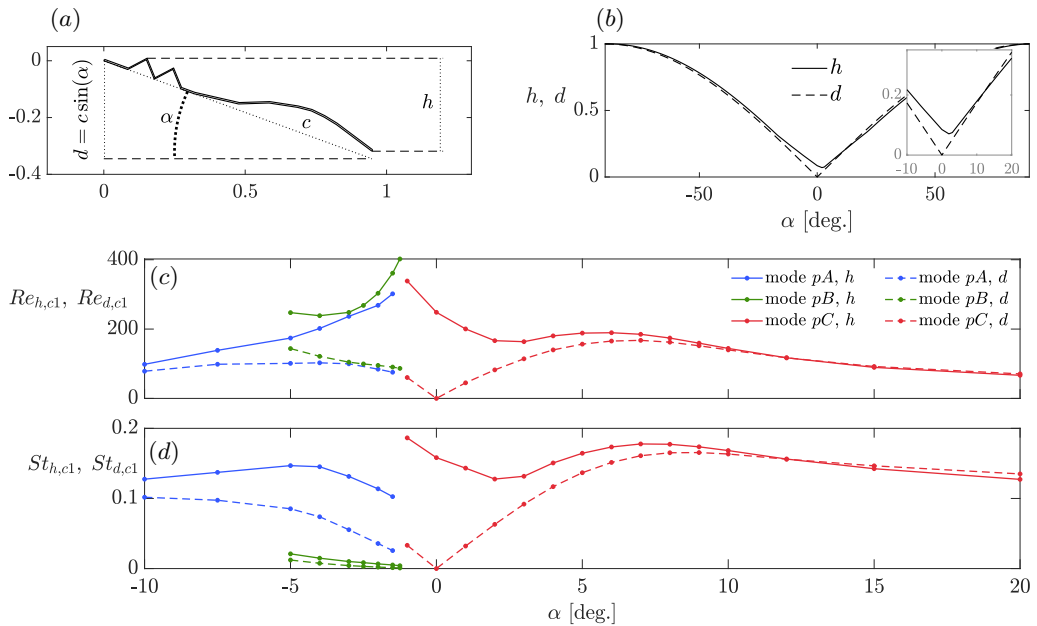
\begin{figure}
\centerline{
\begin{tikzpicture}
\node at (0,4) {\includegraphics[width=\textwidth]{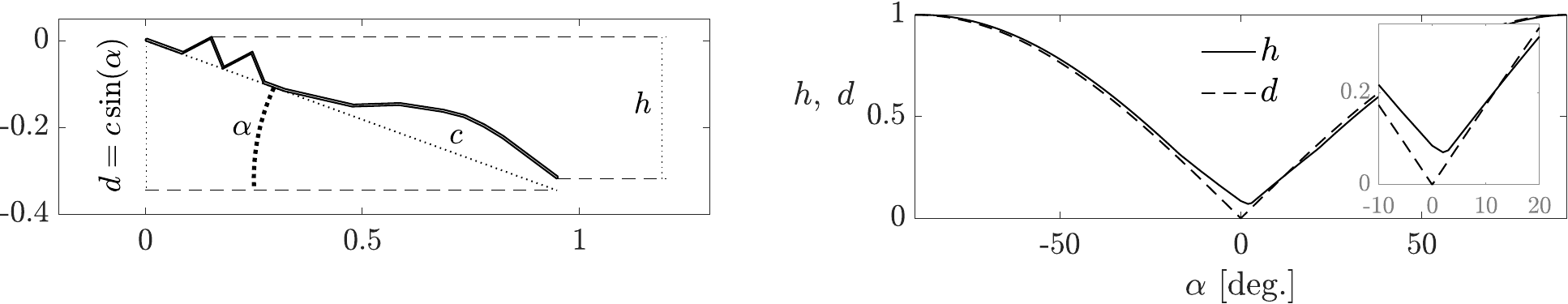}};
\node at (0,0) {\includegraphics[width=\textwidth]{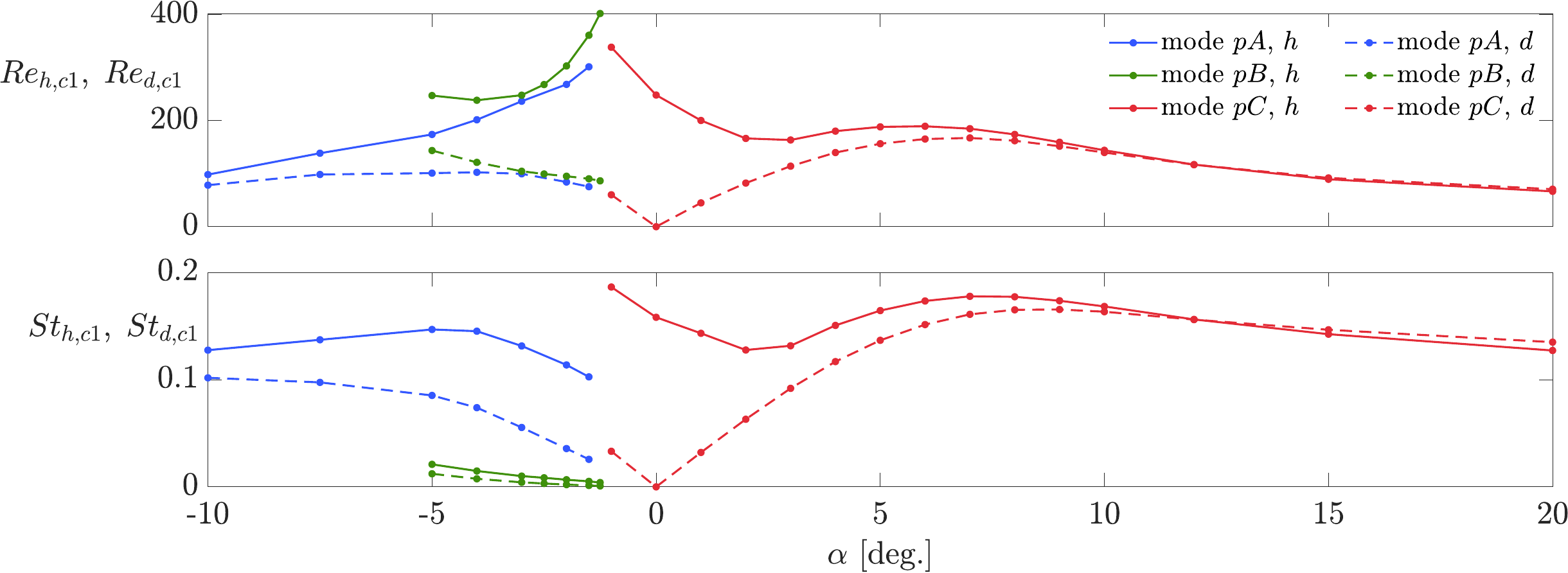}};
\node at (-6,5.5) {$(a)$};
\node at (1.2,5.5) {$(b)$};
\node at (-4.7, 2.1)  {$(c)$};
\node at (-4.7,-0.2){$(d)$};
\end{tikzpicture}
}
\caption{Neutral curves for modes $pA$, $pB$ and $pC$ and corresponding frequency at criticality using alternative definitions of the Reynolds and Strouhal numbers, based on two different measures ($d$ and $h$) of the vertical extent of the body. Panel $(a)$ sketches how $d$ and $h$ are defined. Panel $(b)$ shows the dependence of $d$ and $h$ on $\alpha$. Panels $(c)$ and $(d)$ are as figure \ref{fig:marginal_curves}, but for $Re_d = U_\infty d/\nu$, $St_d = f d/U_\infty$ and $Re_h = U_\infty h / \nu$, $St_h = f h/ U_\infty$.}
\label{fig:neutral_modified}
\end{figure}
Figure \ref{fig:neutral_modified} shows the neutral curves of modes $pA$, $pB$ and $pC$ using alternative definitions of the Reynolds and Strouhal numbers that, in analogy to what is often done in the context of low-$Re$ flows over smooth airfoils and flat plates \citep[see for example][]{fage-johansen-1927}, are based on a measure of the frontal dimension of the body. Here we use $d =c \sin(\alpha)$ and $h$, i.e. the cross-stream projection of the chord and of the airfoil height \citep{levy-seifert-2009}, (see figure \ref{fig:neutral_modified}$(a)$) and introduce the alternative Reynolds and Strouhal numbers as $Re_d = U_\infty c \sin(\alpha)/\nu$ and $Re_h = U_\infty h/\nu$, and $St_d = f c \sin(\alpha)/U_\infty$ and $St_h = f h/U_\infty$. As shown in figure \ref{fig:neutral_modified}(b), $d$ and $h$ are rather different in the considered range of $\alpha$. Interestingly, figure \ref{fig:neutral_modified}$(c)$ shows that for modes $pA$ and $pB$, $Re_{d,c1}$ collapses quite nicely to the same value for the different $\alpha$, i.e. $Re_{d,c1} = 91.46 \pm 10.77$ for mode $pA$ and $Re_{d,c1} = 105.91 \pm 18.67$ for mode $pB$; its relative variation is much lower than that of $Re_{c1}$ (compare figure \ref{fig:marginal_curves}). This shows that $d = c \sin(\alpha)$ is the appropriate length scale to describe the onset of the primary instability for negative $\alpha$, and to predict whether the low-$Re$ steady flow is absolutely and globally unstable to $2D$ perturbations without the need of a stability analysis. In contrast, the same can not be said for mode $pC$. In this case, indeed, at criticality $Re_{d,c1}$ does not collapse to a same value for the different $\alpha$ considered. This is consistent with the fact that the triggering mechanism of mode $pC$ is embedded in the $fV_2$ recirculating region at the TE, and resembles what has been observed for 2D symmetric bluff bodies \citep{chiarini-quadrio-auteri-2022}.

\section{Intermediate $\Rey$: the 2D periodic flow} 
\label{sec:inter-re}

In \S \ref{sec:prim-bif} we have investigated the linear stability of the low-$\Rey$ steady base flow and identified the eigenmodes that yield its primary bifurcation. 
To study the bifurcations the flow undergoes at larger $\Rey$, we (i) perform Floquet linear stability analysis of the ensuing limit cycles, and (ii) use 2D and 3D direct numerical simulations to account for nonlinear effects. Depending on $\alpha$, different limit cycles are possible, each one characterised by a different vortex interaction. Therefore, the nature of the secondary bifurcation changes with $\alpha$. In this section the different 2D limit cycles are characterised, using $\alpha=-5^\circ$, $-1.5^\circ$, $3^\circ$, $7^\circ$ and $10^\circ$ as representative cases. 

\subsection{Large negative angles of attack: $\alpha = -5^\circ$}

\begin{figure}
\centering
\includegraphics[trim={0 0 50 00},clip,width=0.95\textwidth]{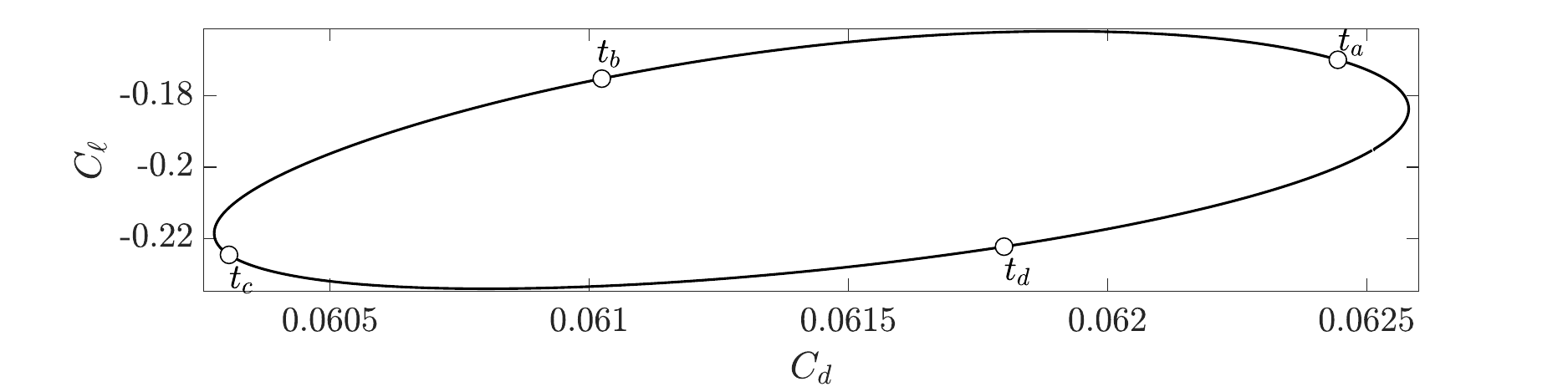}
\includegraphics[trim={0 0 60 30},clip,width=0.49\textwidth]{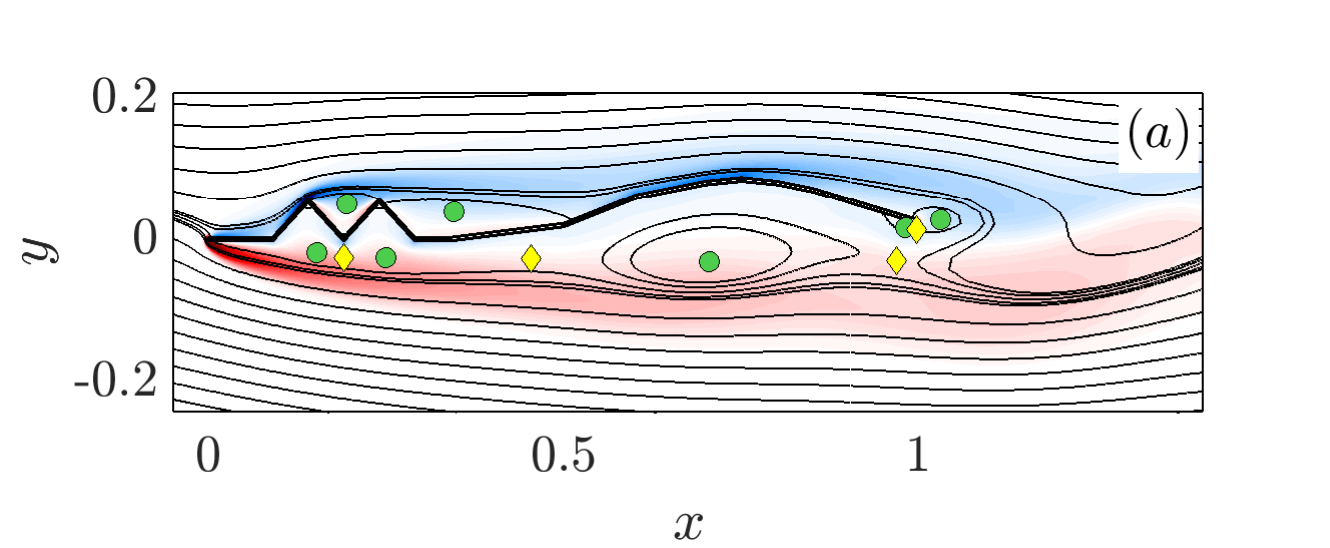}
\includegraphics[trim={0 0 60 30},clip,width=0.49\textwidth]{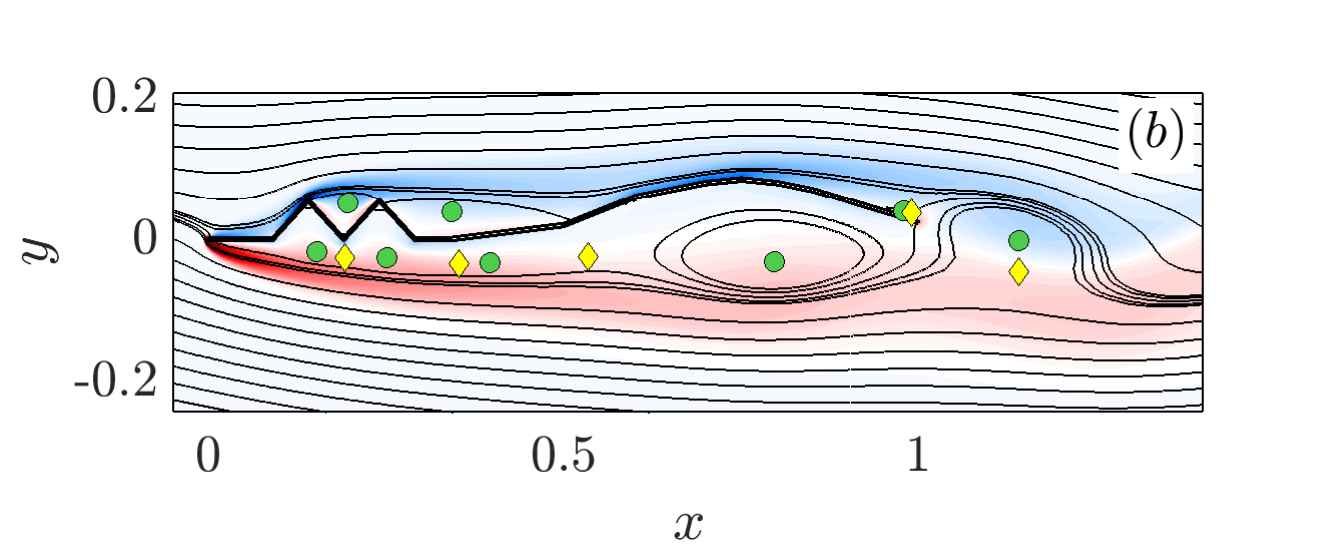}
\includegraphics[trim={0 0 60 30},clip,width=0.49\textwidth]{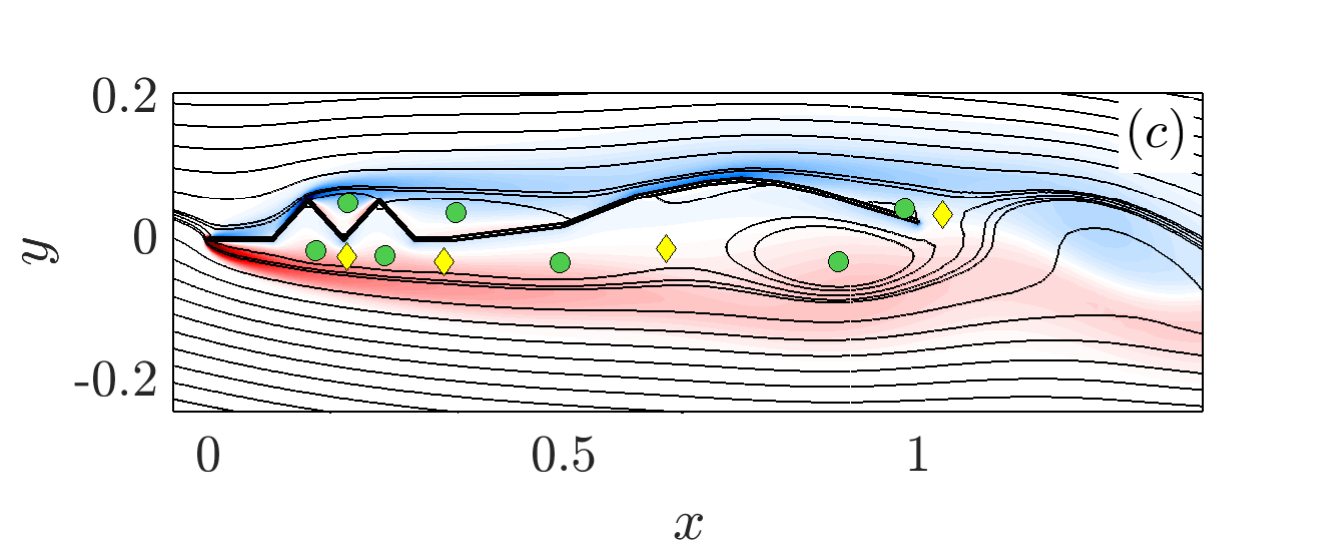}
\includegraphics[trim={0 0 60 30},clip,width=0.49\textwidth]{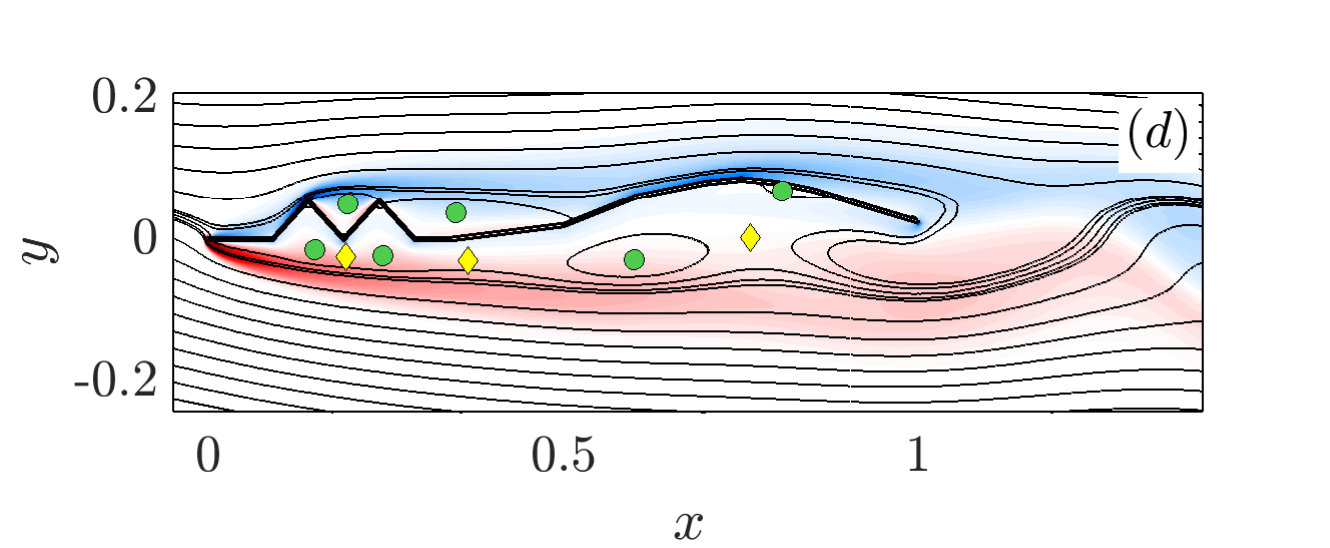}
\caption{Instantaneous flow for $\alpha=-5^\circ$ at $\Rey=1490$, represented with streamlines and vorticity colour maps. The periodic flow has period $T \approx 0.933$. Four temporal instants are takes equally spaced in the shedding period. Green circles indicate elliptical stagnation points, whereas yellow diamonds indicate the hyperbolic stagnation points.}
\label{fig:BF_AoAm5_1490}
\end{figure}

We start examining large negative $\alpha$. Figure \ref{fig:BF_AoAm5_1490} considers $\alpha=-5^\circ$ at $\Rey=1490$, and plots four instantaneous snapshots taken equally spaced within the shedding period. Note that this is the stabilised base flow (by means of the BoostConv algorithm), as the secondary bifurcation occurs at slightly smaller $Re$ (see \S\ref{subsec:sec-bif-naoa}). Streamlines and stagnation points are used to reveal the dynamics of the vortex shedding. As for smaller $\Rey$, several recirculating regions arise due to the presence of the sharp corners. The flow separates at the bottom LE corner but does not reattach, giving rise to a large recirculating region that encompasses the entire bottom side. For these values of $\alpha$ and $\Rey$, the flow unsteadiness is mainly driven by the dynamics of the recirculating regions placed in the rear part of the body. 
We focus first on the top side of the airfoil, and consider the recirculating region that arises at the TE; see the $\Psi_{2}=0$ line separating from the body in panel $a$. As the time advances, this recirculating region enlarges and accumulates negative vorticity. Eventually, a hyperbolic stagnation point arises downstream the bottom TE corner \citep{boghosian-cassel-2016}, and the TE vortex with negative vorticity is shed in the wake (see panel $b$). The shedding of TE vortices with positive vorticity is related with the dynamics of the large recirculating region placed over the bottom side, close to the TE. Along the shedding period, it enlarges until it undergoes vortex splitting (see the hyperbolic stagnation point in panels $d$ and $a$ below the TE), and a new TE vortex with positive vorticity is shed in the wake.

Note that for this $\alpha$ the attractor draws a close line in the $C_\ell-C_d$ map that does not intersect itself: this indicates that the $C_\ell(t)$ and $C_d(t)$ signals are not in phase, and that the phase delay $\phi$ between them is in the $0 \le \phi \le \pi/2$ range.

\subsection{Small negative angles of attack: $\alpha \le -1.25^\circ$}

\begin{figure}
\centering
\includegraphics[trim={0 0 50 00},clip,width=0.95\textwidth]{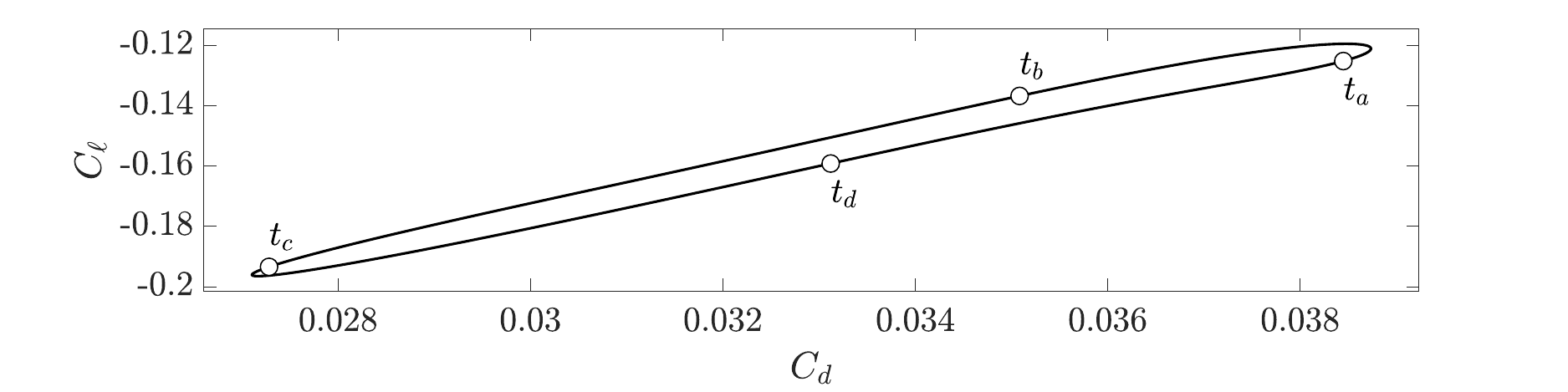}
\includegraphics[trim={0 0 60 30},clip,width=0.49\textwidth]{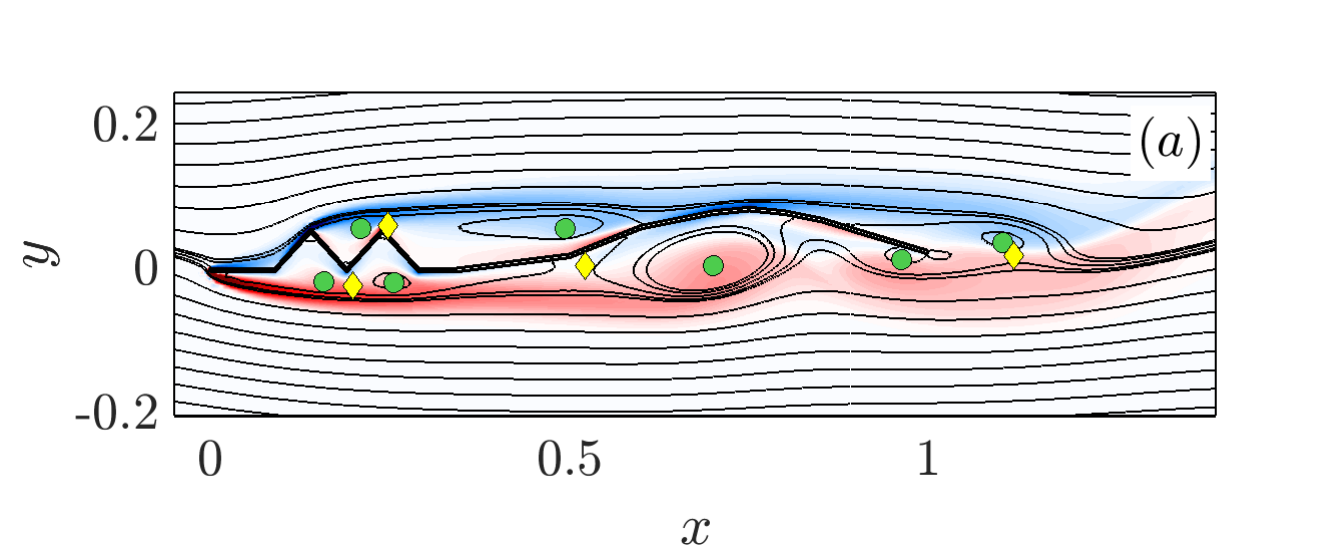}
\includegraphics[trim={0 0 60 30},clip,width=0.49\textwidth]{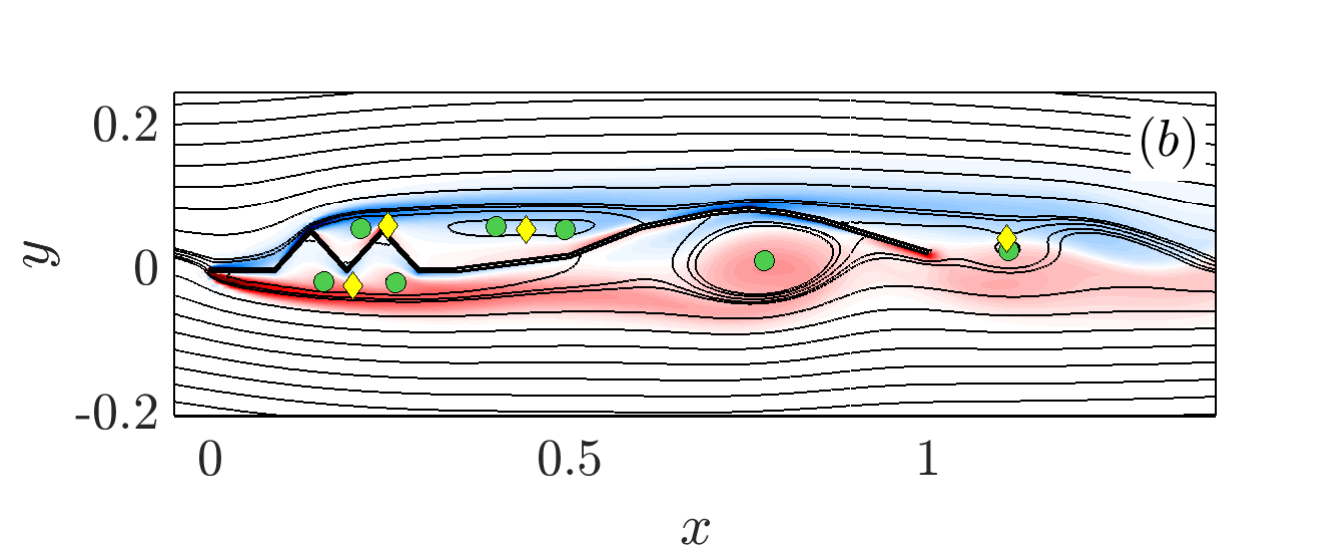}
\includegraphics[trim={0 0 60 30},clip,width=0.49\textwidth]{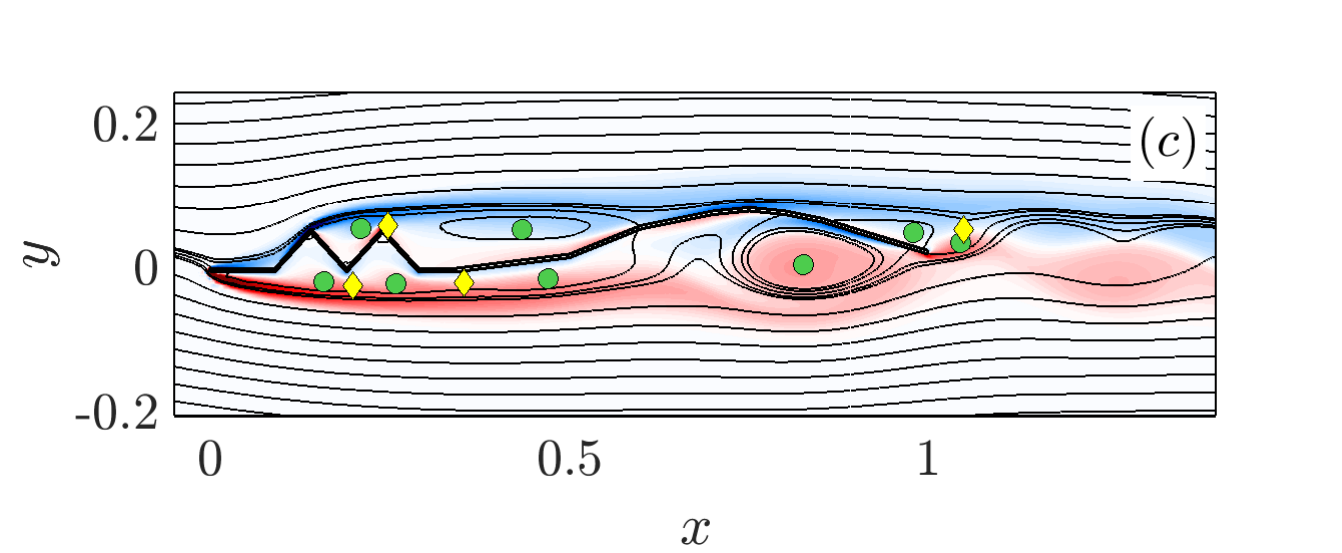}
\includegraphics[trim={0 0 60 30},clip,width=0.49\textwidth]{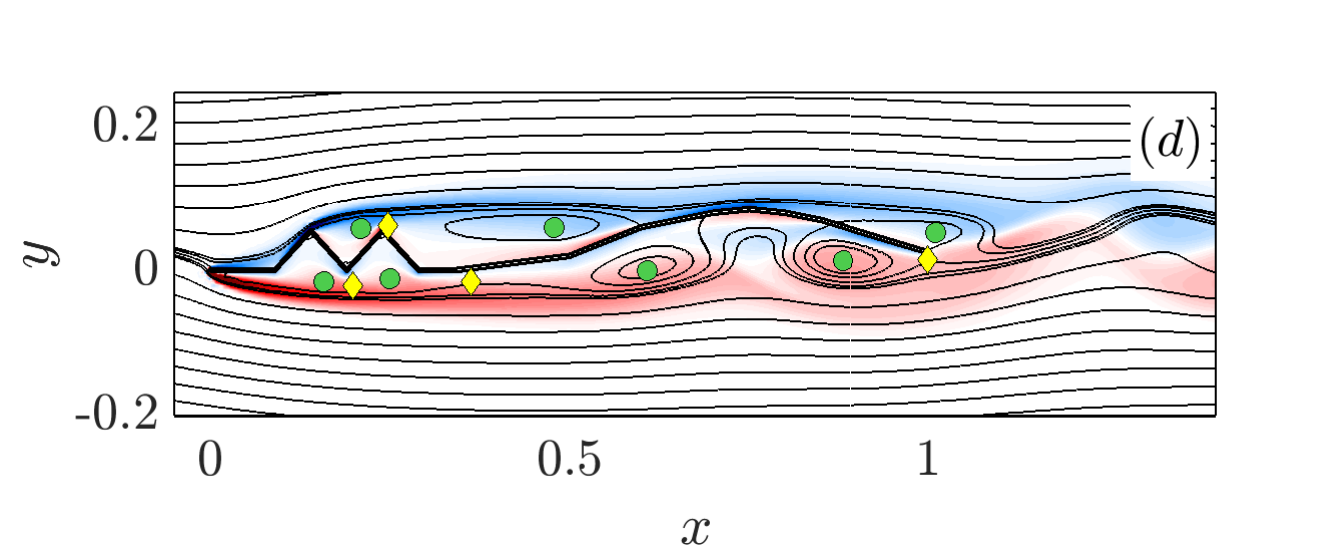}
\caption{Instantaneous flow for $\alpha= -1.5^\circ$ at $\Rey = 3500$, represented with streamlines and vorticity colour maps. The periodic flow has period $T \approx 1.042$. Eight temporal instants are taken equally spaced in the shedding period. Green circles indicate elliptical stagnation points, whereas yellow diamonds indicate the hyperbolic stagnation points.}
\label{fig:BF-AoAm15}
\end{figure}
For less negative $\alpha$ ($\alpha = -1.5^\circ$) the scenario changes; see figure \ref{fig:BF-AoAm15}. After separating at the LE the flow reattaches along the bottom side of the body, giving origin to a recirculating region that enlarges and shrinks, and periodically sheds vortices. This LE vortex shedding locks to the frequency of the vortex shedding from the TE, resembling what found in the flow past 2D elongated rectangular cylinders at intermediate $\Rey$ \citep{okajima-1982,hourigan-thompson-tan-2001}; see also \S \ref{sec:int-alfa}. On the contrary, the recirculating region that arises over the top side downstream the corrugations does not play a central role in the dynamics of the wake for these $\alpha$ and $\Rey$; it enlarges and shrinks, but does not undergo vortex splitting.
We now focus on the bottom side, and start from panel $b$. Advancing in time, the fore recirculating region progressively enlarges and eventually splits, shedding a LE vortex with positive vorticity. This is visualised in panels $c-d$: as the reattaching point moves downstream a new vortex arises, identified by the additional elliptic stagnation point that emerges in panel $c$. The new LE vortex enlarges, until it is shed and travels downstream. When the LE vortex reaches the TE, it gives rise to a TE vortex with positive vorticity, that is eventually shed in the wake. The shedding of TE vortices with negative vorticity, instead, is related with the dynamics of the recirculating region that arises over the rear top side, and its frequency is dictated by the motion of the bottom LE vortex. In fact, a new TE vortex with negative vorticity is shed in the wake when the bottom LE vortex crosses the TE, and a new hyperbolic stagnation point arises there; see panels $d$ and $a$. Interestingly, this mechanism resembles what observed by \cite{chiarini-quadrio-auteri-2022} for the flow past 2D rectangular cylinders, in the regime dominated by the LE vortex shedding.

The shape of the $C_\ell-C_d$ curve differs from what we have found for $\alpha=-5^\circ$: in this case indeed the two $C_\ell(t)$ and $C_d(t)$ signals are substantially in phase. We anticipate that a similar shape is also observed for larger $\alpha$.

\subsection{Small angles of attack: $-1.25^\circ < \alpha \le 3^\circ$}

\begin{figure}
\centering
\includegraphics[trim={0 0 50 00},clip,width=0.95\textwidth]{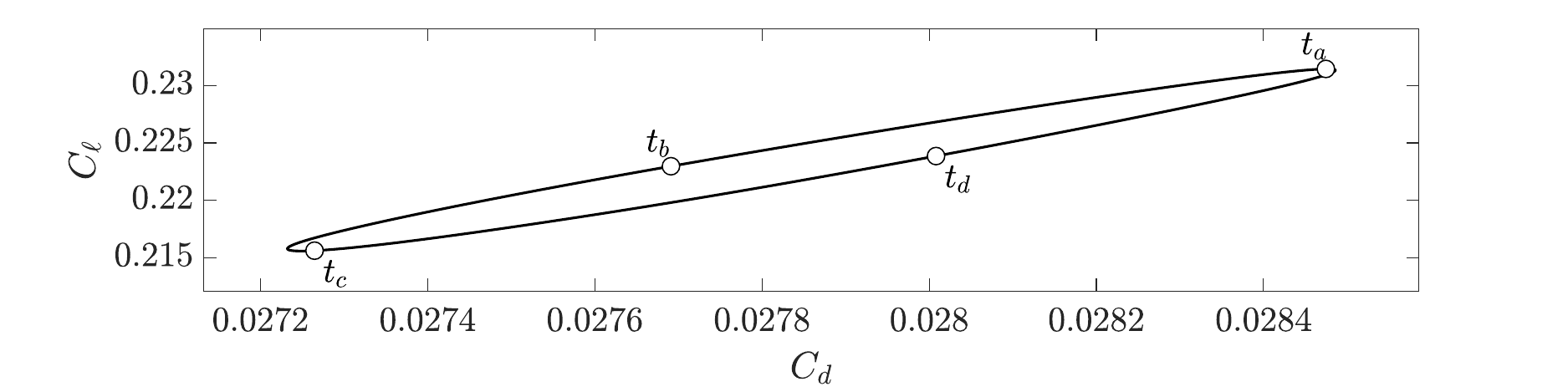}
\includegraphics[trim={0 0 60 30},clip,width=0.49\textwidth]{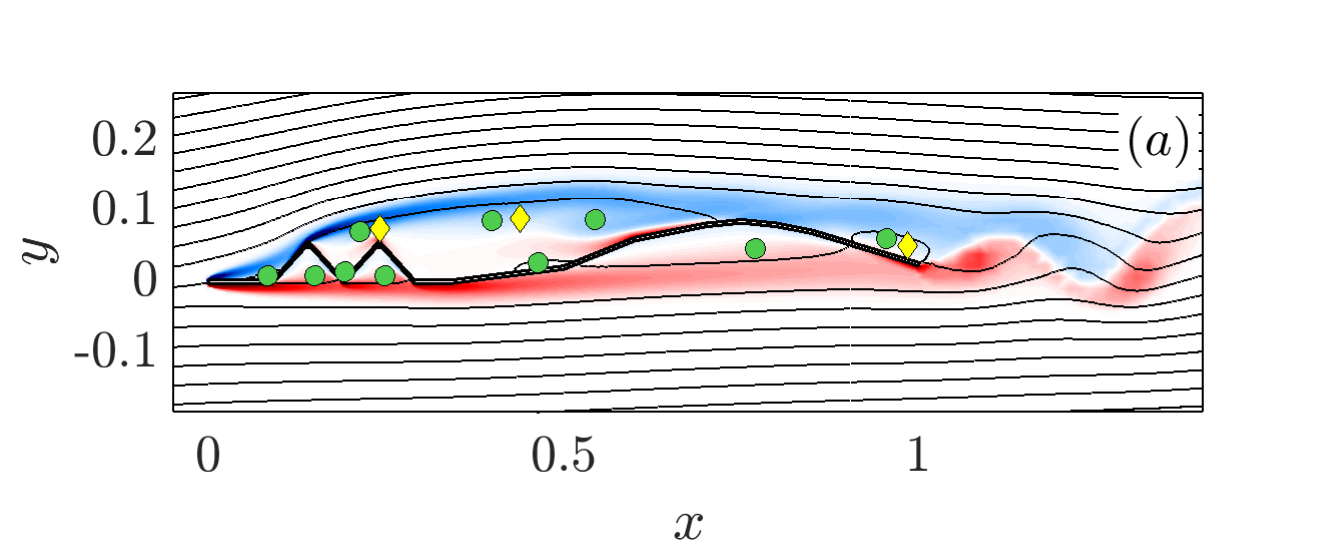}
\includegraphics[trim={0 0 60 30},clip,width=0.49\textwidth]{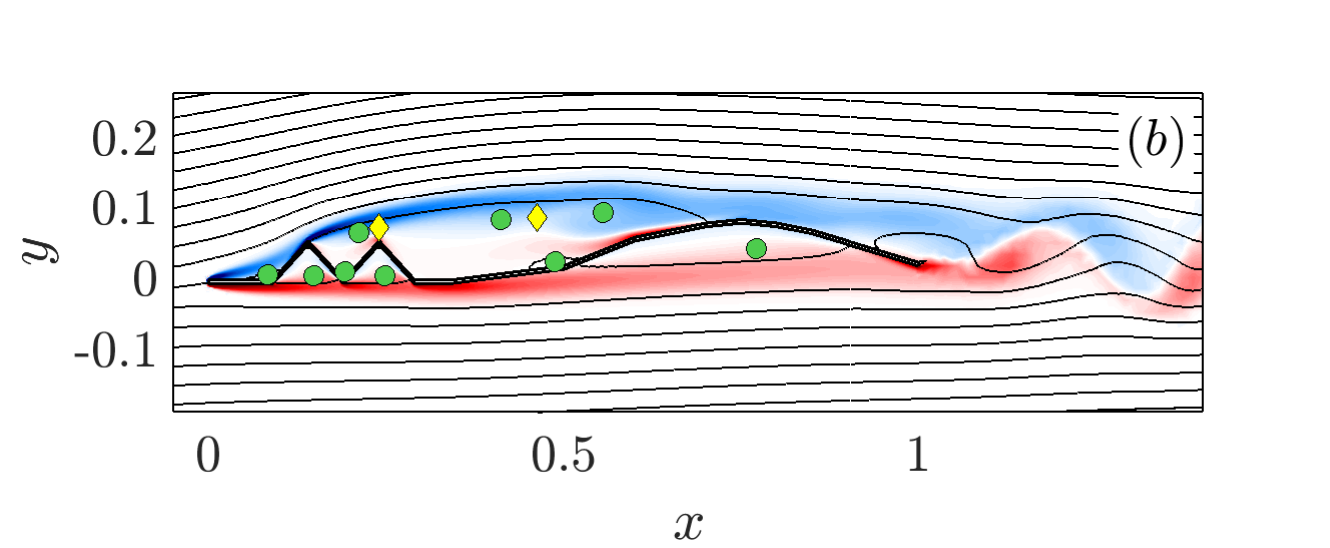}
\includegraphics[trim={0 0 60 30},clip,width=0.49\textwidth]{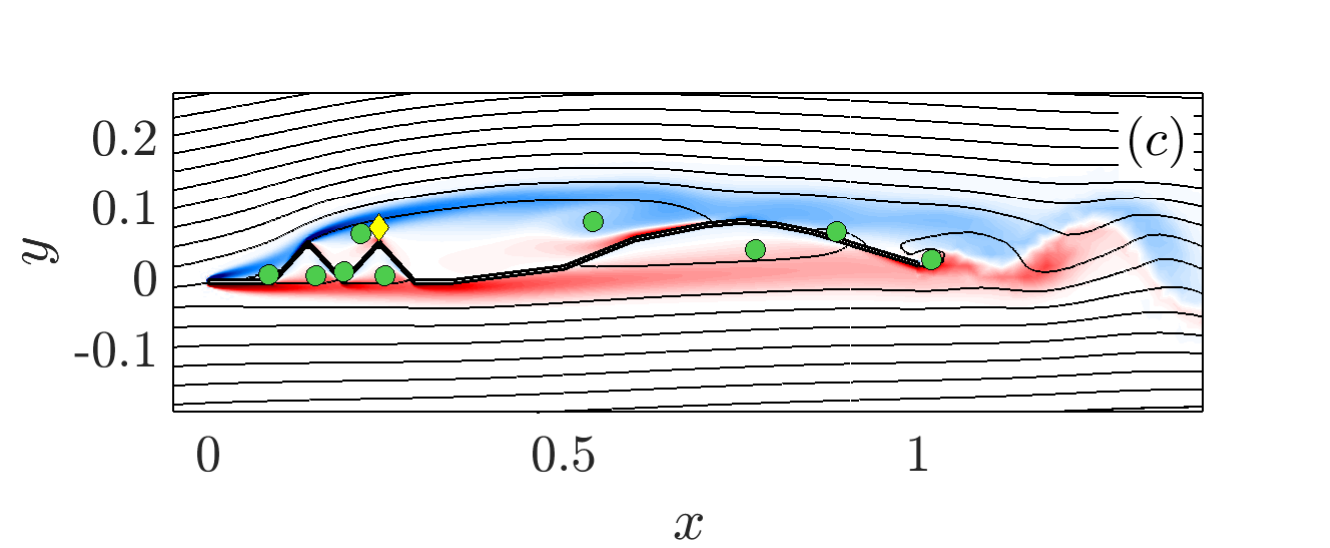}
\includegraphics[trim={0 0 60 30},clip,width=0.49\textwidth]{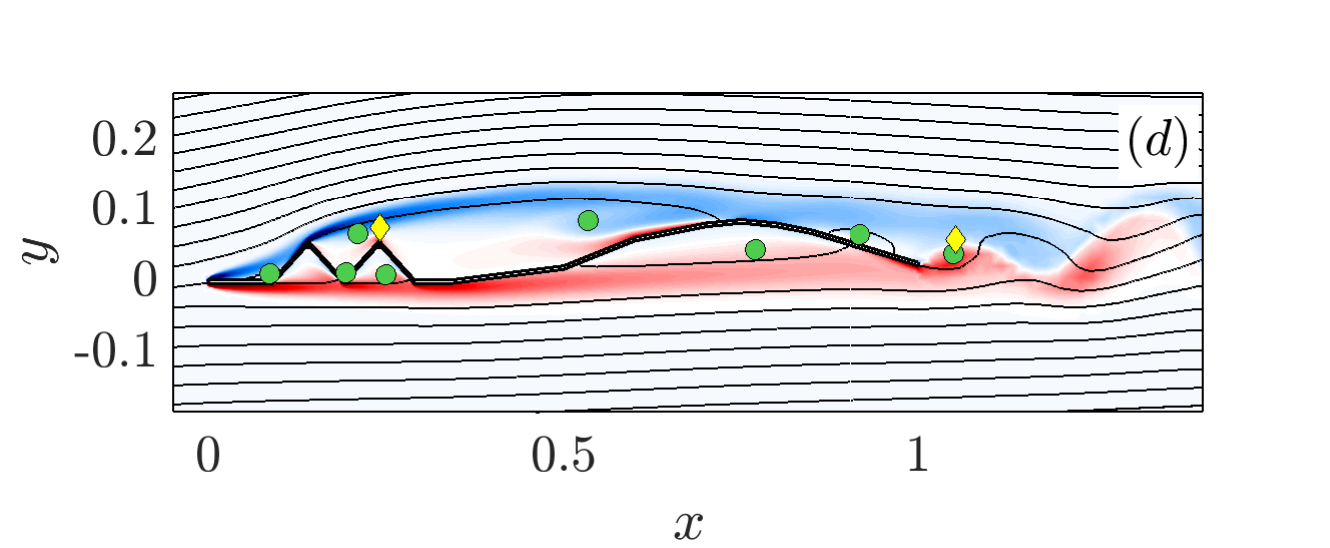}
\caption{Instantaneous flow for $\alpha=3^\circ$ at $\Rey=5000$, represented with streamlines and vorticity colour maps. The periodic flow has period $T \approx 0.367$. Four temporal instants are takes equally spaced in the shedding period. Green circles indicate elliptical stagnation points, whereas yellow diamonds indicate the hyperbolic stagnation points.}
\label{fig:BF_AoA3_5000}
\end{figure}

We now move to small positive and negative $\alpha$ in the $-1.25^\circ < \alpha \leq 3^\circ$ range. Figure \ref{fig:BF_AoA3_5000} considers $\alpha=3^\circ$ at $\Rey=5000$. 

Over the top side, the flow separates at the corner of the upstream groove and reattaches along the rear arc, giving rise to two recirculating regions, separated by a hyperbolic stagnation point. 
However, they do not play a role in the vortex shedding mechanism. In fact, despite the presence of the hyperbolic stagnation point, the downstream recirculating region is not shed, as instead occurs at larger $\alpha$ (see the following discussion). 
Moving downstream, the flow separates again due to the adverse pressure gradient, and a further recirculating region arises close to the TE (see panel $a$). As the time advances, this recirculating region widens, until the reattaching point reaches the TE and a vortex with negative vorticity is shed in the wake (see panel $b$). Once this TE vortex is shed in the wake, a recirculating region with positive vorticity starts accumulating at the bottom TE corner (panel $c$), widens and, eventually, is shed in the wake (panel $d$).

\subsection{Intermediate positive angles of attack: $ 4^\circ \le \alpha \le 7^\circ$}
\label{sec:int-alfa}

\begin{figure}
\centering
\includegraphics[trim={0 0 50 00},clip,width=0.95\textwidth]{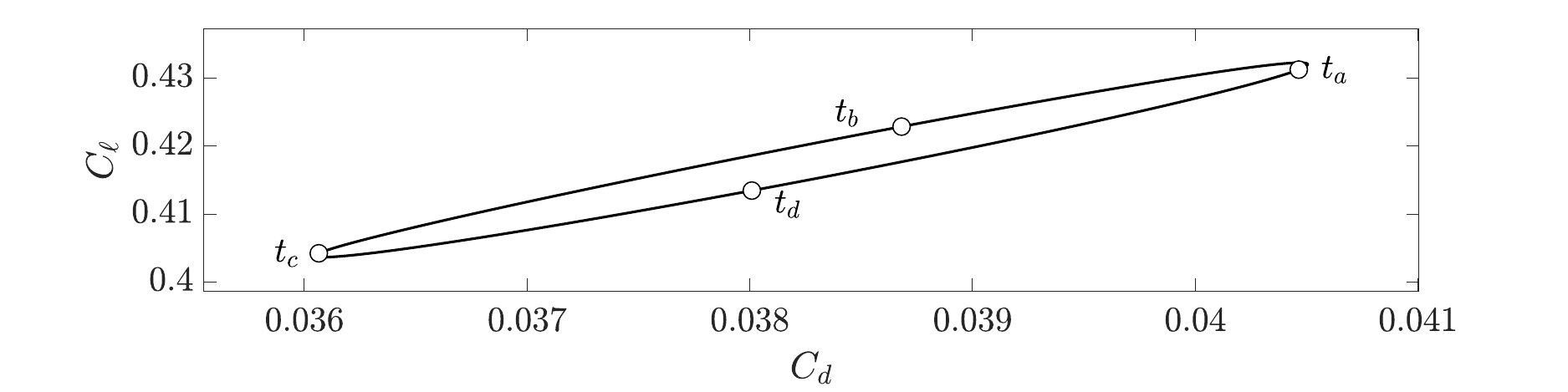}
\includegraphics[trim={0 0 60 30},clip,width=0.49\textwidth]{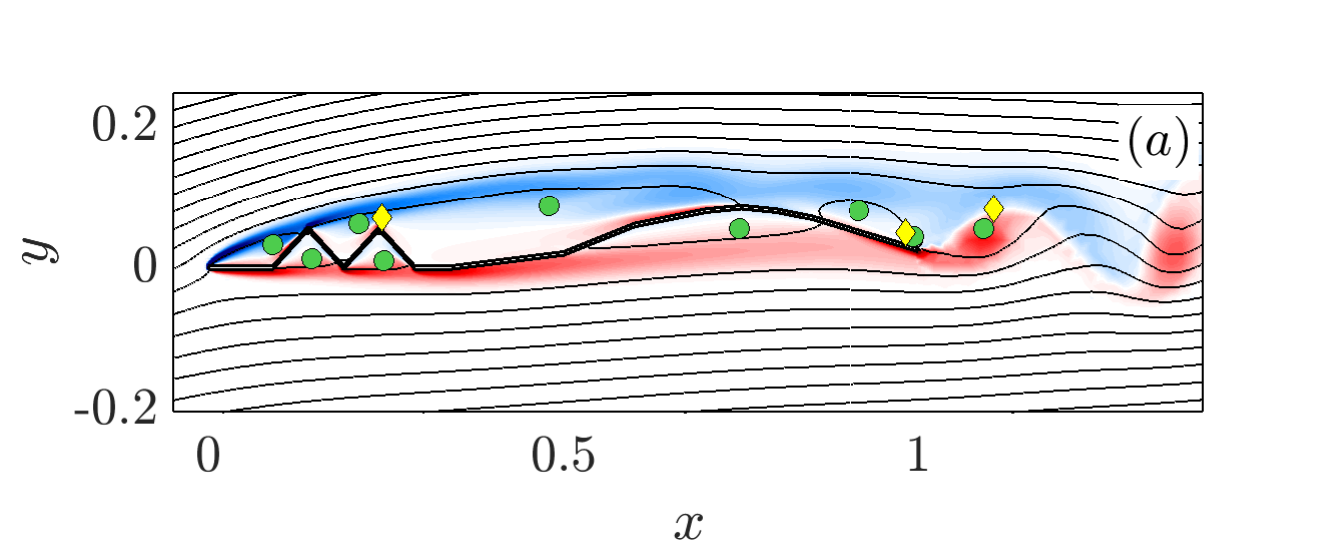}
\includegraphics[trim={0 0 60 30},clip,width=0.49\textwidth]{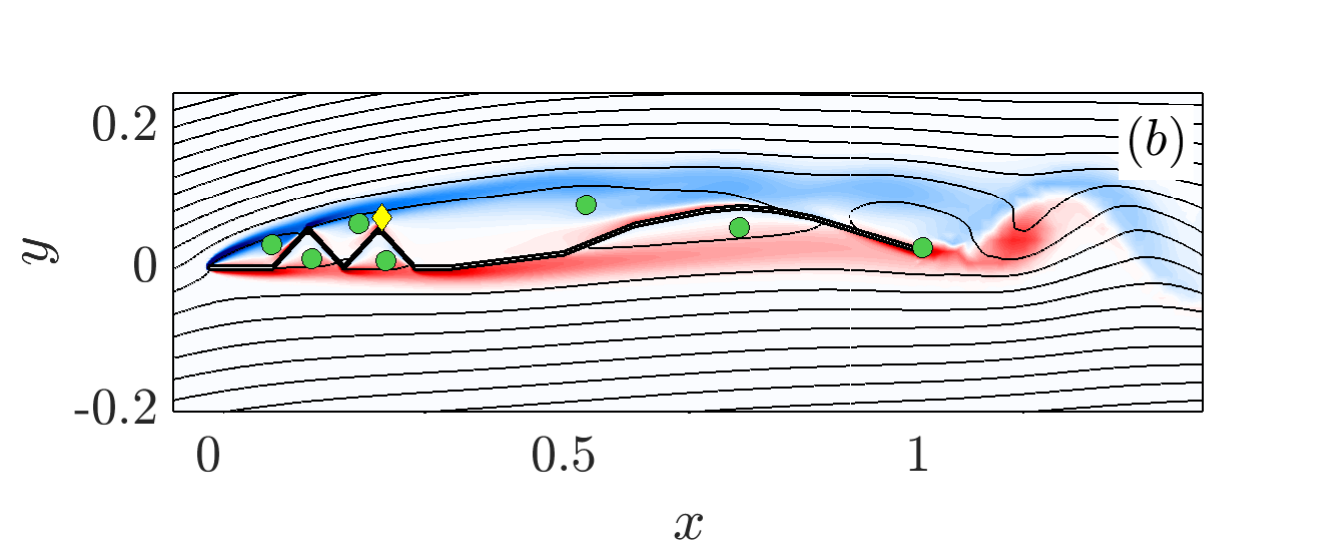}
\includegraphics[trim={0 0 60 30},clip,width=0.49\textwidth]{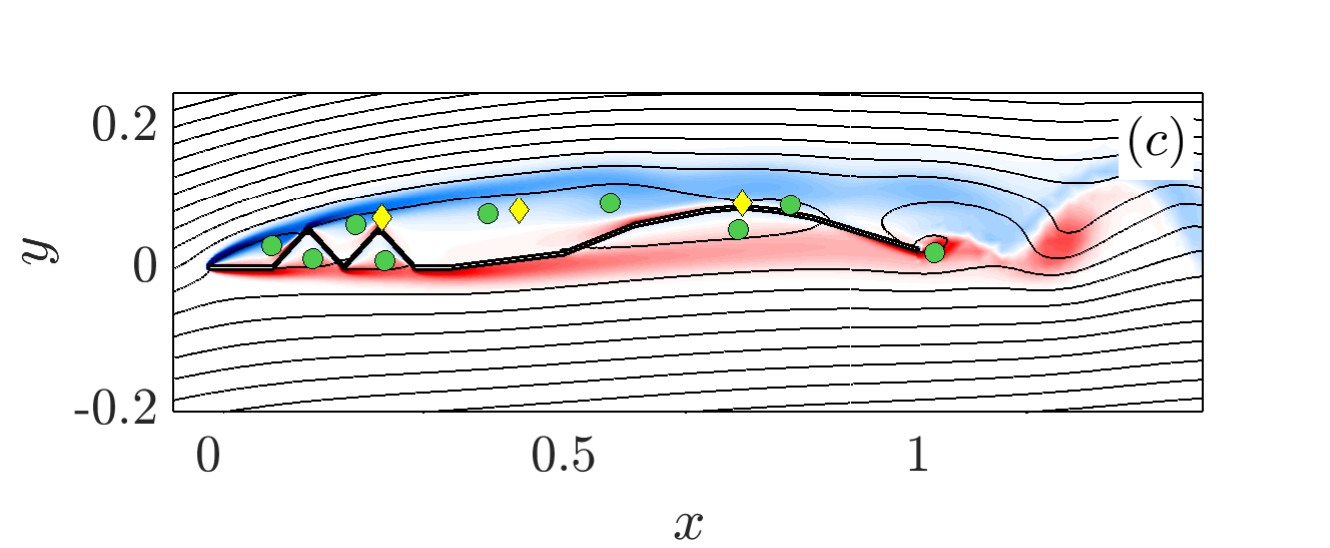}
\includegraphics[trim={0 0 60 30},clip,width=0.49\textwidth]{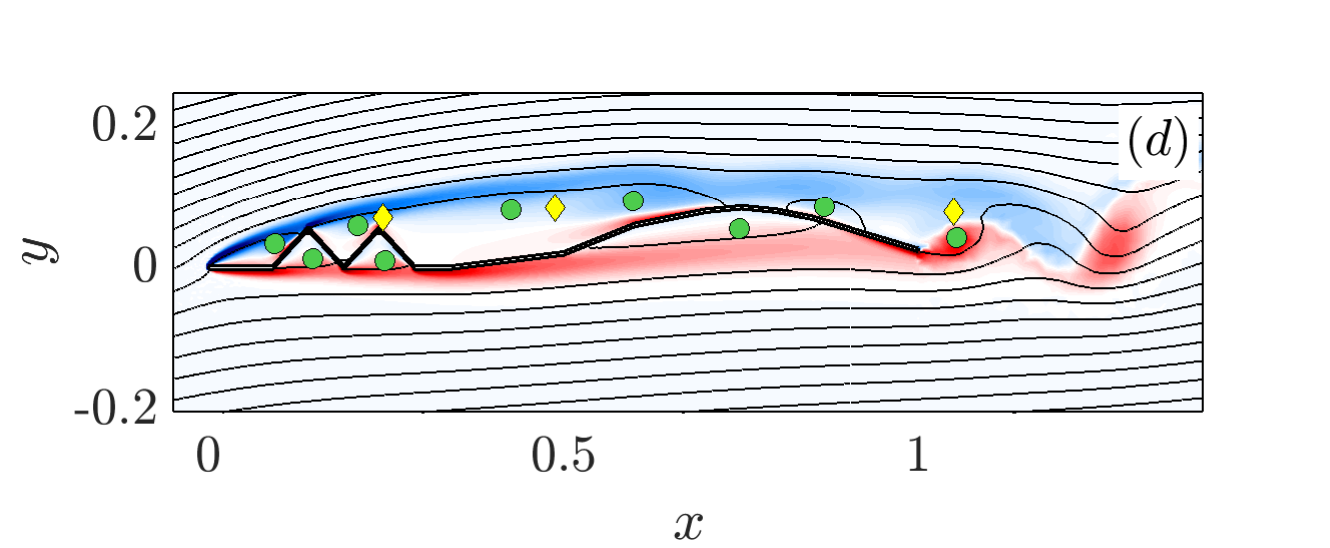}
\caption{Instantaneous flow for $\alpha=7^\circ$ at $\Rey=4700$, represented with streamlines and vorticity colour maps. The periodic flow has period $T \approx 0.419$. Four temporal instants are takes equally spaced in the shedding period. Green circles indicate elliptical stagnation points, whereas yellow diamonds indicate the hyperbolic stagnation points.}
\label{fig:BF_AoA7_4700}
\end{figure}

Figure \ref{fig:BF_AoA7_4700} considers $\alpha=7^\circ$ at $\Rey=4700$, being representative of $\alpha$ in the $4^\circ \le \alpha \le 7^\circ$ range. 
Unlike for smaller $\alpha$, here the recirculating region that arises after the flow separation at the upstream groove corner %($fV_1$) 
plays a role in the vortex shedding dynamics. This recirculating region periodically splits, releasing vortices that are advected downstream and give rise to TE vortices. This resembles the periodic vortex shedding observed in the flow past elongated rectangular cylinders, with $\AR = L/D>3$ (here $L$ and $D$ are the streamwise and vertical sizes of the cylinder). In that case, indeed, due to the so-called impinging leading-edge vortex (ILEV) instability, vortices are periodically shed from the LE and TE shear layers, and the two phenomena lock to the same frequency \citep{okajima-1982,nakamura-nakashima-1986,mills-etal-1995,hourigan-thompson-tan-2001,chiarini-quadrio-auteri-2022}. The mechanism of the ILEV instability is the following. A vortex is shed from the LE shear layer and is advected downstream. When the LE vortex passes over the TE, it creates a perturbation that alters the flow circulation and triggers thus the shedding of a new LE vortex. At the same time, the interaction of the LE vortex with the TE induces vortex shedding in the wake. The vortex dynamics of the limit cycle at these intermediate $\alpha$ is similar. We consider the top side of the body, and start from panel $c$ of figure \ref{fig:BF_AoA7_4700}. Within the $\Psi_2=0$ line that separates at the upstream groove corner, three recirculating regions are detected, being separated by hyperbolic stagnation points. As the time advances, the recirculating region widens and then shrinks, splitting into two parts: a LE vortex is shed (panel $c$) and travels downstream (panel $d$). When the LE vortex reaches the TE (panel $a$), a hyperbolic stagnation point arises at the bottom side, and a new TE vortex with negative vorticity is shed in the wake (panel $b$). At the same time, a recirculating region with positive vorticity starts accumulating at the bottom TE corner (panel $b$ and panel $c$), being eventually shed in the wake half period later (panel $d$). More specifically, the vortex dynamics over the top side of the geometry closely resembles the flow past rectangular cylinders in the case where the preferred TE shedding frequency is permitted \citep{chiarini-quadrio-auteri-2022}.

\subsection{Large positive $\alpha$}

\begin{figure}
\centering
\includegraphics[trim={0 0 50 00},clip,width=0.95\textwidth]{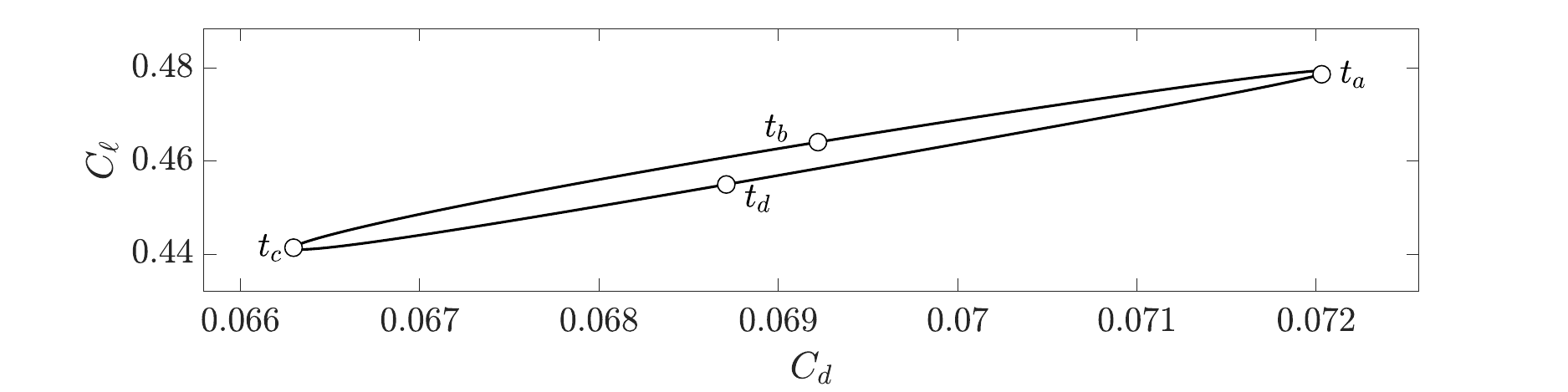}
\includegraphics[trim={0 0 60 30},clip,width=0.49\textwidth]{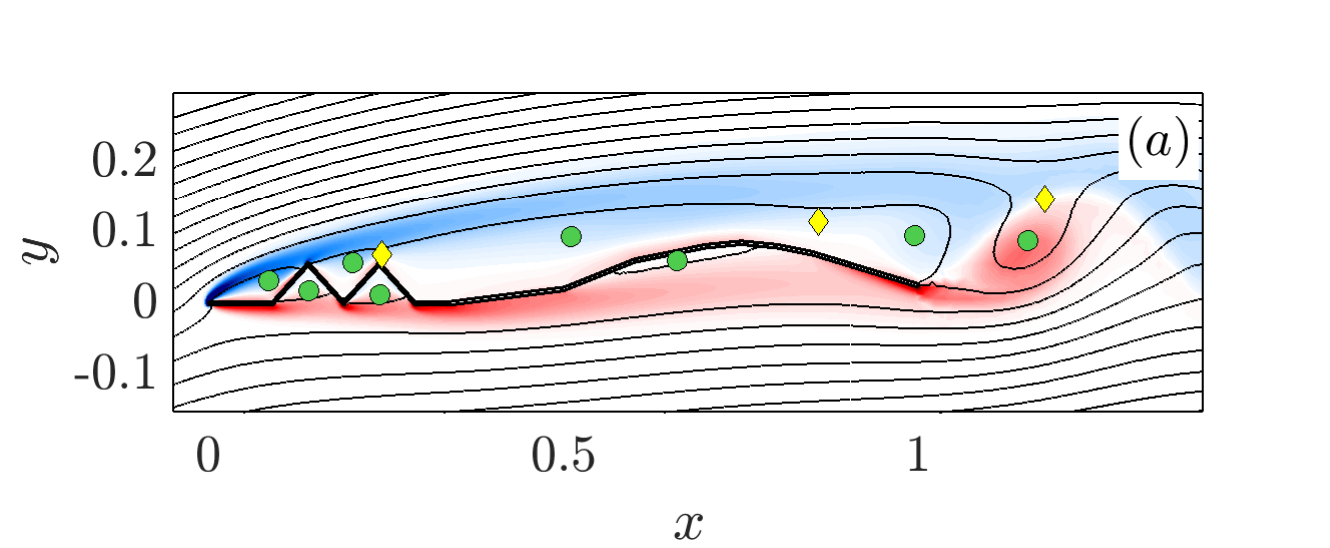}
\includegraphics[trim={0 0 60 30},clip,width=0.49\textwidth]{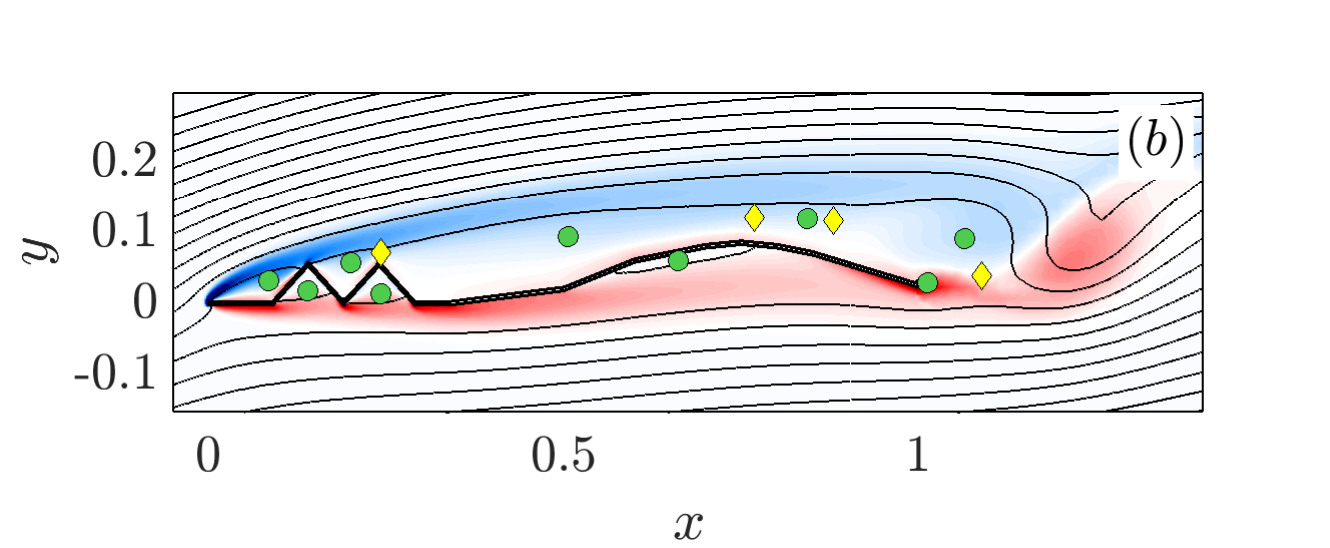}
\includegraphics[trim={0 0 60 30},clip,width=0.49\textwidth]{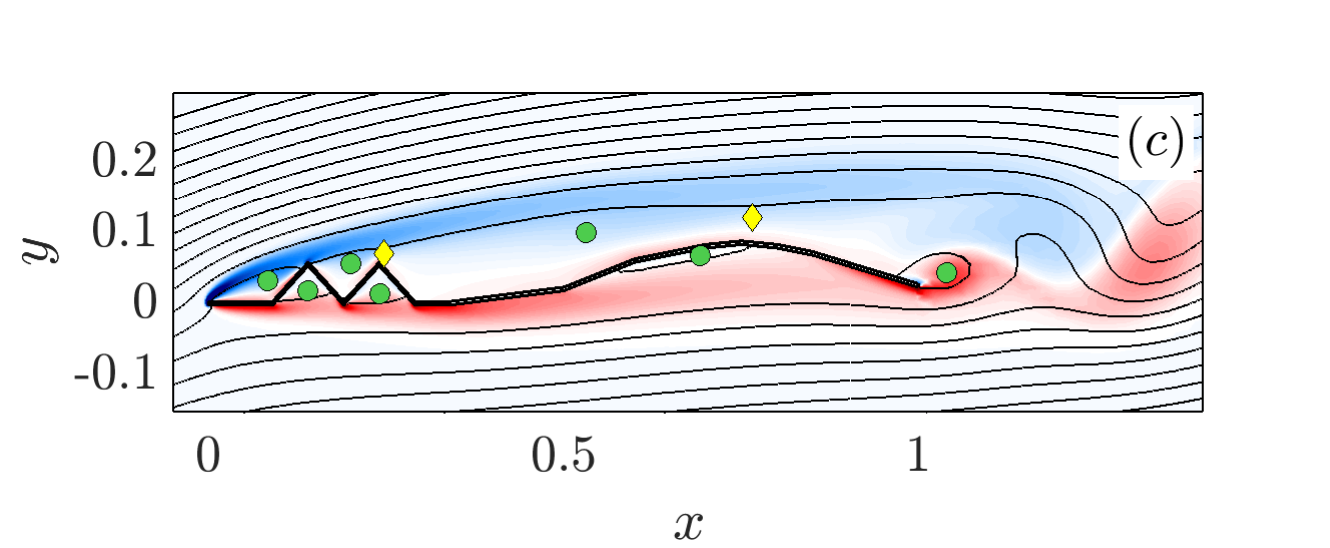}
\includegraphics[trim={0 0 60 30},clip,width=0.49\textwidth]{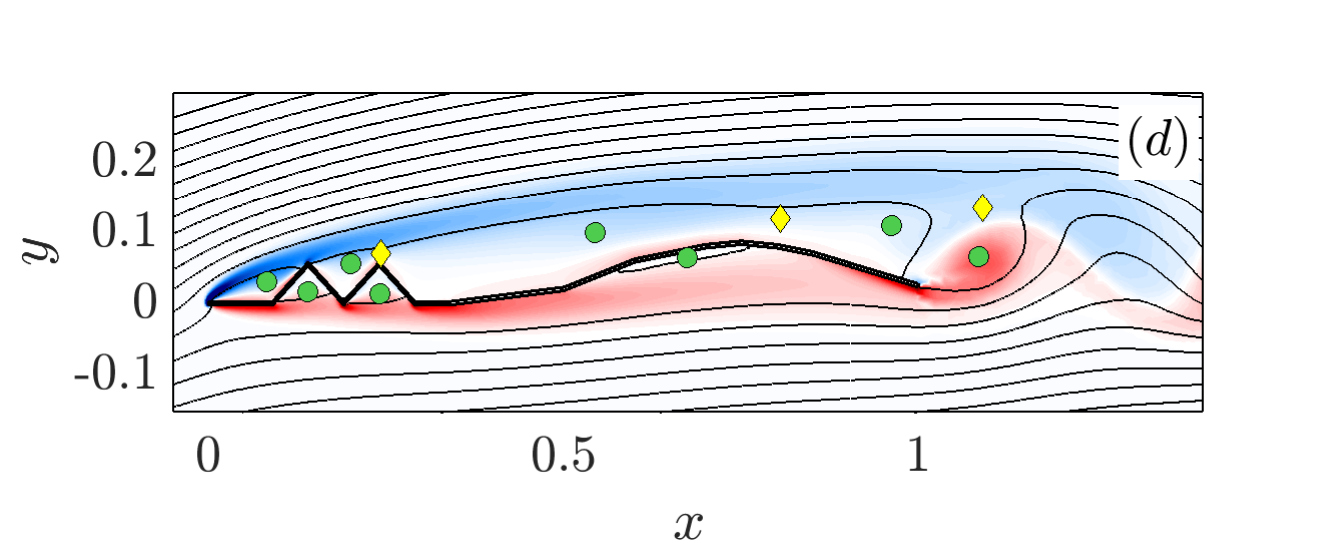}
\caption{Instantaneous flow for $\alpha=10^\circ$ at $\Rey=2000$, represented with streamlines and vorticity colour maps. The periodic flow has period $T \approx 0.691$. Four temporal instants are takes equally spaced in the shedding period. Green circles indicate elliptical stagnation points, whereas yellow diamonds indicate the hyperbolic stagnation points.}
\label{fig:BF_AoA10_2000}
\end{figure}

We now move to $\alpha \ge 8^\circ$, and use $\alpha=10^\circ$ at $\Rey=2000$ as representative case; see figure \ref{fig:BF_AoA10_2000}.
For these values of $\alpha$ and $\Rey$, the flow separates at the corner of the upstream corrugation, and only intermittently reattaches on the top body surface. We start from panel $a$. Here the flow reattaches at $x \approx 1$, and an elliptic stagnation point detects a recirculating region close to the TE. As the time advances the flow separates, and a TE vortex with negative vorticity is shed in the wake (panel $b$). At the same time, a recirculating region with positive vorticity starts accumulating at the bottom TE corner (panel $b$), it widens (panel $c$) and, eventually, it is shed in the wake (panel $d$). When this shedding occurs, the flow reattaches over the rear top side, and a new recirculating region with negative vorticity forms close to the TE. 

\section{The secondary bifurcation}
\label{sec:sec-bif}

Having characterised the periodic limit cycles at intermediate $Re$, we now address the secondary bifurcation of the flow by means of Floquet analysis. Both 2D and 3D nonlinear simulations are used to corroborate the results. For negative $\alpha$ the secondary instability is 2D. To reiterate (see figure \ref{sec:flow-reg}), for intermediate negative $\alpha$ the secondary flow instability consists of a 2D Neimark-Sacker bifurcation and introduces a new frequency that matches that of mode $pB$ described in \S \ref{sec:prim-bif}. 
For large negative $\alpha$, instead, the secondary instability consists of a 2D subharmonic bifurcation and leads to a new limit cycle. For positive $\alpha$ the secondary bifurcation is 3D. Different growing modes are detected with the Floquet analysis. For $0^\circ \le \alpha \le 7^\circ$ the leading mode is synchronous and the secondary instability yields a 3D pitchfork bifurcation. For larger $\alpha$, instead, the most amplified mode is of subharmonic nature.

\subsection{The 2D secondary bifurcation for $-5^\circ \le \alpha \le -1.5^\circ$}
\label{subsec:sec-bif-naoa}

\begin{figure}
\centerline{
\begin{tikzpicture}
\node at (1.25,8.4) {\includegraphics[width=0.49\textwidth]{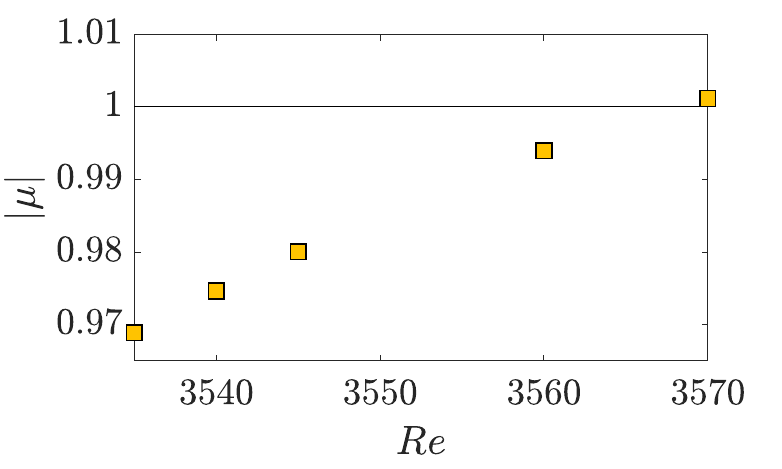}};
\node at (7.75,8.4) {\includegraphics[width=0.49\textwidth]{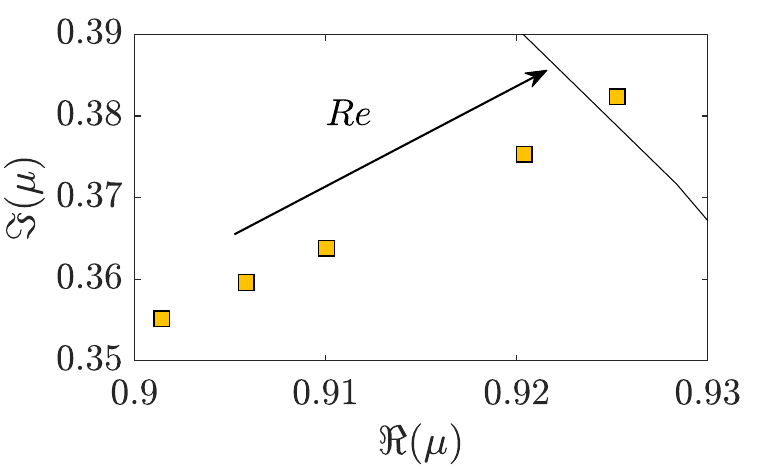}};
\node at (4.5,4.55) {\includegraphics[trim={50 0 50 0},clip,width=1\textwidth]{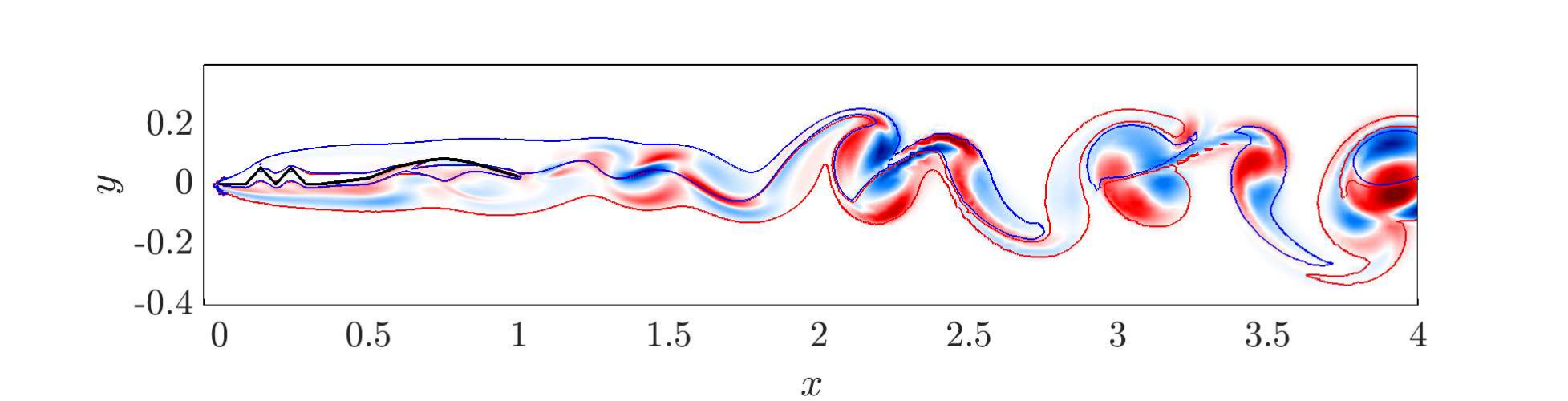}};
\node at (-0.6,9.8)   {$(a)$};
\node at (5.9,9.8)    {$(b)$};
\node at (-0.75,5.55) {$(c)$};
\end{tikzpicture}
}
\caption{Secondary bifurcation for $\alpha = -1.5^\circ$. Top: dependence of the leading branch of Floquet multipliers on $Re$. Left: dependence of $|\mu|$ on $Re$. Right: Floquet multipliers in the complex plane; the arrow indicate the direction increase of $Re$. Bottom: Floquet mode responsible for the secondary bifurcation at $\Rey=3500$. The colourmap represents contours of spanwise vorticity of the perturbation field, while the red and blue solid lines are isocontours $\Omega_{z,2} = \pm 0.5$ of the periodic base flow.}
\label{fig:vort-AoAm15-Re3750}
\end{figure}
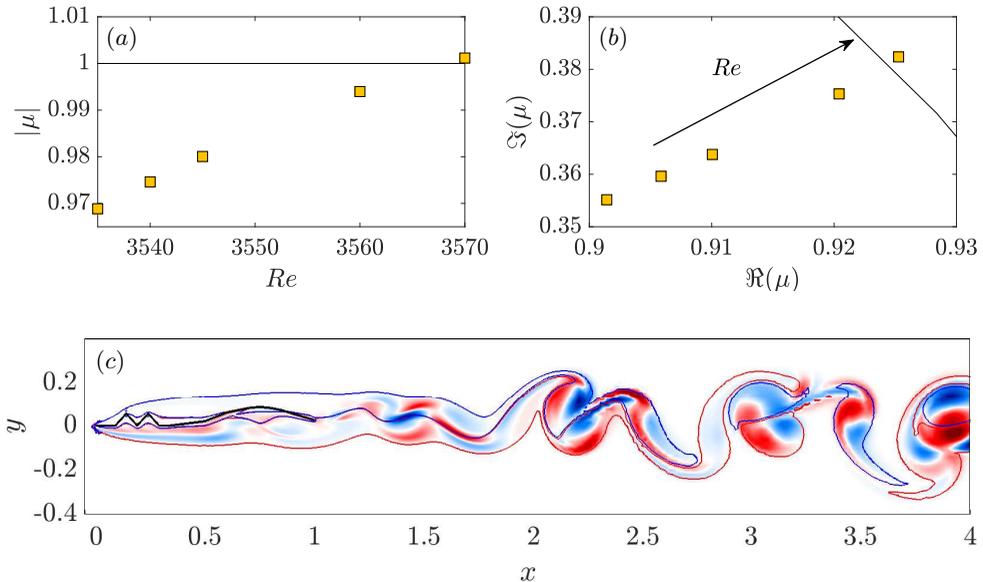

We start considering small negative $\alpha$, using $\alpha = -1.5^\circ$ as representative case. As anticipated, for $-5^\circ < \alpha \le -1.25^\circ$ the limit cycle loses its stability at $Re = Re_{c2}$ by means of a 2D Neimark-Sacker bifurcation. For $Re \ge Re_{c2}$ ($Re_{c2} \approx 3568$ for $\alpha = -1.5^\circ$) the nonlinear simulations reveal that a new frequency incommensurate with that of the limit cycle emerges in the spectrum, and a torus replaces the limit cycle in the phase space. Notably, the new frequency matches reasonably well that of the unsteady mode $pB$ of the low-$Re$ steady base flow. For $\alpha = -1.5^\circ$ and $Re = 3750$, for example, from the nonlinear simulations we measure $St \approx 0.05$, that has to be compared with the value of $f_{c1} \approx 0.046$ found for mode $pB$ at criticality with the linear stability analysis (figure \ref{fig:marginal_curves}). It is also worth noticing that the value of $Re_{c2}$ is very close to the critical Reynolds number found for $pB$, that for $\alpha =-1.5^\circ$ is $Re \approx 3475$. 

To further characterise this bifurcation, we have investigated the linear stability of the limit cycle via Floquet analysis. The results are shown in figure \ref{fig:vort-AoAm15-Re3750}: panels $a$ and $b$ highlight the dependence of the Floquet multipliers on $Re$, while panel $c$ depicts the spatial structure of the unstable Floquet mode. A leading branch of multipliers with $\beta = 0$ has been detected. It corresponds to a pair of complex conjugate multipliers that near the unit cycle as $\Rey$ increases, with positive real part and non null imaginary part, in agreement with a 2D secondary Hopf bifurcation. We refer to this mode as mode $sQPb$. At $Re=3570$ the multipliers are $\mu = (0.926 \pm 0.382)$, that correspond to Floquet exponents of $\sigma_2 = (0.0011,\pm 0.3737)$, and therefore to a frequency of $f \approx 0.0595$ that is close to the results of the nonlinear simulations. Note that in agreement with mode $pB$ of the low-$Re$ steady base flow, the magnitude of mode $sQPb$ is non null along the bottom side of the airfoil; see figure \ref{fig:vort-AoAm15-Re3750}$(c)$. 

\begin{figure}
    \centering
    \includegraphics[trim={0 0 0 0},clip,width=\textwidth]{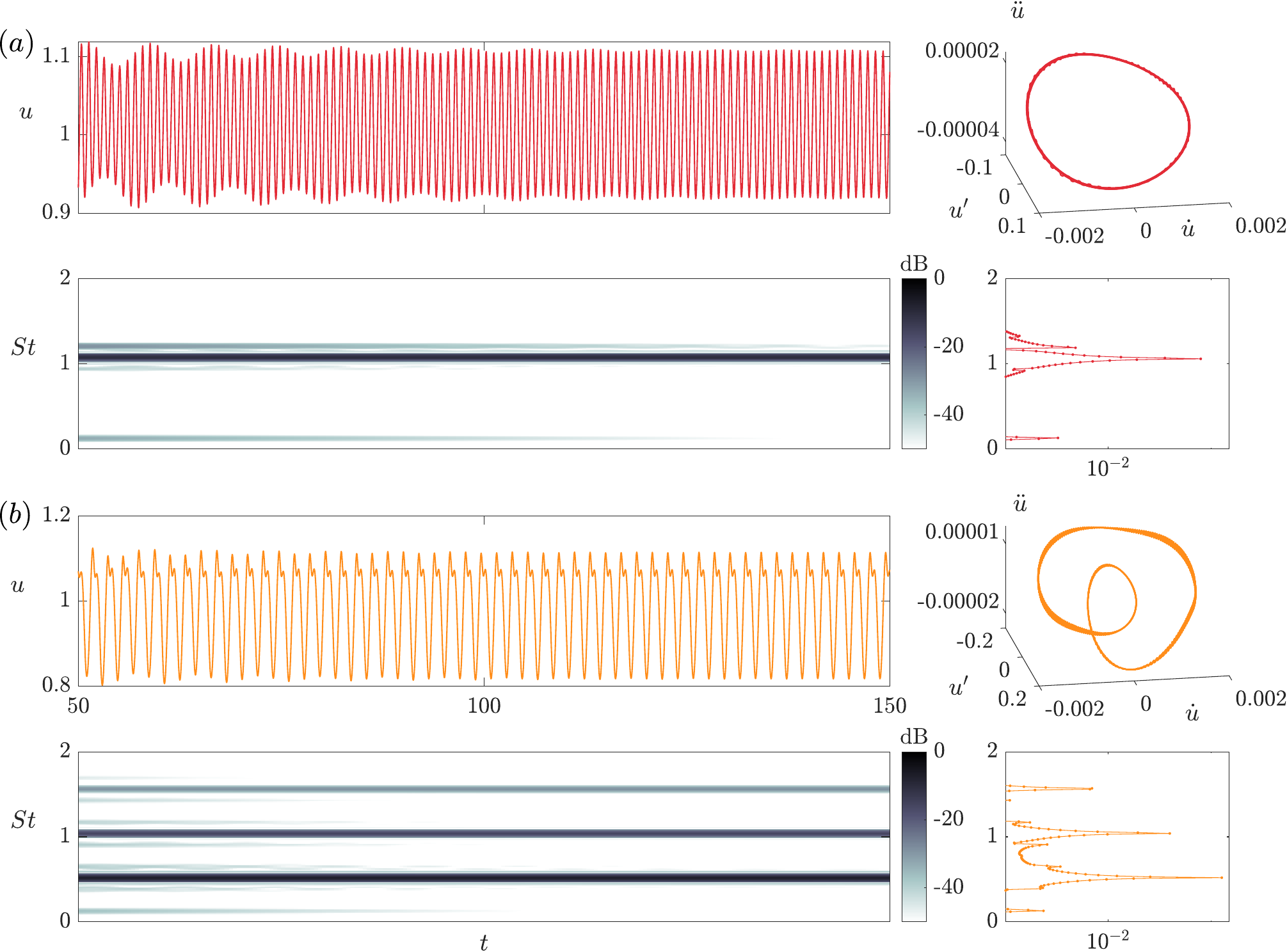}
    \caption{Results from DNS at $\alpha=-5^\circ$ for $\Rey=1450$ (\textit{a}) and $\Rey=2000$ (\textit{b}). Top left: streamwise velocity signal $u$ at $(x,\,y,\,z) = (1.3,\,0.17,\,0.5)$. Top right: flow attractor for $t > 150$. Bottom left: sliding FFT. Bottom right: FFT corresponding to the streamwise velocity signal figure at top left.}
    \label{fig:dns_AoAm5}
\end{figure}

For large negative $\alpha$ the scenario is more convoluted. This is visualised in figure \ref{fig:dns_AoAm5} by means of the results of the 3D nonlinear simulations for $\alpha=-5^\circ$. We find that at $Re= 1200$ the flow is periodic and the flow dynamics is driven by mode $pA$ of the low-$Re$ steady base flow. When increasing $Re$ up to $Re=1450$, an additional peak at a frequency that matches that of $pB$ arises in the frequency spectrum, but only in the initial transient; see figure \ref{fig:dns_AoAm5}(a). For long integration times, indeed, the peak disappears, the corresponding flow oscillations are attenuated and the flow recovers its 2D periodic behaviour ($St \approx 1.06$). 
At $Re=2000$, instead, the nonlinear simulations show that the flow remains 2D, but settles into a limit cycle with a frequency of $St \approx 0.56$ which is half of that of the original one; see figure \ref{fig:dns_AoAm5}(b). For large negative $\alpha$ the secondary bifurcation is thus 2D and subharmonic.
\begin{figure}
\centerline{
\begin{tikzpicture}
\node at (4.5,8.4) {\includegraphics[width=1\textwidth]{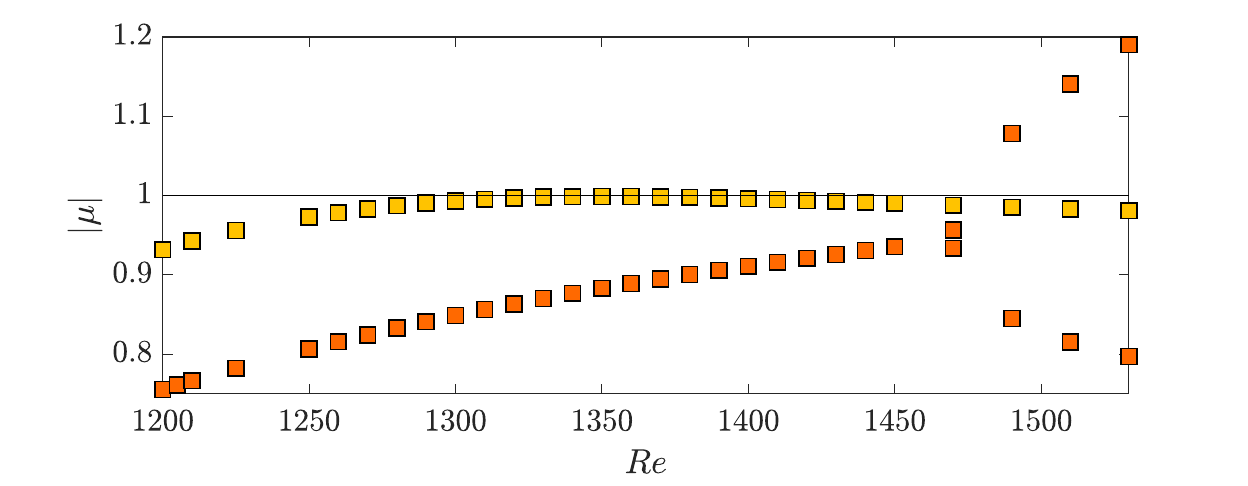}};
\node at (1.25,3.8) {\includegraphics[width=0.49\textwidth]{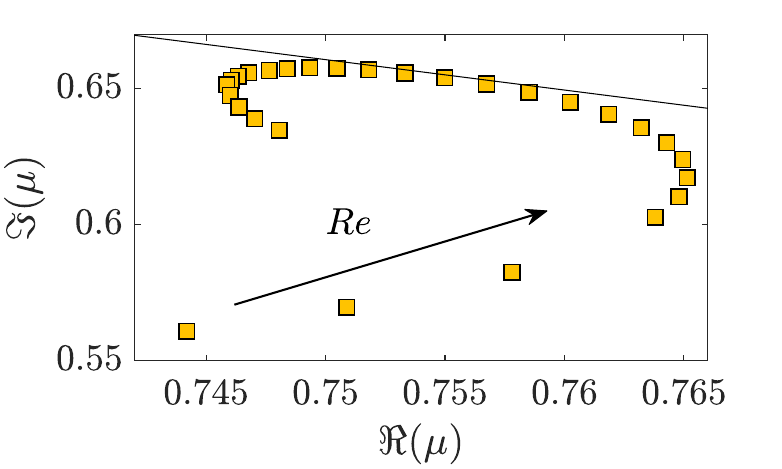}};
\node at (7.75,3.8) {\includegraphics[width=0.49\textwidth]{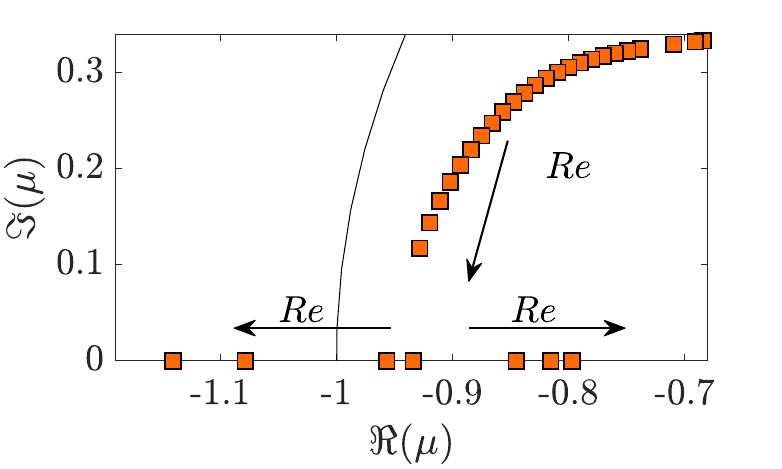}};
\node at (4.5,-0.1) {\includegraphics[trim={50 0 50 0},clip,width=1\textwidth]{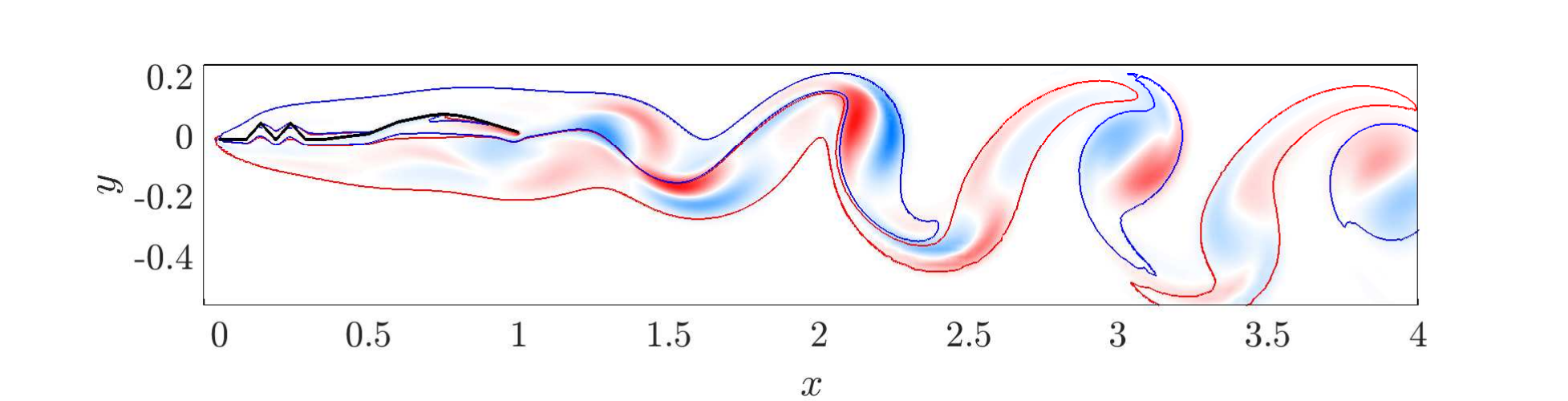}};
\node at (-0.1,10.3) {$(a)$};
\node at (-0.65,5.1) {$(b)$};
\node at (5.7,5.1)  {$(c)$};
\node at (-0.7,0.9)   {$(d)$};
\end{tikzpicture}
}
\caption{Secondary bifurcation for $\alpha = -5^\circ$. Panel $(a)$: dependence of the modulus of the Floquet multipliers on $Re$. Panel $(b)$: Floquet multipliers associated with the mode $sQPb$ in the complex plane. Panel $(c)$: Floquet multipliers associated with the mode $sSb$ responsible for the secondary bifurcation. Panel $(d)$: Floquet mode responsible for the secondary bifurcation at $Re = 1510$. The colourmap represents contours of spanwise vorticity of the perturbation field, while the red and blue solid lines are isocontours $\Omega_{z,2} =  \pm 0.5$ of the periodic base flow.}
\label{fig:Aoam5_sec}
\end{figure}
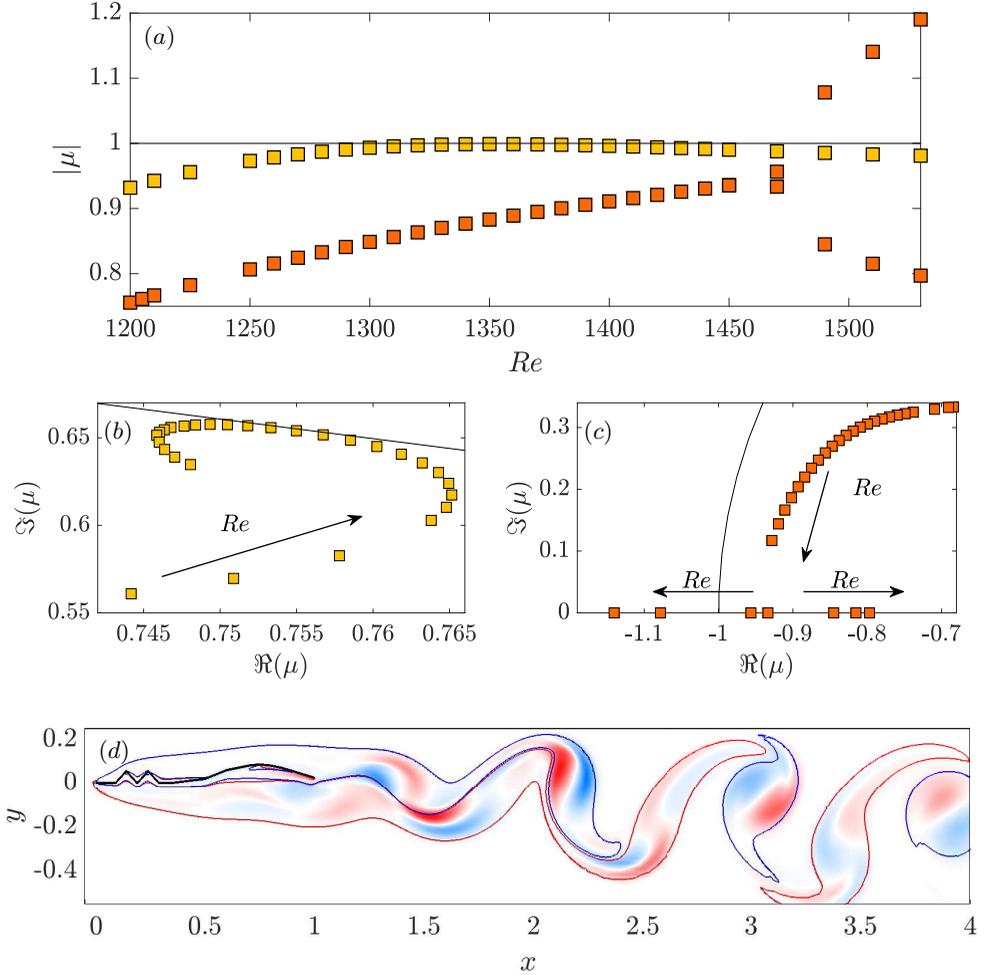

The Floquet analysis supports this picture. In fact, as shown in figure \ref{fig:Aoam5_sec}, for $\alpha = -5^\circ$ mode $sQPb$ does not exhibit a positive amplification rate (i.e. $|\mu|<1$) in the range of $Re$ investigated. The modulus of the corresponding Floquet multipliers first increases with $Re$, with the multiplier almost crossing the unit circle for $Re =1350$ (\ie $|\mu | \approx 1$), and then decays for larger $Re$. For $Re = 1350$ the pair of complex conjugate multipliers is $ \mu = (0.755,\pm 0.654)$, corresponding to Floquet exponents of $\sigma_2 = (-0.0011,\pm 0.7606)$ and to a frequency of $f \approx 0.12$, that is very close to the $ St \approx 0.13$ value detected in the transient of the nonlinear simulation and found for mode $pB$ at criticality. Note that unlike for $\alpha = -1.5^\circ$, for $\alpha = -5^\circ$ mode $sQPb$ exhibits a quasi-zero growth rate (marginal stability) at a Reynolds number that is smaller than the $Re_{c,pB} \approx 1650$ found with the stability analysis of the low-$Re$ steady base flow. In agreement with the nonlinear simulations, the Floquet analysis detects an additional branch of multipliers with $\beta = 0$ that indeed becomes amplified at $Re \gtrapprox 1480$. For $Re < 1470$, this branch consists of a pair of complex conjugate multipliers with negative real part and a non null imaginary part that decreases as $Re$ increases. At $Re \approx 1470$ the two multipliers collapse on the real axis and become real. Then, further increasing $Re$ the two multipliers move along the real axes and one crosses the unit circumference at $Re \approx 1480$ and the corresponding mode exhibits a positive growth rate: this confirms that the secondary bifurcation of the flow is 2D and of subharmonic nature. Figure \ref{fig:Aoam5_sec} shows the structure of the Floquet mode responsible for the bifurcation, referred to as mode $sSb$. Similarly to mode $sQPb$, this mode has large values in the wake and over the bottom side of the body. In the wake the perturbation field synchronises with the base flow, with perturbation dipoles being placed within the base flow monopoles. The orientation of these dipoles changes from one period to the next one accordingly with the subharmonic nature of the mode.

\subsection{The 3D synchronous bifurcation for $-1.25^\circ < \alpha \le 7^\circ$}
\label{sec:3Dbif}

For $-1.25^\circ < \alpha \le 7^\circ$ the secondary bifurcation leads to a 3D flow that preserves the same temporal periodicity of the periodic base flow; see figure \ref{fig:St_Re-alpha_plane}. In this section we detail this bifurcation by considering $\alpha=3^\circ$ and $\alpha=7^\circ$ as representative cases. We only investigate the leading modes of the periodic limit cycles, i.e. the ones that drive the secondary instability of the flow.

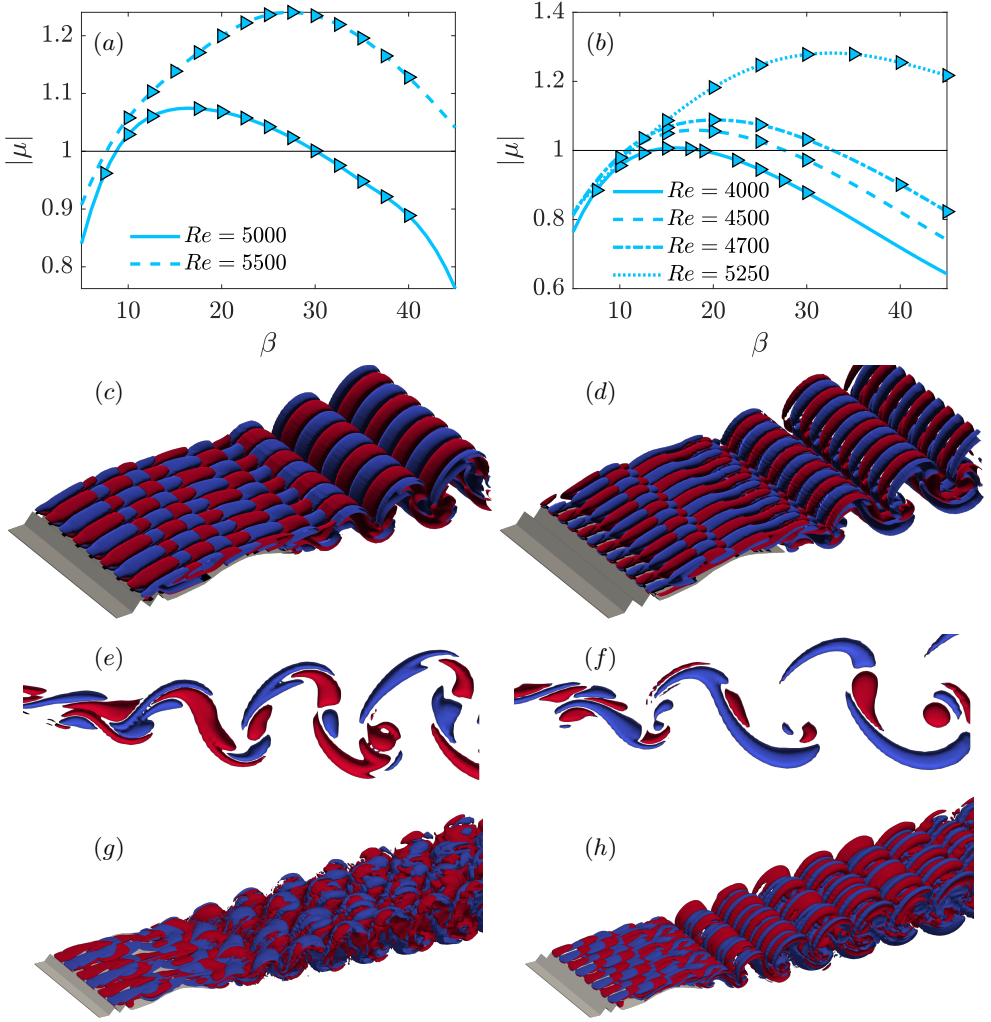
\begin{figure}
  \centerline{
  \begin{tikzpicture}
  \centering
  \node at (0,8.4)   {\includegraphics[width=0.49\textwidth]{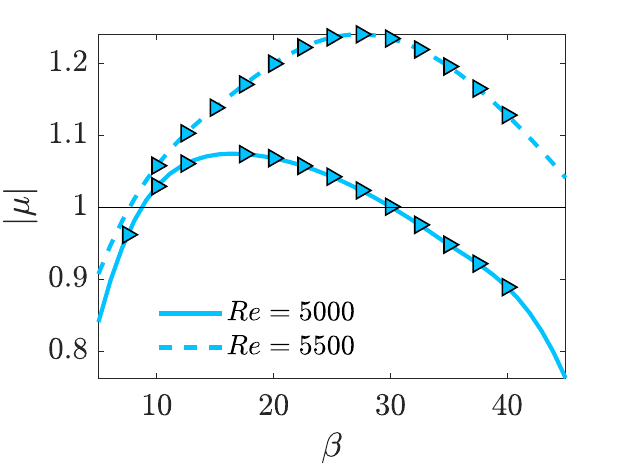}};
  \node at (6.5,8.4) {\includegraphics[width=0.49\textwidth]{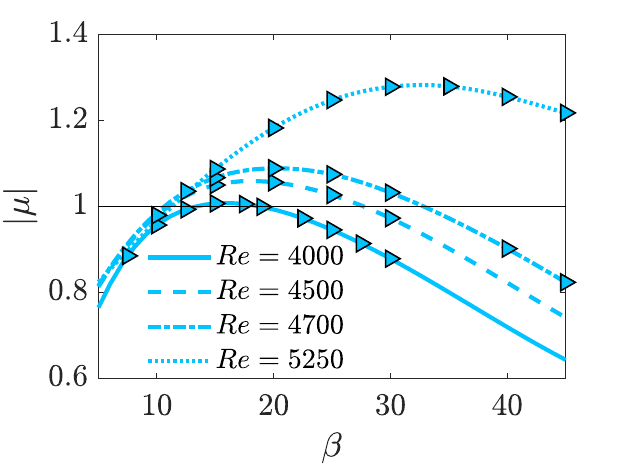}};
  \node at (0,4.2)   {\includegraphics[width=0.49\textwidth]{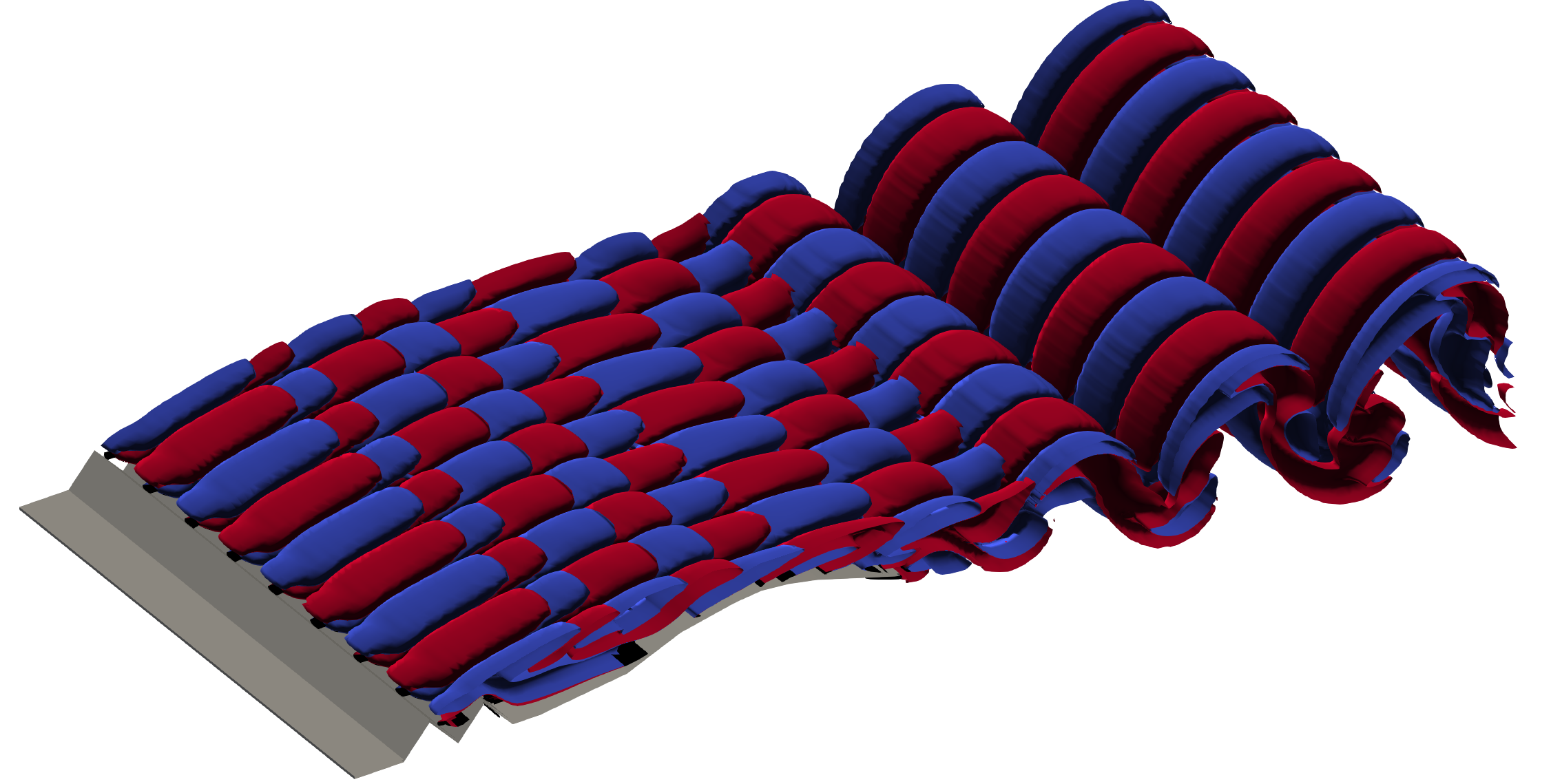}};
  \node at (6.5,4.2) {\includegraphics[width=0.49\textwidth]{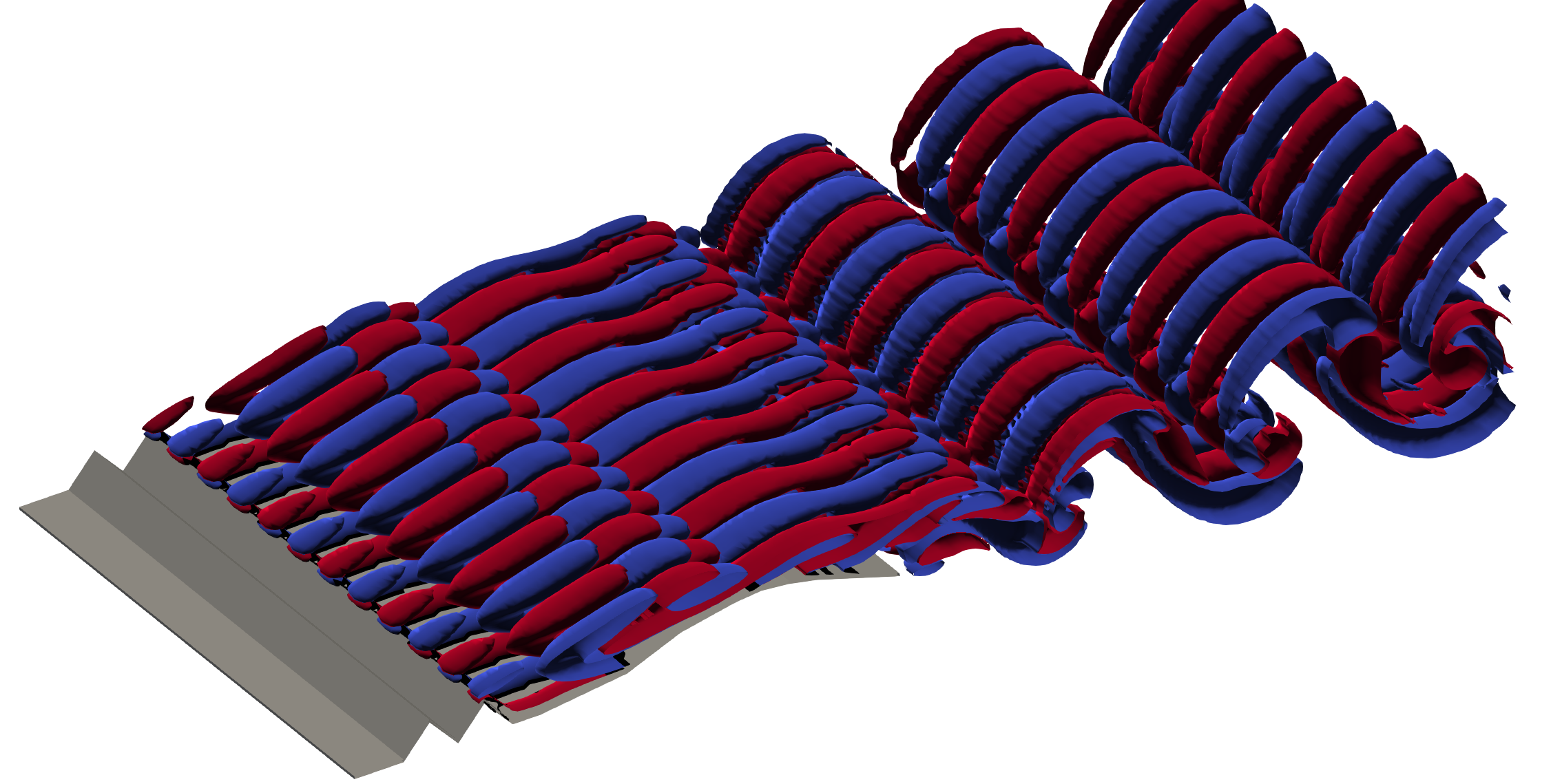}};
  \node at (0,1.3)     {\includegraphics[trim={650 350 0 250},clip,width=0.45\textwidth]{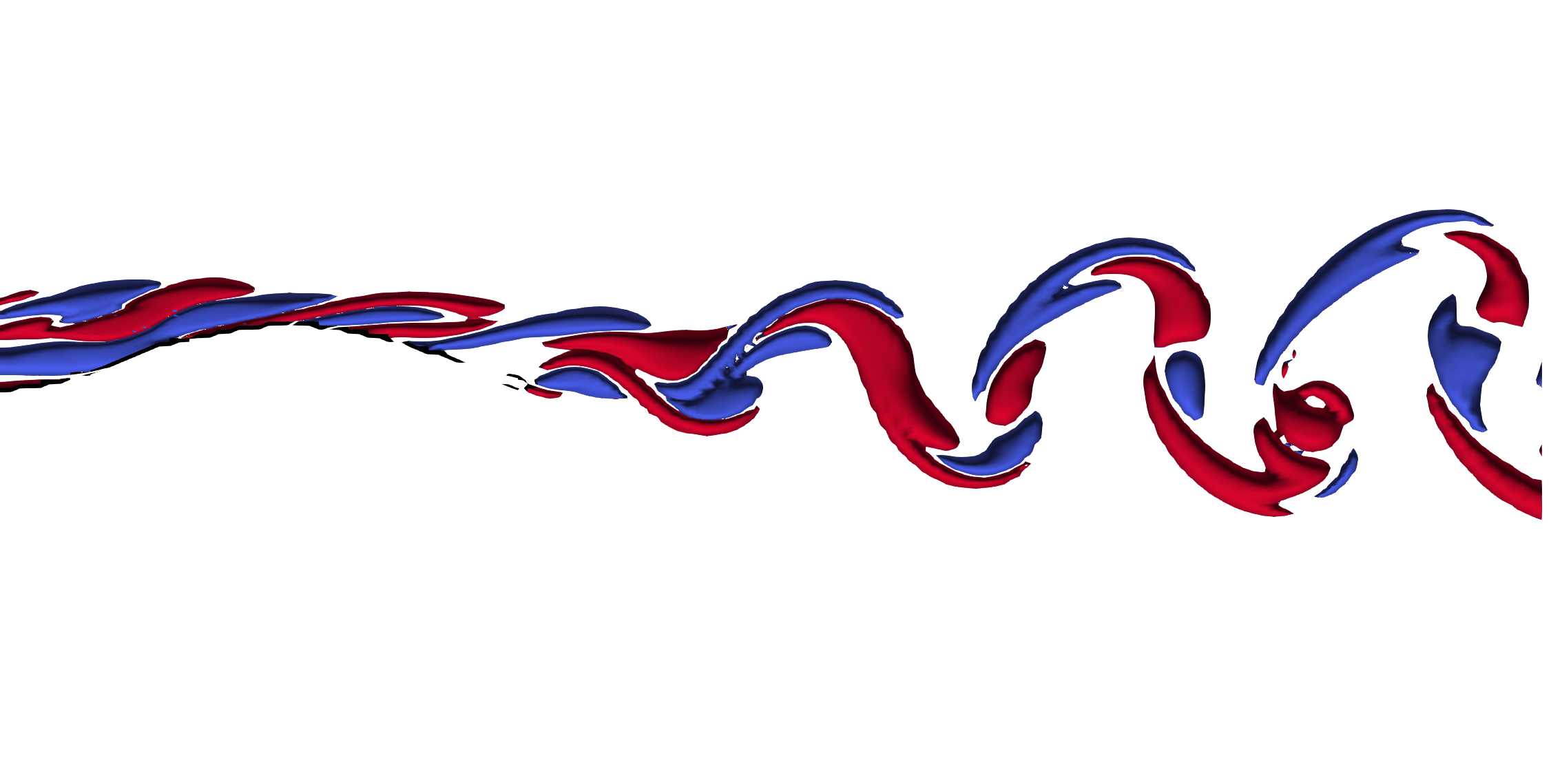}};
  \node at (6.5,1.3)   {\includegraphics[trim={650 350 0 250},clip,width=0.45\textwidth]{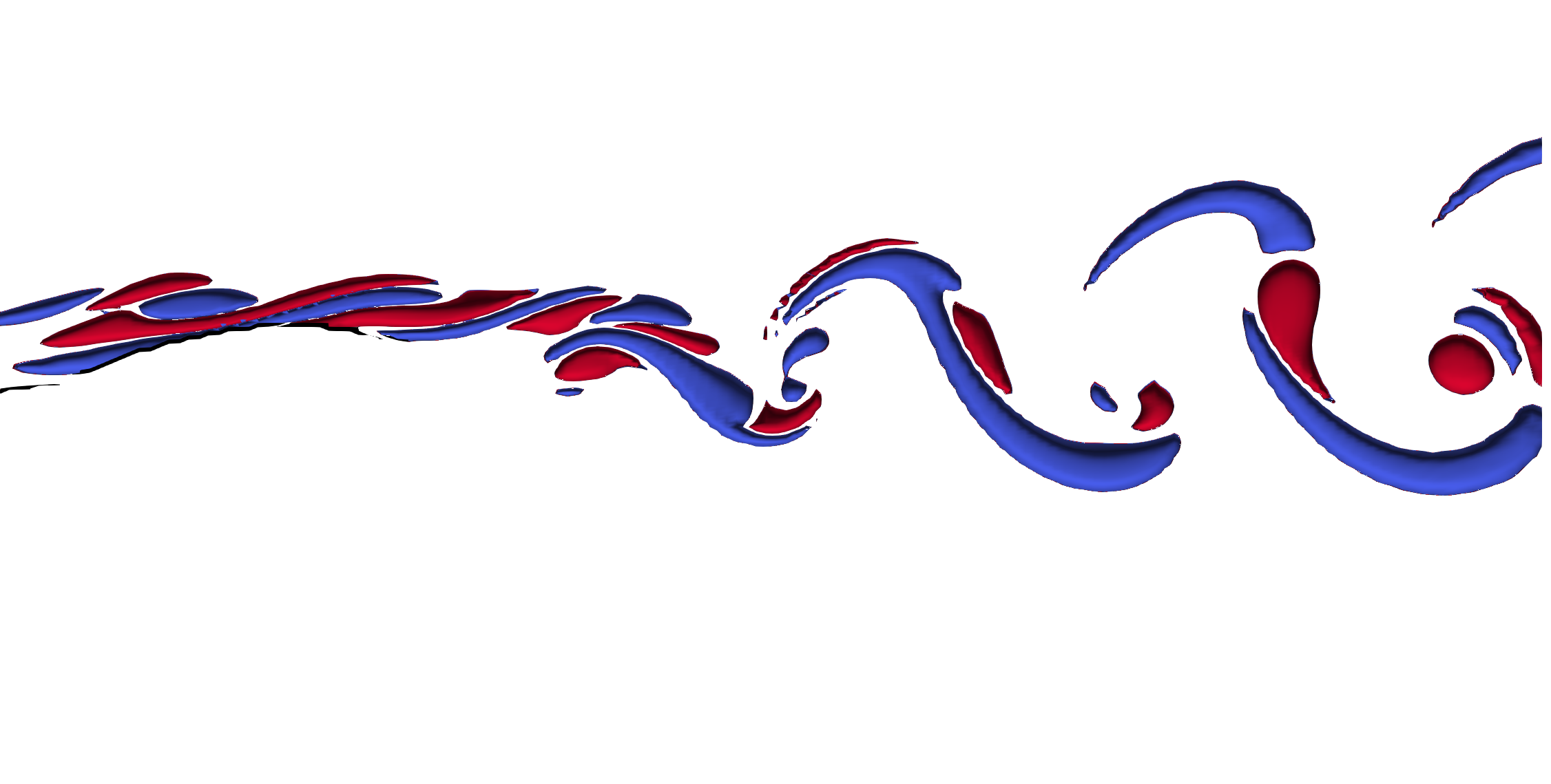}};    
  \node at (0,-1.3)  {\includegraphics[trim={0 0 0 0},clip,width=0.45\textwidth]{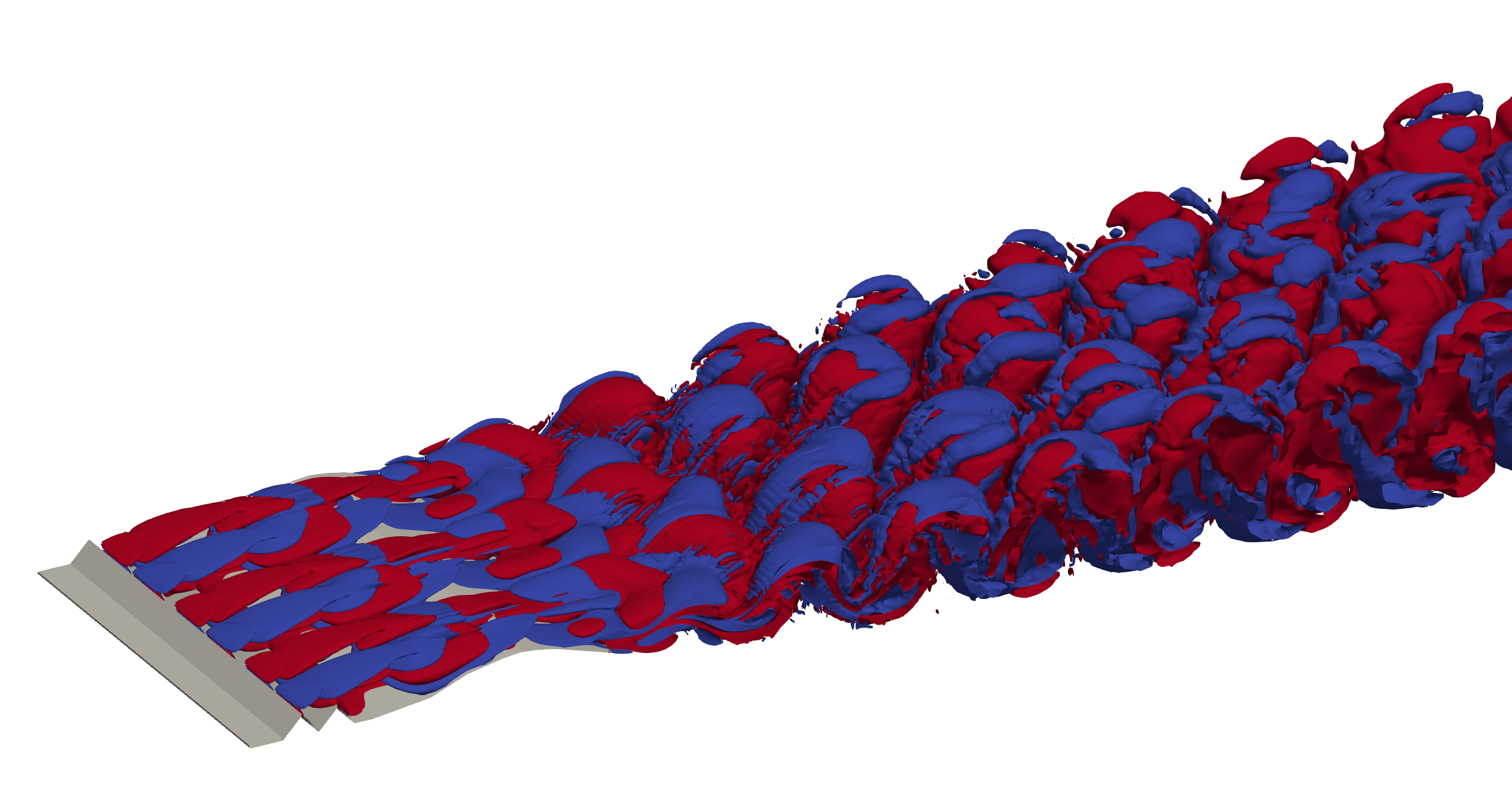}};
  \node at (6.5,-1.3){\includegraphics[trim={0 0 0 0},clip,width=0.45\textwidth]{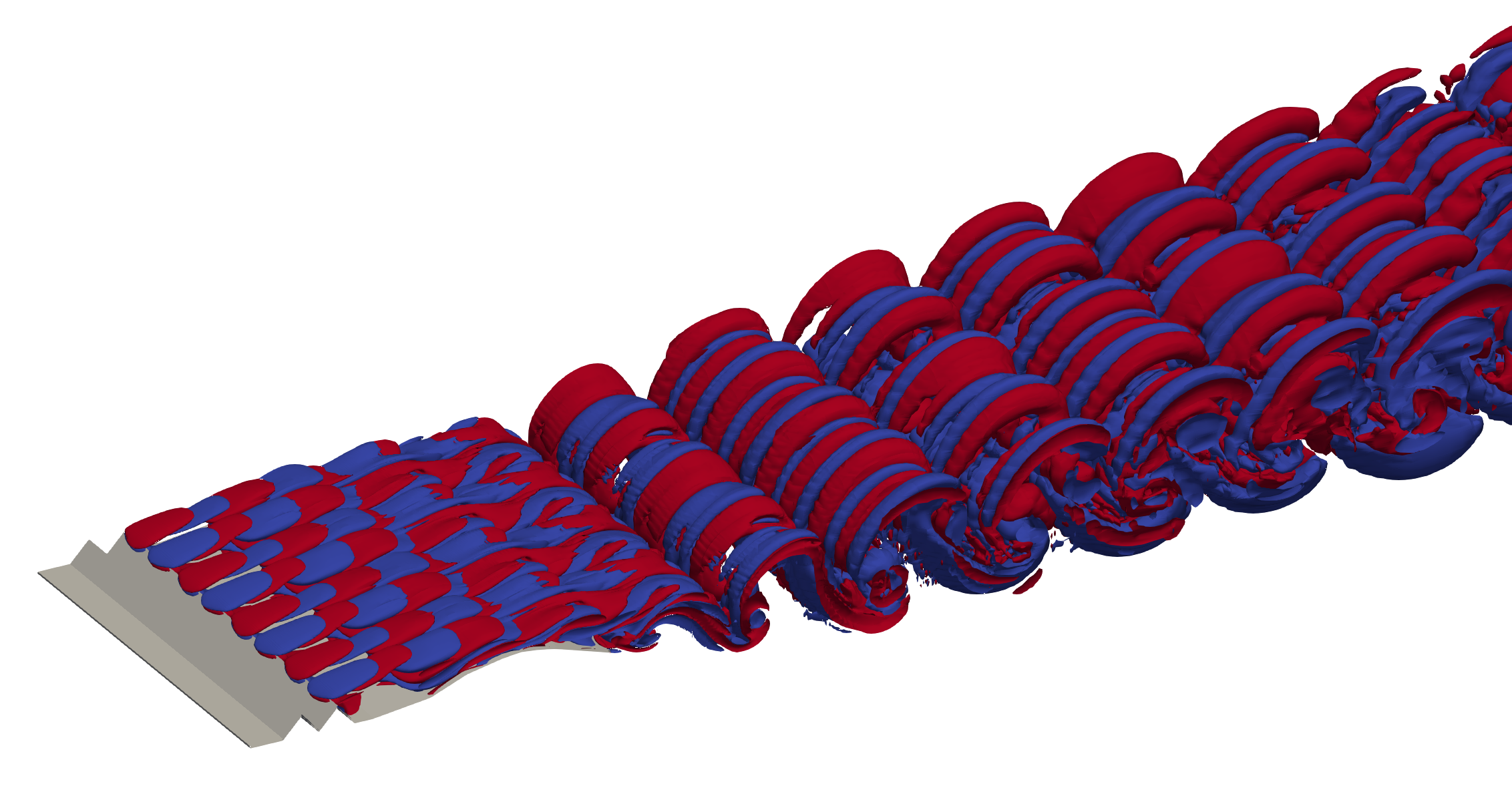}};
  \node at (-1.9,10.1) {$(a)$};
  \node at (4.6, 10.1) {$(b)$};
  \node at (-1.9,5.5)  {$(c)$};
  \node at (4.6, 5.5)  {$(d)$};
  \node at (-1.9,2.0)  {$(e)$};
  \node at (4.6, 2.0)  {$(f)$};
  \node at (-1.9,-0.5) {$(g)$};
  \node at (4.6, -0.5) {$(h)$};
  \end{tikzpicture}
  }  
  \caption{Secondary bifurcation for $\alpha=3^\circ$ (left) and $\alpha=7^\circ$ (right) via mode $sA$. Panels $(a,b)$: Modulus of the Floquet multipliers associated with the most amplified mode as a function of $\beta$ and $Re$. Here all the multipliers are real and positive. Panels $(c,d)$: $3D$ reconstruction of the Floquet mode over a spanwise extent of $L_z=c$ for $\alpha=3^\circ$, $Re=5500$ and $\beta=25$ (left) and $\alpha=7^\circ$, $Re=5250$ and $\beta=35$ (right). The red/blue colour denotes positive/negative isosurfaces of the streamwise vorticity with value $\hat{\omega}_{x,2} = \pm 5$. Panels $(e,f)$ are a lateral view and a zoom in the wake region of panels $(c,d)$. Panels $(g,h)$: instantaneous snapshot from the nonlinear 3D simulations for $\alpha=3^\circ$ and $Re=5500$ (left), and $\alpha=7^\circ$ and $Re=5250$ (right). The red/blue colour denotes positive/negative (total) streamwise vorticity with value $\omega_x = \pm 0.5$.}
  \label{fig:Secondary_AoA3_AoA7}
\end{figure}

For all $\alpha$ in the $-1.25^\circ < \alpha \le 7^\circ$ range, the secondary instability consists of a pitchfork bifurcation that leads to a 3D state with the same periodicity of the non bifurcated 2D limit cycle. Like the well-known cases of the circular and square cylinders \citep{barkley-henderson-1996,robichaux-balachandar-vanka-1999}, the 2D periodic flow is linearly unstable to synchronous 3D perturbations. Figure \ref{fig:Secondary_AoA3_AoA7} characterises the bifurcation for $\alpha=3^\circ$ (left panels) and $\alpha=7^\circ$ (right panels) using results from Floquet stability analysis (top and central panels) and the 3D nonlinear simulations (bottom panels). For all cases, the Floquet analysis shows that the first multiplier to cross the unit circle is real and positive. The critical Reynolds number $\Rey_{c,2}$ decreases with $\alpha$, being slightly less than $\Rey=5000$ for $\alpha=3^\circ$ and $Re_{c2} \approx 4500$ for $\alpha=7^\circ$, in agreement with the results of the nonlinear simulations shown in figure \ref{fig:St_Re-alpha_plane}. For $\Rey>\Rey_{c2}$ a band of amplified wavenumbers appears and widens as $\Rey$ increases, with the leading branch moving towards larger $\beta$; see figure \ref{fig:Secondary_AoA3_AoA7}$(a,b)$. The critical wavenumber is $\beta_c \approx 15 - 20$ at $\Rey \approx \Rey_{c2}$, and slightly increases with $\alpha$. Hereafter, we refer to this mode as mode $sA$. Notably, this perfectly matches with the results of the 3D nonlinear simulations; see figure \ref{fig:Secondary_AoA3_AoA7}$(g,h)$. Indeed, the wavenumbers emerging after to the establishment of the secondary mode are $\beta = 2 \pi/\lambda_z \approx 19$ for $\alpha=3^\circ$ and $Re=5500$, and $\beta = 2 \pi/\lambda_z \approx 25$ for $\alpha=7^\circ$ and $Re=5250$; note that at these $Re$ the flow is already the non-periodic regime for both $\alpha$ (see figure \ref{fig:St_Re-alpha_plane}). The small discrepancy with the $\beta_c$ values detected with the stability analysis is due to the considered $Re$ that is above the $Re_{c2}$ threshold, and to the periodic boundary conditions that enforce the wavelength of the disturbance to be a sub-multiplier of $\ell_z$. 

Overall, these results show that the route to three-dimensionalisation for non-smooth airfoils differs from what found for smooth airfoils. Indeed, in the present case, due to the complex vortex interaction on the top side of the airfoil (see the following discussion), the flow three-dimensionalisation is driven by the synchronous mode $sA$, while \cite{gupta-etal-2023} found that for NACA airfoils the first 3D mode to become amplified is of subharmonic nature for all $\alpha$. 

The central panels of figure \ref{fig:Secondary_AoA3_AoA7} show the spatial structure of mode $sA$ (reconstructed in the spanwise direction between $0 \le z \le 1$) for $\alpha=3^\circ$ (panels $c$ and $e$) and $\alpha=7^\circ$ (panels $d$ and $f$). 
The perturbation fields are maximum near the vortex cores of the base flow, as typical for the flow past bluff bodies \citep[see for example][]{thompson-leweke-williamson-2001}. Unlike for short bluff bodies \citep{barkley-henderson-1996,robichaux-balachandar-vanka-1999,choi-yang-2014} or for elongated bluff bodies with streamlined LE \citep{ryan-thompson-hourigan-2005}, here the perturbations are not localised in the wake, but large values are detected also over the top side of the body. The same structure is observed also with the 3D nonlinear simulations (panels $g$ and $h$). This recalls the mode $QS$ of the flow past elongated rectangular cylinders \citep{chiarini-quadrio-auteri-2022d}, and indicates that this is not a growing mode of the wake, but it is related to the dynamics of the vortices over the lateral side of the airfoil, as shown in the following with the structural sensitivity. Note that the symmetry is such that the sign of the streamwise vorticity does not change from one period to the next, in agreement with the synchronous nature of the mode. The spatial structure of mode $sA$ changes when $\alpha$ increases, in agreement with the change of the base flow; see figure \ref{fig:Secondary_AoA3_AoA7}$(e,f)$. Unlike for larger $\alpha$ ($\alpha=7^\circ$), for small $\alpha$ ($\alpha=3^ \circ$), the stremwise vorticity in the braid regions that connect the vortex cores changes sign from one half period to the next one. 

\begin{figure}
\centering
\includegraphics[width=1\textwidth]{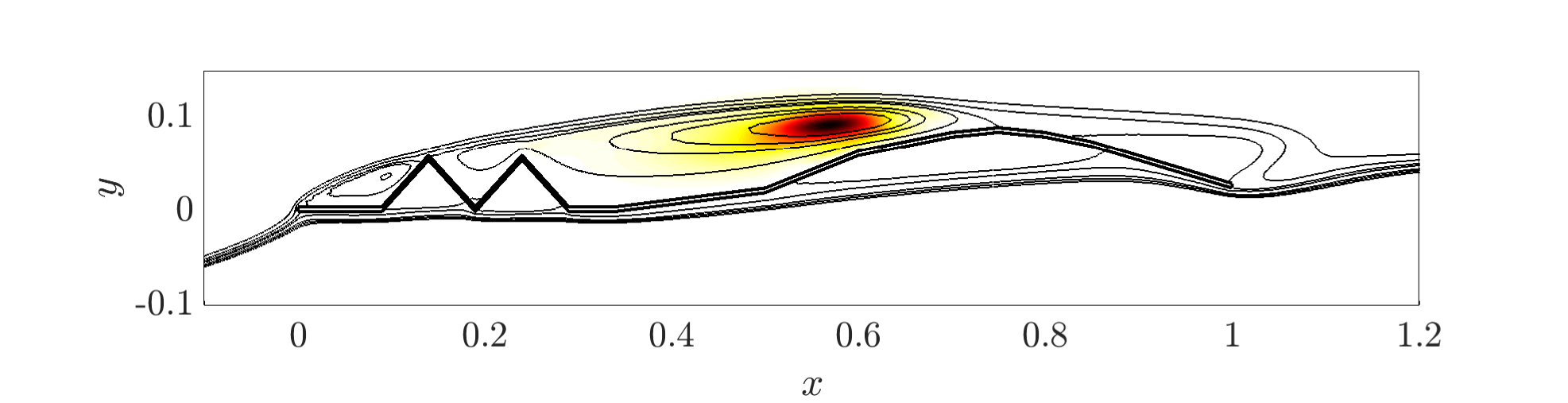}
\caption{Time averaged structural sensitivity for mode $sA$ for $\alpha=7^\circ$, $Re=5250$ and $\beta=35$.}
\label{fig:sAsens_tav}
\end{figure}
Figure \ref{fig:sAsens_tav} shows the spectral norm of the time-averaged structural sensitivity tensor \citep{giannetti-camarri-luchini-2010}
\begin{equation}
S(x,y,\beta) = \frac{\int_t^{t+T} \hat{\bm{u}}^\dag(x,y,\beta,t) \hat{\bm{u}}(x,y,\beta,t) \text{d} t}
{ \int_t^{t+T} \int_\Omega \hat{\bm{u}}^\dag \cdot \hat{\bm{u}} \text{d}\Omega \text{d}t}
\end{equation}
for mode $sA$. Here we consider $\alpha=7^\circ$, but the same distribution has been found also for smaller $\alpha$.
The norm $\Vert S \Vert$ of the sensitivity tensor vanishes almost everywhere, except near the top side of the body, indicating a localised wavemaker region. In agreement with the spatial structure of the direct modes, this differs from what observed for the classical modes $A$ and $B$ of the circular cylinder case, as the structural sensitivity does not peak in the wake region, but within the recirculating regions that arises downstream the two grooves. This confirms that the triggering mechanism is localised in this region, and is related to the dynamics of the top side recirculating region. Despite the different vortex dynamics in the base flow, the map of $\Vert S \Vert$ has a similar spatial distribution for both small ($-1.25^\circ \le \alpha \le 3^\circ$) and intermediate ($4 \le \alpha \le 7^\circ$) $\alpha$, indicating that the vortex shedding over the lateral top side of the body observed for the latter (see \S\ref{sec:int-alfa}) does not play any role in the triggering mechanism. In particular, $\Vert S \Vert$ peaks close to the elliptic stagnation point that identifies the downstream recirculating region within the average $\Psi_2=0$ streamline that separates from the corrugations. 

\subsection{The 3D subharmonic bifurcation for $ 8^\circ \le \alpha \le 10^\circ$}
\label{sec:2Dbif}

\begin{figure}
  \centerline{
  \begin{tikzpicture}
  \node at (3.4,8.4)   {\includegraphics[width=0.9\textwidth]{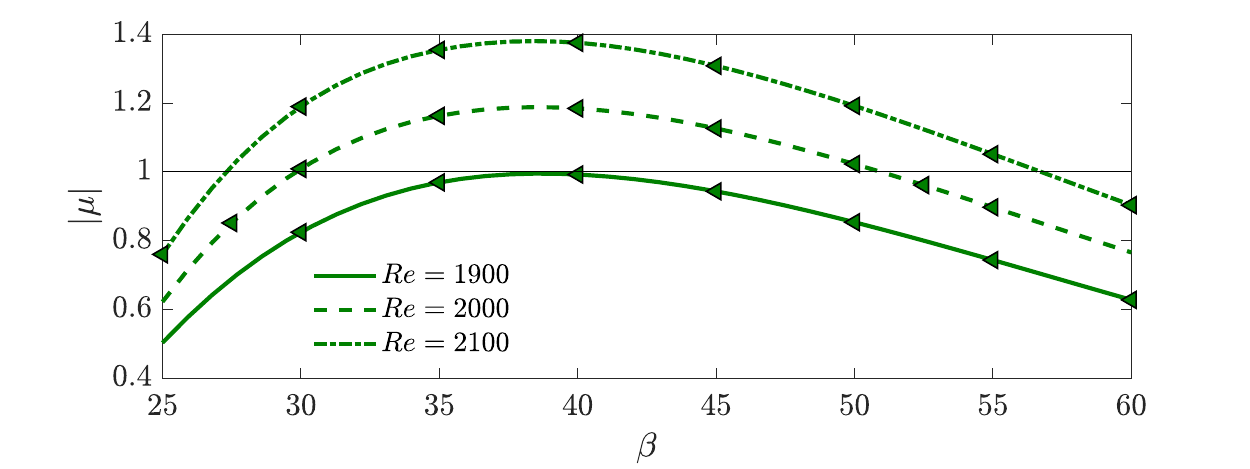}};
  \node at (0.0,4.1)   {\includegraphics[trim={120 0 120 0},clip,width=0.49\textwidth]{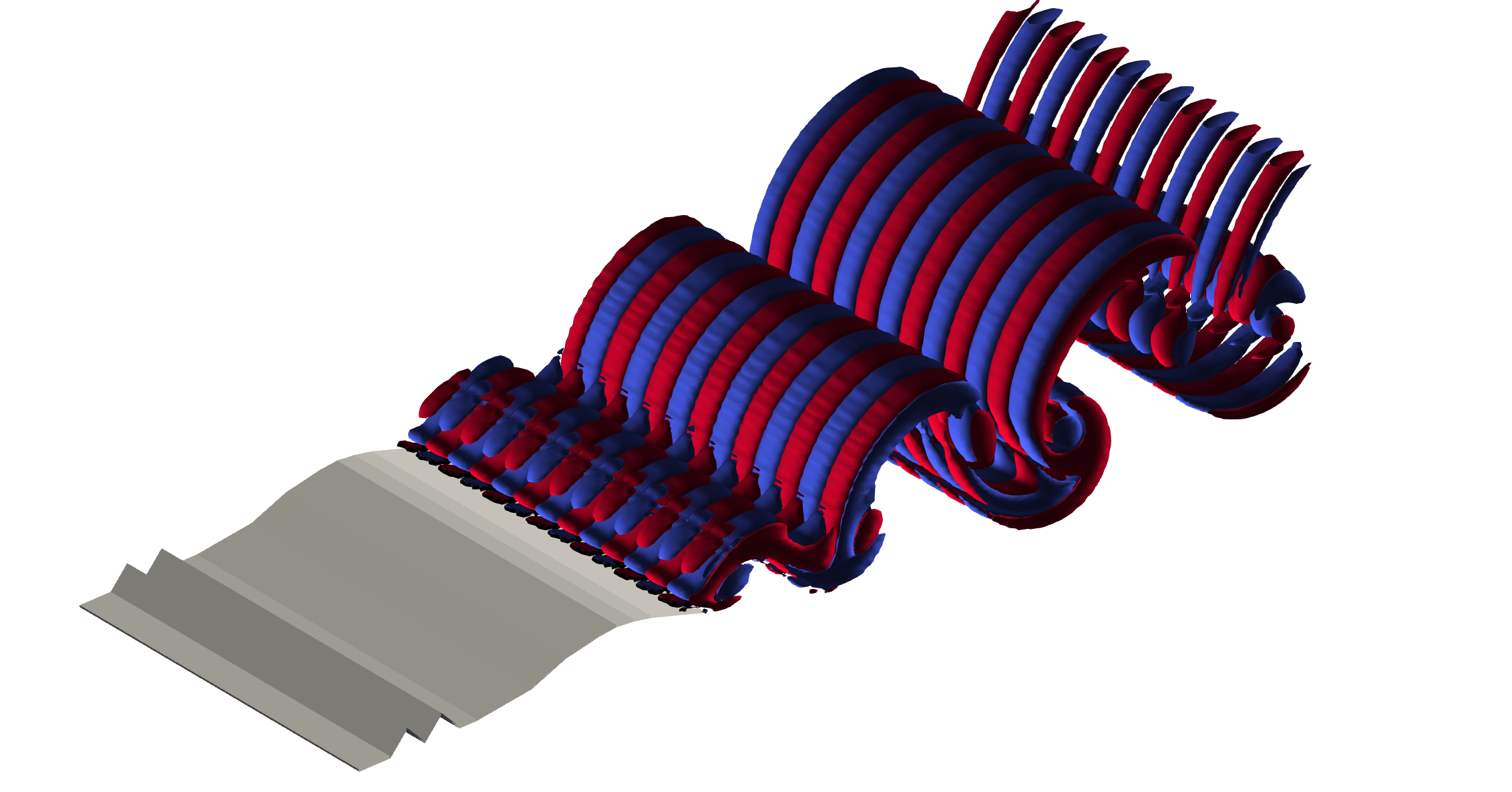}};
  \node at (6.5,4.1)   {\includegraphics[trim={120 0 120 0},clip,width=0.49\textwidth]{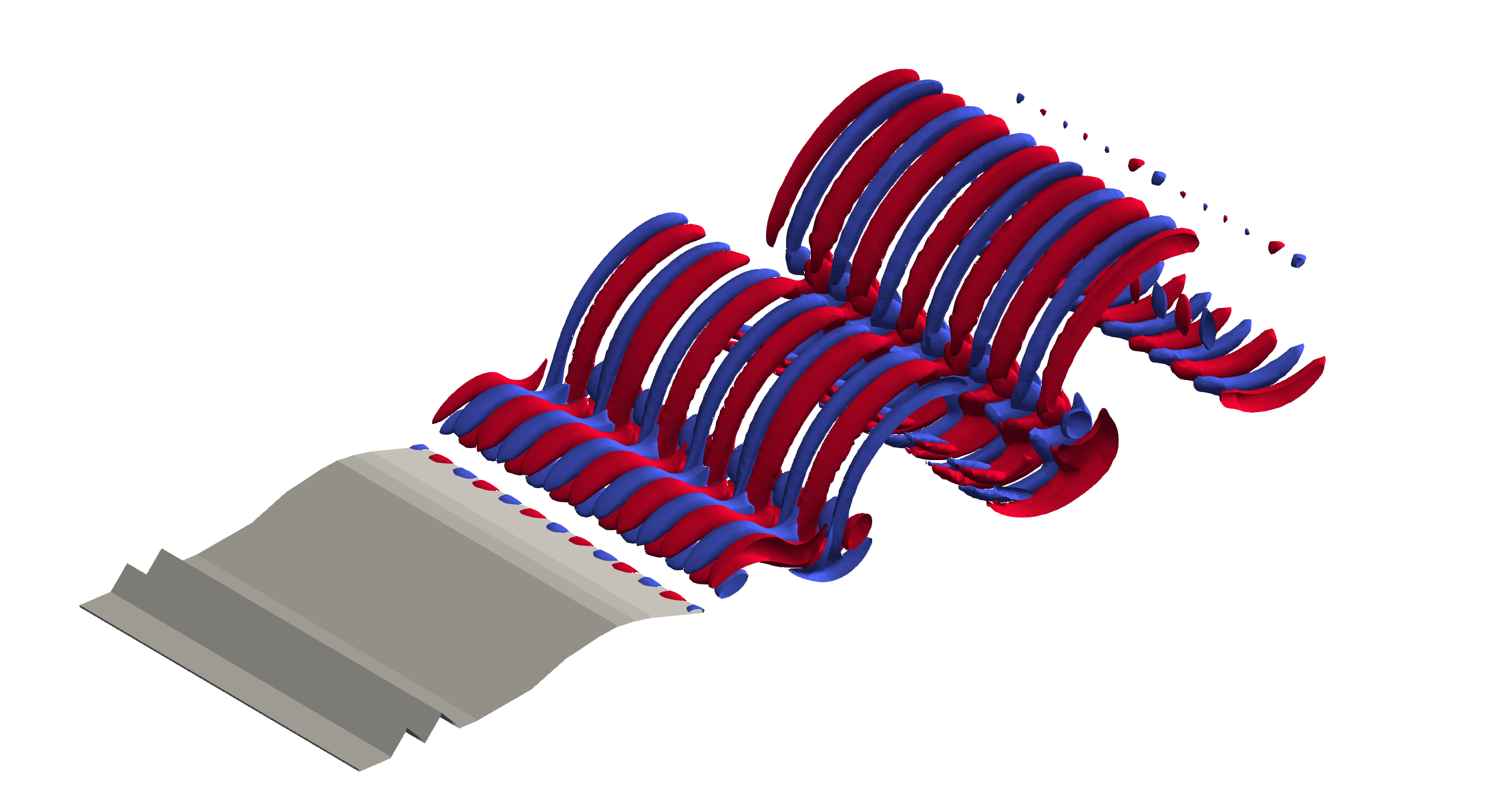}};
  \node at (-0.75,10.1) {$(a)$};
  \node at (-1.9,5.1) {$(b)$};
  \node at (4.5,5.1) {$(c)$};
  \end{tikzpicture}
  } 
  \caption{Secondary bifurcation for $\alpha=10^\circ$ via mode $sS$. Panel $(a)$: Modulus of the Floquet multipliers associated with the most amplified mode as a function of $\beta$ and $Re$. Here all the multipliers are real and negative. Panel $(b)$: $3D$ reconstruction of the Floquet mode for $Re=2100$ and $\beta=40$. The red/blue colour denotes positive/negative isosurfaces of the streamwise vorticity with value $\hat{\omega}_{x,2} \pm 5$. Bottom: instantaneous snapshot from the nonlinear 3D simulations at $Re=2100$. The red/blue colour denotes positive/negative (total) streamwise vorticity with value $\omega_x \pm 2$.}
  \label{fig:Secondary_AoA10}
\end{figure}
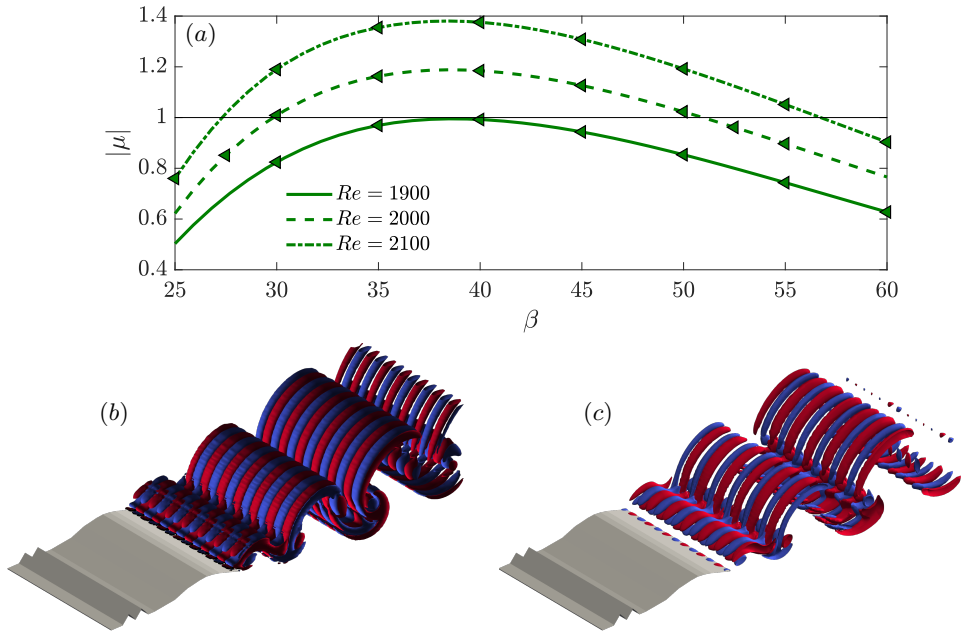 

We now focus on large positive $\alpha$.
For $8^\circ \le \alpha \le 10^\circ$ we find that the secondary flow instability consists of a 3D subharmonic bifurcation (see figure \ref{fig:flow_attractors_AoA10}). Here we characterise this bifurcation and use $\alpha=10^\circ$ as representative case; see figure \ref{fig:Secondary_AoA10}. Like in the previous section, we use both Floquet stability analysis (panels a and b) and 3D nonlinear simulations (panel c).

The Floquet stability analysis reveals the existence of a growing multiplier crossing the unit circle at $\mu = (-1,0)$ for $Re = Re_{c2} \approx 1900$, which is consistent with a subharmonic bifurcation; see figure \ref{fig:Secondary_AoA10}$(a)$. Hereafter, we refer to the corresponding mode as $sS$. The first multiplier with $|\mu|>1$ appears at $\beta \approx 37.5$, denoting a wavelength of $\lambda_z = 2 \pi / \beta \approx 0.175$ which is smaller with respect to that of mode $sA$. As $Re$ increases, a band of amplified wavenumbers appears, with $\beta_{\max}$ however only marginally changing with $Re$.

The subharmonic nature of mode $sS$ is conveniently visualised in figure \ref{fig:Secondary_AoA10}$(b)$, where the 3D spatial structure of the Floquet mode is reconstructed: the sign of $\hat{\omega}_{x,2}$ changes from one period to the next one (see the alternate red/blue isosurfaces in the wake along the streamwise direction). Notably, figure \ref{fig:Secondary_AoA10}$(b)$ also shows that, unlike $sA$, mode $sS$ is a wake mode: in this case the recirculating regions that arise over the top side of the airfoil do not play a role in the triggering mechanism. 
This scenario is corroborated by the 3D nonlinear simulations. The instantaneous streamwise vorticity field in figure \ref{fig:Secondary_AoA10}$(c)$, indeed, shows that the 3D perturbations grow downstream the body in the near wake, exhibiting a wavenumber of $\beta =  12 \pi \approx 37.7$, which is in perfect agreement with that provided by the Floquet analysis. 

Similarly to what found for the primary bifurcation and for mode $sA$, the critical Reynolds number $Re_{c2}$ of mode $sS$ decreases with $\alpha$. According to the Floquet stability analysis, indeed, $Re_{c2} \approx 3200$ for $\alpha=8^\circ$, $Re_{c2} \approx 2400$ for $\alpha=9^\circ$ and $Re_{c2} \approx 1900$ for $\alpha=10^\circ$ (not shown).
Interestingly, the secondary instability for these large $\alpha$ agrees with the results of \cite{yang-etal-2013} for inclined flat plates at $20^\circ \le \alpha \le 30^\circ$, and with those of \cite{gupta-etal-2023} for a smooth NACA0012 airfoil. \cite{gupta-etal-2023} indeed showed that for smooth airfoils the flow three-dimensionalisation at these $\alpha$ ($9^\circ \lessapprox \alpha \lessapprox 10^\circ$) is driven by a subharmonic mode of the wake with spanwise wavelength of $\lambda_z \approx 0.2$, that is close to our findings.

\section{Aerodynamic performance}
\label{sec:aero-perf}

Variations in the flow regime evidently affect the aerodynamic forces of the dragonfly-inspired geometry.
\cite{levy-seifert-2009} showed that the corrugated airfoil shows superior performance compared to other smooth airfoils (for example the low-$Re$ Eppler-E61 airfoil). In particular, they found that despite the lifting capabilities of the dragonfly-inspired airfoil are generally lower, the large decrease of the mean drag coefficient leads to an enhanced aerodynamic efficiency at all $\alpha$; for additional details we refer to appendix \ref{sec:comp-literat}, where the comparison is replicated for validation purposes. In this section, we discuss the aerodynamic performances of the corrugated airfoil in the considered portion of the $\alpha-Re$ parameter space, and relate them with the detected flow regimes.

\begin{figure}
    \centerline{
    \begin{tikzpicture}
    \node at (0,4.5) {\includegraphics[trim={0 0 0 0},clip,width=\textwidth]{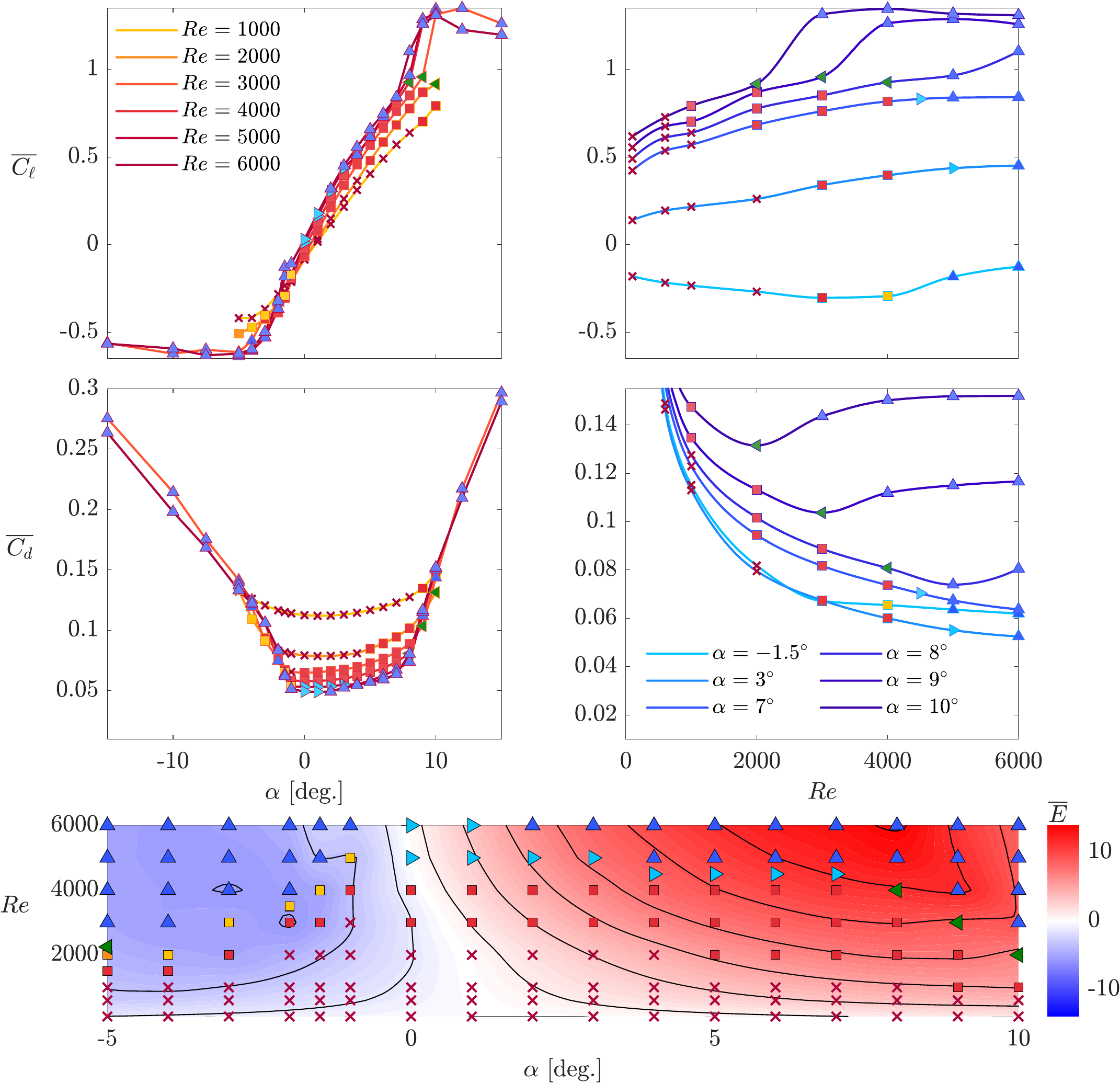}};
    \node at (-6.4,10.95) {$(a)$};
    \node at (0.1,10.95)  {$(b)$};
    \node at (-6.4,6.35)  {$(c)$};
    \node at (0.1,6.35)   {$(d)$};
    \node at (-6.4,1.35)  {$(e)$};
    \end{tikzpicture}
    }
    \caption{Mean aerodynamic forces. Top: mean lift coefficient as a function of $\alpha$ (panel $a$) and $Re$ (panel $b$). Centre: mean drag coefficient as a function of $\alpha$ (panel $c$) and of $Re$ (panel $d$). Bottom (panel $e$): mean lift-to-drag ratio in the $\alpha-\Rey$ parameter space. Symbols as in figure \ref{fig:St_Re-alpha_plane}.}
    \label{fig:CL-CD_Re-alpha_plane}
\end{figure}
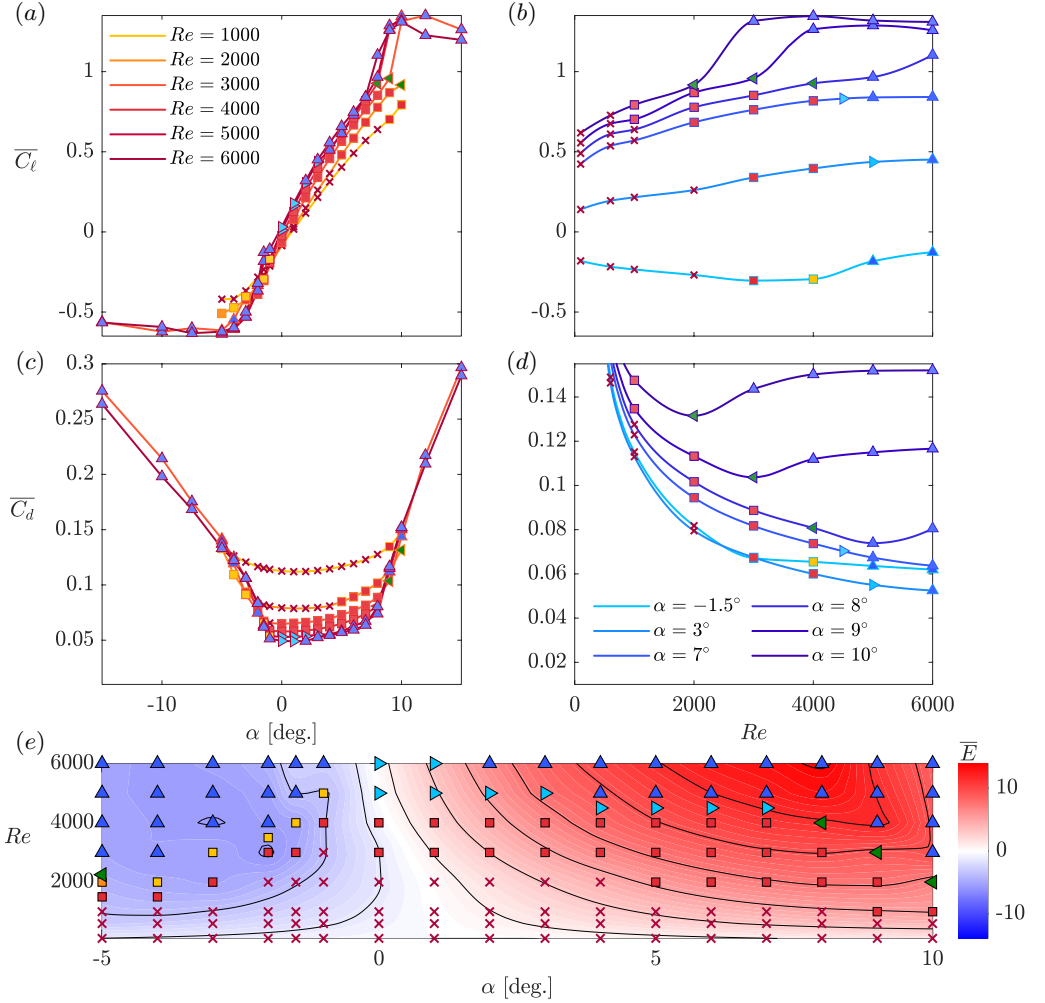

Figures \ref{fig:CL-CD_Re-alpha_plane}$(a,b,c,d)$ depict the mean lift and drag coefficients as a function of $\alpha$ and $Re$. The mean lift-to-drag ratio $\overline{C_\ell/C_d}$ in the whole $\alpha-\Rey$ parameter space is also shown in figure \ref{fig:CL-CD_Re-alpha_plane}$(e)$; here and hereafter the $\overline{\cdot}$ operator indicates time average. The symbols are the same as in figure \ref{fig:St_Re-alpha_plane}, and are used to highlight the flow regimes.
As shown in figure \ref{fig:CL-CD_Re-alpha_plane}$(a)$, the mean lift coefficient $\overline{C_\ell}$ shows a trend similar to that of traditional airfoils. It increases until reaching the stall which occurs at $\alpha \approx 12^{\circ}$ for $Re=3000$ and $\alpha \approx 10^{\circ}$ for $Re=6000$. Therefore, in the considered range of $Re$, stall incidence decreases with the Reynolds number. Interestingly, the strong increase of $\overline{C_\ell}$ before the stall at $Re=3000$ occurs immediately after the secondary 3D bifurcation of subharmonic nature. The mean drag coefficient, instead, has a non-monotonic dependence on $\alpha$ for all $Re$, and shows the classical parabolic shape for $\Rey \lesssim 2000$ only; figure \ref{fig:CL-CD_Re-alpha_plane}$(b)$. An increase of $\Rey$ leads to a growth of the scale selectivity~\citep{nastro2020}, and therefore to an increase of the flow sensitivity to the strongly-asymmetrical corrugated geometry. As a consequence, the $\overline{C_\ell}-\alpha$ and $\overline{C_d}-\alpha$ curves exhibit an accentuation of their asymmetry for increasing $\Rey$.

Figures \ref{fig:CL-CD_Re-alpha_plane}$(b,d)$ offer a complementary picture, with the $\overline{C_\ell}$ and $\overline{C_d}$ shown as a function of $Re$ for fixed values of $\alpha$. 
For $\alpha \le 0^\circ$, when fixing $\alpha$, $\overline{C_\ell}$ has a non-monotonic dependence on $Re$. 
For $\alpha>0^\circ$, instead, $\overline{C_\ell}$ increases with $Re$, albeit for $\alpha \geq 9^\circ$ it stops growing at $Re \gtrapprox 4000$. 
When fixing $\alpha$, the mean drag coefficient has a non-monotonic dependence on $Re$ for $\alpha>7^\circ$: $\overline{C_d}$ decreases for small $Re$ and then increases reaching a plateau for the largest $Re$. 
In this range, the minimum $\overline{C_d}$ increases with $\alpha$ and moves towards smaller $Re$.
Interestingly, this minimum value corresponds, or is very close, to the  3D subharmonic bifurcation of the limit cycle, \ie $\Rey \approx 4000$ for $\alpha = 8^\circ$, $\Rey \approx 3000$ for $\alpha = 9^\circ$ and $\Rey \approx 2000$ for $\alpha = 10^\circ$ (see the green triangles).
On the other hand, for $\alpha\leq 7^\circ$ $\overline{C_d}$ monotonically decreases with $Re$ in the considered range of parameters.
For $0^\circ \leq \alpha \leq 7^\circ$ the secondary 3D bifurcation via the synchronous mode (see the light blue triangles) does not induce a change in the trend of time-averaged aerodynamic coefficients when $\Rey$ is increased.

As shown in figure \ref{fig:CL-CD_Re-alpha_plane}(e), the combined effect of $Re$ on $\overline{C_\ell}$ and $\overline{C_d}$ leads to a monotonic increase (decrease) of the aerodynamic efficiency with $Re$ for $\alpha \geq -1^\circ$ ($\alpha \leq -1.5^\circ$).
For $\Rey = 6000$, the maximum lift-to-drag ratio is obtained at $\alpha \approx 8^\circ$, in agreement with the results of \citet{bauerheim-chapin-2020}.
At these $Re$ and $\alpha$ the lift experiences a significant increase, while the drag has not yet experienced the large increase as for larger $\alpha$.
As found also by \citet{bauerheim-chapin-2020} for $\Rey = 6000$, the largest lift and drag coefficients, indeed, are reached at $\alpha \approx 10^\circ$.

\begin{figure}
    \centerline{
    \begin{tikzpicture}
    \node at (0,4.5) {\includegraphics[trim={0 0 0 0},clip,width=\textwidth]{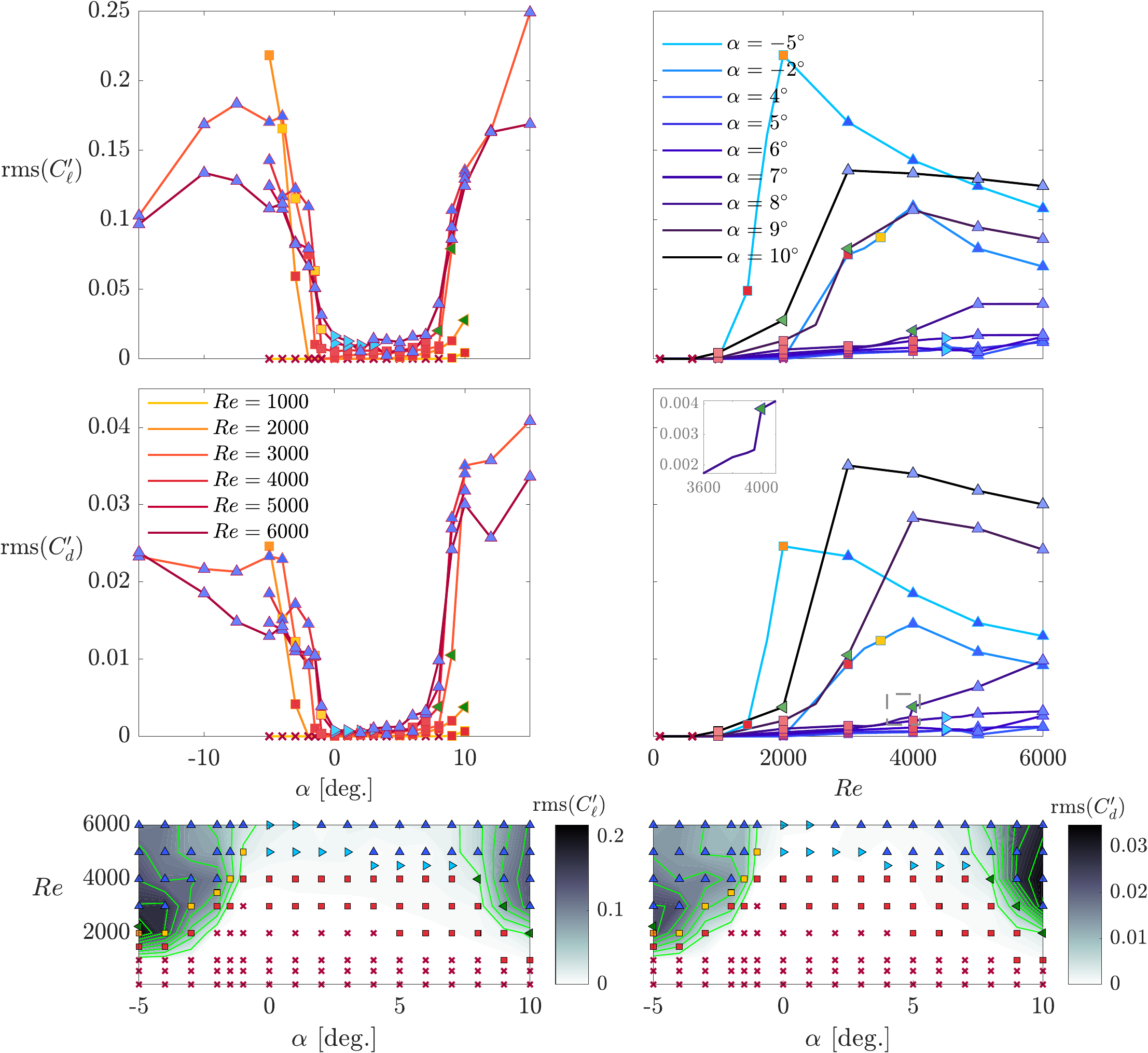}};
    \node at (-6.0,10.75) {$(a)$};
    \node at (0.6,10.75)  {$(b)$};
    \node at (-6.0,6.15)  {$(c)$};
    \node at (0.6,6.15)   {$(d)$};
    \node at (-6.0,1.35)  {$(e)$};
    \node at (0.6,1.35)   {$(f)$};
    \end{tikzpicture}
    }    
    \caption{Root mean square of the aerodynamics forces. Top: lift coefficient root mean square (rms) as a function of $\alpha$ (panel $a$) and $Re$ (panel $b$). Centre: drag coefficient rms as a function of the angle of $\alpha$ (panel $c$) and $Re$ (panel $d$). Bottom: lift (panel $e$) and drag (panel $f$) coefficient rms in the $\alpha-\Rey$ parameter space. The close-up inset in panel $d$ evidences the steep increase in rms near the secondary bifurcation at $\alpha = 8^\circ$. Symbols as in figure \ref{fig:St_Re-alpha_plane}.}
    \label{fig:CL-CD_rms_Re-alpha_plane}
\end{figure}

Figure \ref{fig:CL-CD_rms_Re-alpha_plane} depicts the root mean square (rms) of the lift and drag coefficients as a function of $\alpha$ (panels $a$ and $c$) and $\Rey$ (panels $b$ and $d$).
The lift and drag oscillations in the whole $\alpha-\Rey$ parameter space are also reported in figures \ref{fig:CL-CD_rms_Re-alpha_plane}$(e,f)$ for a comprehensive visualisation.
Figures \ref{fig:CL-CD_rms_Re-alpha_plane}$(a,c)$ show that for all $\Rey$ the lift- and drag-coefficient fluctuations are low in the $0^\circ \leq \alpha \leq 7^\circ$ range, while they become significant for negative and large positive $\alpha$. 
As shown in \S\ref{sec:inter-re}, this correlates well with the different nature of the secondary bifurcation.
The fluctuations are indeed found to strongly intensify when the flow experiences a secondary bifurcation that leads to a change in the primary frequency of the limit cycles ($\alpha \le -1.5^\circ$ and $\alpha > 7^\circ$), while a synchronous secondary bifurcation seems to affect less the lift/drag fluctuations ($-1.5^\circ < \alpha \le 7^\circ$). 
The maximum lift fluctuation indeed occurs at $\alpha = -5^\circ$ for $\Rey = 2000$ when the flow experiences a secondary bifurcation due to the 2D subharmonic mode, while the maximum drag fluctuation is found for $\alpha = 10^\circ$ at $\Rey \approx 4000$ after the 3D subharmonic secondary bifurcation of the flow.
This aspect is even more evident in figures \ref{fig:CL-CD_rms_Re-alpha_plane}$(b,d)$.
For example, at $\alpha = -5^\circ$ $\text{rms}(C_{\ell}')$ ($\text{rms}(C_{d}')$) increases from $0.05$ ($0.0015$) to $0.22$ ($0.025$) passing from $\Rey = 1450$ to $\Rey=2000$, and then slightly decreases for higher $\Rey$.
For $0^\circ \leq \alpha \leq 7^\circ$, instead, the fluctuations grow very weakly with $Re$ and $\alpha$: the maximum values, \ie $\text{rms}(C_{\ell}') = 0.017$ and $\text{rms}(C_{d}') = 0.0032$, are found at $\Rey = 6000$ and $\alpha = 7^\circ$.

A last comment regards the largest $Re$ and $\alpha$ considered. At $\alpha = 8^\circ$, $\text{rms}(C_d')$ continues to increase with $\Rey$ after the 3D subharmonic bifurcation, while $\text{rms}(C_\ell')$ reaches a plateau.
For $\alpha \geq 9^\circ$, instead, after the wake three-dimensionalisation the lift and drag oscillations reach a maximum and then weakly drop. This non-monotonic behaviour with $Re$ at large $|\alpha|$, also found by \cite{levy-seifert-2009}, is related to the complex vortex dynamics of this dragonfly-inspired geometry.

\section{Conclusions and perspectives}
\label{sec:conc_pers}

In this study we have investigated the sequence of bifurcations of the laminar flow past a dragonfly-inspired airfoil, using the geometry introduced by \cite{newman-etal-1977}. The angle of attack is varied between $-5^\circ \le \alpha \le 10^\circ$, and $Re$ is increased up to $Re=6000$. We have used linear stability analyses to describe the primary and secondary bifurcations, and 2D and 3D direct numerical simulations to fully account for the nonlinear effects and detail the bifurcation scenario at larger $Re$. The aim of this work is to characterise the flow regimes that arise in the $\alpha-Re$ space of parameters, characterise the flow bifurcations, and provide insights on the physics underlying the enhanced low-$Re$ aerodynamic performance of corrugated airfoils compared to more common smooth airfoils.

The linear stability analysis shows that for all $\alpha$ the flow experiences first a Hopf bifurcation, becoming unstable to oscillatory 2D perturbations that lead to alternating vortex-shedding in the wake. Two different modes, namely mode $pA$ and mode $pC$, are found to drive the primary instability for negative and positive $\alpha$ respectively. The two modes are characterised by a different frequency and a distinct spatial structure. The structural sensitivity \citep{giannetti-luchini-2007} shows that for mode $pC$ the wavemaker is localised in two lobes located downstream the trailing-edge in the wake; mode $pC$ is an unstable mode of the wake and resembles the classical von K\'{a}rm\'{a}n mode observed for common smooth airfoils \citep{nastro-etal-2023,gupta-etal-2023}. Mode $pA$, instead, is rather different. In this case the structural sensitivity does not show a localised peak and the mode is relatively large along the bottom side of the airfoil. This suggests the existence of a non-local feedback in the triggering mechanism, that involves the recirculating regions that originate from the shear layer that separates at the bottom leading edge. The critical $Re$ of the two modes decreases as $|\alpha|$ increases, and for large positive and negative $\alpha$ it is found to scale as a power law.

At intermediate $Re$ the flow approaches a periodic limit cycle for all $\alpha$ considered. Due to the complex geometry, different limit cycles arise, each one being characterised by a different vortex interaction. For positive $\alpha$, the vortex shedding in the wake is governed by the dynamics of the recirculating regions that arise over the top side of the airfoil. For negative $\alpha$, instead, the vortex shedding is driven by the dynamics of the recirculating regions placed over the bottom side. Interestingly, for intermediate positive/negative $\alpha$ the flow experiences vortex shedding from both the top/bottom LE and TE shear layers, and the two phenomena are frequency locked. This closely resembles what occurs for elongated rectangular cylinders \citep{okajima-1982,chiarini-quadrio-auteri-2022}. 

The nature of the secondary bifurcation changes with $\alpha$, in agreement with the different 2D limit cycles. Our results show that the secondary instability of the flow resembles what found for smooth airfoils \citep[see][]{gupta-etal-2023} for large positive $\alpha$ only, where the wake dynamics dominates. For smaller $\alpha$, instead, the nature of the secondary instability of the flow changes, being triggered by the complex dynamics of the vortices placed along the sides of the airfoil. The Floquet stability analysis reveals that the secondary bifurcation is 2D for negative $\alpha$, and 3D for positive ones. Like for the primary instability, the critical $Re$ decreases as $|\alpha|$ increases. For intermediate negative $\alpha$, the secondary bifurcation is associated with a pair of complex conjugate Floquet multipliers, which lead to a Neimark--Sacker bifurcation of the limit cycle: the flow remains 2D but loses the temporal periodicity. For large negative $\alpha$, instead, this branch of multipliers does not cross the unit circle, and the secondary bifurcation is driven by an additional branch of real negative multipliers. For these $\alpha$ the secondary instability is 2D and subharmonic. For positive $\alpha$ several modes of different nature arise. For small $\alpha$ the leading secondary mode is 3D and synchronous. This is similar to what found for bluff bodies, like circular \citep{barkley-henderson-1996} and square \citep{robichaux-balachandar-vanka-1999} cylinders. However, unlike modes $A$ and $B$ of the circular and square cylinder case, the structural sensitivity localises the wavemaker of this mode in the recirculating region placed over the top side, indicating that the triggering mechanism is localised there, and that it is related with the dynamics of the top-side vortices. For large positive $\alpha$, instead, the secondary instability of the flow consists of a 3D subharmonic bifurcation of the wake, similarly to what found for smooth airfoils \citep{gupta-etal-2023}.

The aerodynamic performance of the airfoil strongly depends on the flow regime.
The mean lift coefficient is found to increase with $\alpha$ for all $\Rey$, whereas the mean drag coefficient shows a non monotonic behaviour with respect to $\alpha$, with the classical parabolic shape observed for small $Re$ only.
The largest lift-to-drag ratio is achieved at $\alpha = 8^\circ$ and $\Rey = 6000$. At these $Re$ and $\alpha$ the lift has largely increased, while the drag has not experienced yet the large increase as for higher $\alpha$. The highest lift and drag coefficients are instead obtained for $Re = 6000$, but at the larger $\alpha$ of $\alpha \approx 10^\circ$.
Notably, we found that the most significant changes in the time-averaged lift/drag values and in their corresponding fluctuations occur when the 2D periodic wake transitions via a quasi-periodic or subharmonic secondary mode, i.e. the $sSb$ and $sQb$ modes for negative $\alpha$ and the $sS$ mode for high positive $\alpha$. The wake transition via the synchronous $sA$ mode for $-1.25^\circ < \alpha \leq 7^\circ$, instead, does not entail a significant change of the mean and fluctuations of the aerodynamic forces. In this range of intermediate $\alpha$, the dragonfly-inspired airfoil provides thus a better capability to limit integral force oscillations with respect to other smooth airfoils \citep{levy-seifert-2009}.

The present study will serve as a starting point to further investigation. A natural extension of this work, of particular interest in the field of UAV and MAV, may include the finite wing effects and tandem wings. 
Another interesting extension would be to consider a moving corrugated airfoil, to reproduce a flapping flight. This can be done in the spirit of the works by \cite{jallas-marquet-fabre-2017}, which investigated the flow regimes past a pitching smooth airfoil, and \cite{fujita-iima-2023}, that considered a corrugated airfoil and simplify the motion focusing on the unsteady lift generation by translating it from rest. Moreover, of particular interest would be to assess the performance of the corrugated airfoils under unsteady conditions by simulating a gust encounter. During their operation, indeed, small-scale UAVs and MAVs are like to encounter gusts that may also be of the order of the vehicle reference scale \citep{jones-etal-2022}. In this regard, also the effect of stochastic inflow conditions on the flow regimes, and on the aerodynamic performances as well, should be studied in a systematic manner. 
We also mention that a better understanding of the origin of the enhanced low-$Re$ aerodynamic performances of these corrugated airfoils may be obtained by performing a force element analysis \citep{chang-1992}, which allows to detects the origin of the aerodynamic loads \citep[see also][]{ribeiro-etal-2022}.
Eventually, of particular interest would be to include the dynamics of the structure in the analysis. 

\section*{Acknowledgements}
G. Nastro would like to acknowledge M. Bauerheim for introducing the dragonfly-inspired geometry considered in this work to him on his arrival at ISAE-SUPAERO.
Three-dimensional direct numerical simulations were performed using resources from GENCI [CCRT-CINES-IDRIS] (Grant A0122A07178) and from CALMIP (Grant 2022-p1425).

\section*{Funding} 
This research received no specific grant from any funding agency, commercial or not-for-profit sectors.

\section*{Declaration of Interests} 
The authors report no conflict of interest.

\appendix
\section{Comparison with the work of \cite{levy-seifert-2009}}
\label{sec:comp-literat}

\begin{figure}
    \centerline{
    \begin{tikzpicture}
    \node at (0,0) {\includegraphics[trim={0 0 0 0},clip,width=\textwidth]{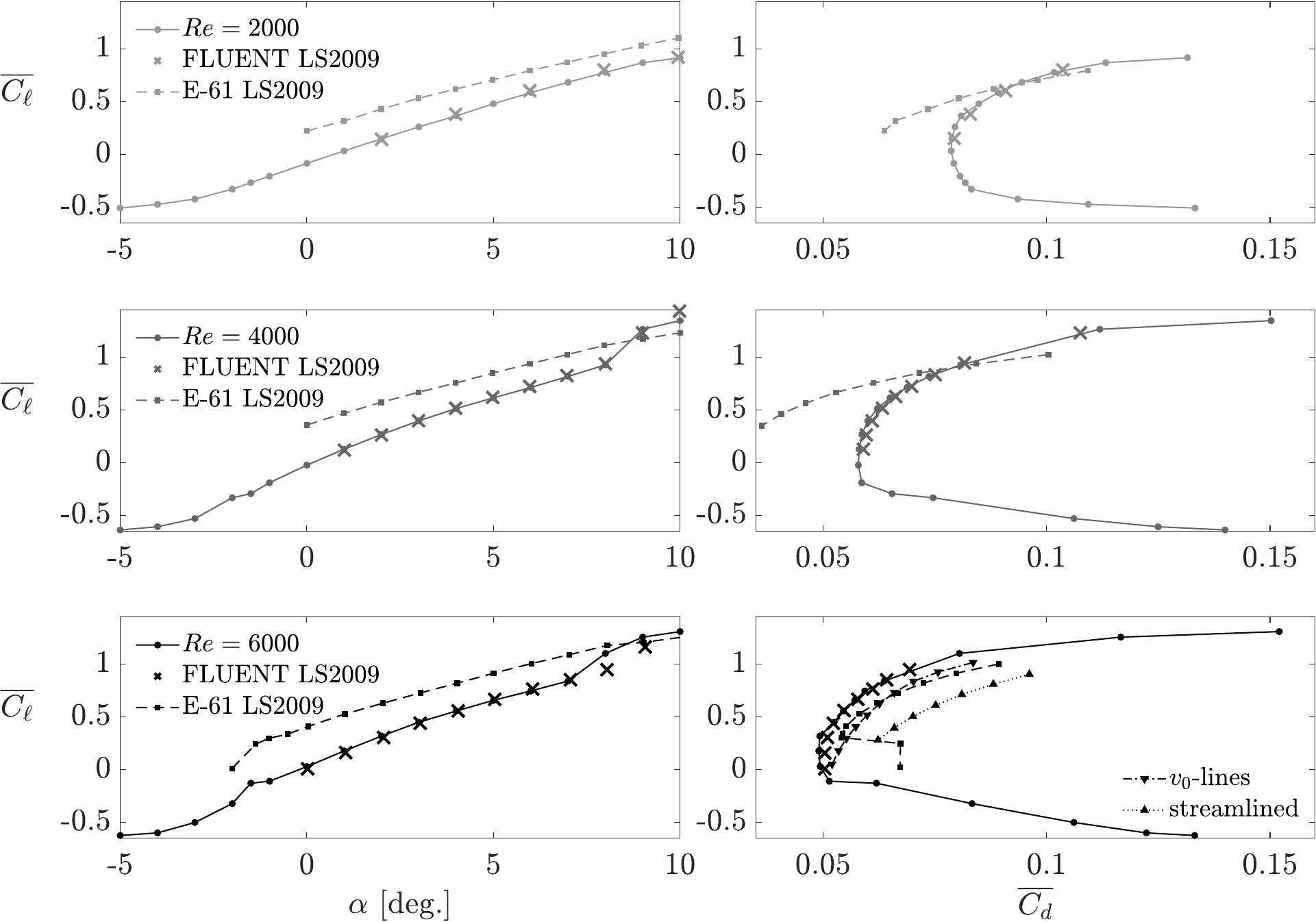}};
    \node at (-5.8,5) {$(a)$};
    \node at (0.7,5)  {$(b)$};
    \node at (-5.8,1.8) {$(c)$};
    \node at (0.7,1.8)  {$(d)$};
    \node at (-5.8,-1.4)  {$(e)$};
    \node at (0.7,-1.4) {$(f)$};
    \end{tikzpicture}
    }
    \caption{Panels $(a,c,e)$: Mean lift coefficient as a function of the angle of attack $\alpha$. Panels $(b,d,f)$: polar curves $\overline{C_{\ell}}-\overline{C_d}$. Panels $(a,b)$: $\Rey = 2000$. Panels $(c,d)$: $\Rey=4000$. Panels $(e,f)$: $\Rey=6000$. Comparison with the results of \citet{levy-seifert-2009} (referred as LS2009).}
    \label{fig:polar_curves}
\end{figure}
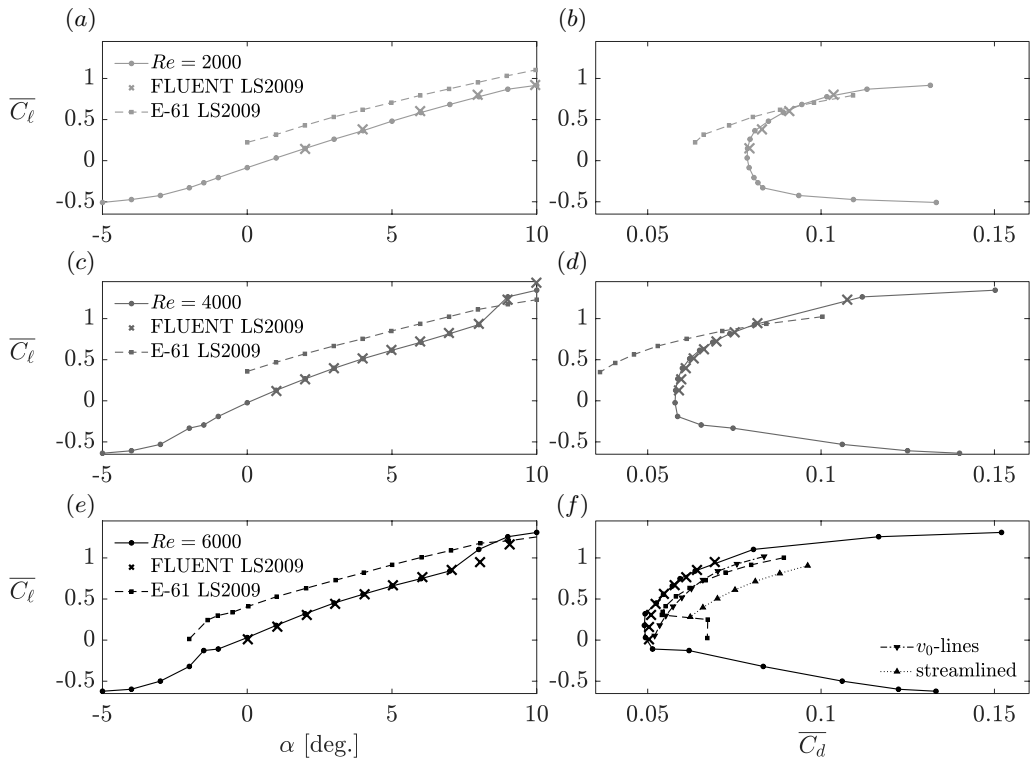

The dependence of the mean lift and drag coefficients on $\alpha$ is presented in figure \ref{fig:polar_curves} for $\Rey=2000,4000,6000$.
Results are compared to those obtained by \citet{levy-seifert-2009} (referred as LS2009).
In particular, the results on the dragonfly-inspired geometry are compared to those of three smooth airfoils: the Eppler-E61 airfoil whose aft-upper region surface is similar to the rear-upper part of the corrugated airfoil~\citep[see figure 1\textit{c} of][]{levy-seifert-2009}, a ``streamlined" airfoil based on the closest streamline that circles the corrugated airfoil, and the so-called ``$v_0$-lined" airfoil based on the contours created by the collection of points with a zero mean streamwise velocity, encompassing the separated flow regions \citep{levy-seifert-2009}.

The $\overline{C_\ell}$--$\alpha$ curves (left panels in figure \ref{fig:polar_curves}) show that the mean lift of the Eppler-E61 is approximately linear in the range of $-0.7^\circ < \alpha < 8$ regardless of the Reynolds number, while for the dragonfly-inspired geometry increasing $\Rey$ accentuates the deviation from the linear behaviour in the same range of $\alpha$.
As observed by \citet{levy-seifert-2009}, the mean lift slope at $0^\circ < \alpha < 2^\circ$ is even larger than $2 \pi \, \text{rad}^{-1}$, contrary to that of the Eppler-E61 ($\approx \pi \, \text{rad}^{-1}$).
The passing of the potential limit of $2 \pi \, \text{rad}^{-1}$ for the dragonfly-inspired periodic wing can be explained by the progressive growth of the separation regions in the forward part with $\alpha$ (both between and downstream of the two corrugations).
This causes an effective increase in the airfoil camber with the angle of attack, resulting in overall higher lifting capability \citep{buckholtz1986}.

Interestingly, despite the lower lifting capabilities for a given $\alpha$, figure \ref{fig:polar_curves} shows that the lift-to-drag ratio of the dragonfly-inspired wing might be superior to that of the smooth airfoils.
At $\Rey=6000$, for mean lift coefficient between $0$ and $1$, all smooth configurations considered exhibit a higher mean drag coefficient $\overline{C_D}$ with respect to the dragonfly-inspired wing which thus provides higher lift-to-drag ratio.

\section{Sensitivity to the computational grid}
\label{sec:sens_grid_res}

In this section the sensitivity of the results to the domain size and grid resolution is investigated for both the primary and secondary bifurcations.

\subsection{Primary bifurcation}
\label{sec:sens_grid_res_pb}

We have examined the influence of the grid resolution on the frequency $f_{\text{DNS}}$ from nonlinear simulations, the growth rate $\sigma_1$ and the frequency $f_1 = \omega_1/ 2\pi$ of the leading primary instability developing on the dragonfly-inspired airfoil at different $\alpha$ and $\Rey$.
Polynomial order convergence on a two-dimensional computational domain extending for $(\ell_x,\ell_y) =( 20c,20c)$ has been verified to assess the mesh requirement, and some results are summarised in table \ref{tab:table_sens_grid_res_pb}. 
\begin{table}
\centering
\begin{tabular}{cccccccccccccc}
$\alpha$   & & $Re$  & & $P$ & & $N_{\text{dof}}$ & & $St_{\text{DNS}}$ & & $\sigma_1$ & & $f_1 = \omega_1/ 2\pi$ & \\
$10^\circ$ & & $815$ & & $4$ & & $3.58 \times 10^6$   & & $0.9441$  & & $1.7021 \times 10^{-2}$  & & $0.9417$ & \\
$10^\circ$ & & $815$ & & $6$ & & $8.06 \times 10^6$   & & $0.9436$  & & $1.6882 \times 10^{-2}$   & & $0.9418$ & \\
$10^\circ$ & & $815$ & & $8$ & &$1.43 \times 10^7$   & & $0.9434$ & & $1.6801 \times 10^{-2}$  & & $0.9418$ &\\
\hline
$ 5^\circ$ & & $2000$ & & $4$ & & $3.58 \times 10^6$   & & $1.7140$ & & $1.8242 \times 10^{-1}$  & & $ 1.6107 $ & \\
$ 5^\circ$ & & $2000$ & & $6$ & & $8.06 \times 10^6$   & & $1.6954$ & & $1.8266 \times 10^{-1}$  & & $1.6107$   & \\
$ 5^\circ$ & & $2000$ & & $8$ & &$1.43 \times 10^7$   & & $1.6948$ & & $1.8271 \times 10^{-1}$  & & $1.6107$ & \\
\hline
$ 0^\circ$ & & $2925$ & & $4$ & & $3.58 \times 10^6$ & & $1.8672$  & & $3.3571 \times 10^{-3}$ & & $ 1.8612$  &\\
$ 0^\circ$ & & $2925$ & & $6$ & & $8.06 \times 10^6$ & & $1.8635$  & & $3.5395 \times 10^{-3}$ & & $ 1.8610$  &\\
$ 0^\circ$ & & $2925$ & & $8$ & &$1.43 \times 10^7$ & & $1.8618$ & & $ 3.5709  \times 10^{-3} $  & & $1.8610$ &\\
\hline
$-5^\circ$ & & $1200$ & & $4$ & & $3.58 \times 10^6$   & & $1.0009$     & & $ 6.2568 \times 10^{-2}$ & & $0.9825$ & \\
$-5^\circ$ & & $1200$ & & $6$ & & $8.06 \times 10^6$   & & $0.9969$     & & $ 6.0730 \times 10^{-2}$ & & $0.9822$ & \\
$-5^\circ$ & & $1200$ & & $8$ & &$1.43 \times 10^7$   & & $0.9968$ & & $6.0227 \times 10^{-2}$  & & $0.9821$ &\\
\end{tabular}
\caption{Frequency $St_{\text{DNS}}$ from nonlinear simulations, growth rate $\sigma_1$ and frequency $f_1 = \omega_1/ 2\pi$ of the leading primary instability developing on the dragonfly-inspired airfoil at different $\alpha$ and $\Rey$ for different polynomial orders $P$ on a two-dimensional computational domain extending for $(\ell_x,\ell_y) =( 20c,20c)$.}
\label{tab:table_sens_grid_res_pb}
\end{table}
The total number of degrees of freedom $N_{\text{dof}}$ is $3.58 \times 10^6$ for $P=4$,  $N_{\text{dof}} = 8.06 \times 10^6$ for $P=6$ and $N_{\text{dof}} = 1.43 \times 10^7$ for $P=8$.
The analysis of the polynomial order on the grid shows that the numerical convergence is substantially reached since the relative error is always lower than $\sim 1 \%$.

\subsection{Secondary bifurcation}
\label{sec:grid-sec}

\begin{figure}
\centering
\includegraphics[width=1\textwidth]{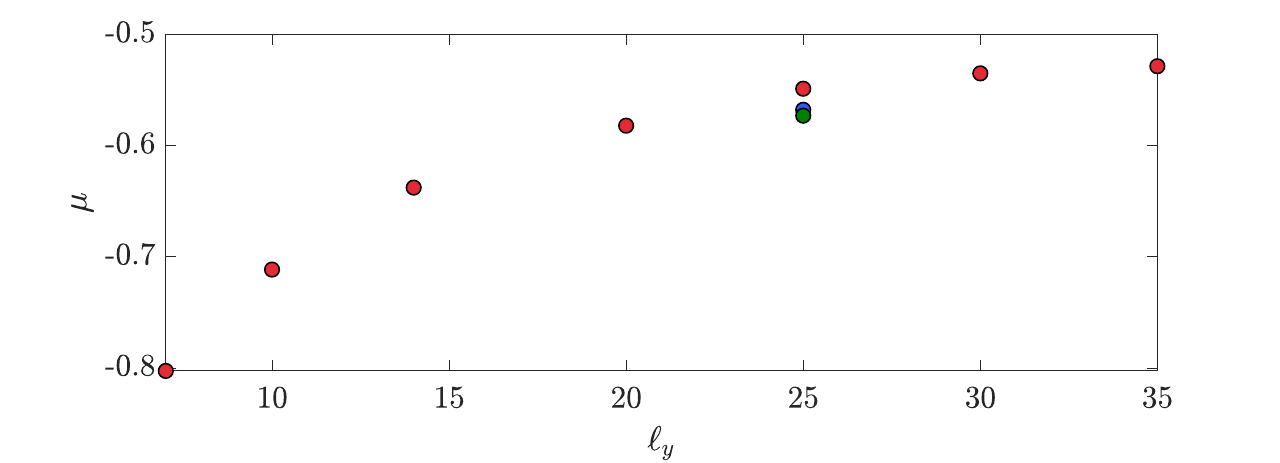}
\caption{Dependence of the Floquet multiplier on the size of the domain for $\alpha = 10^\circ$, $Re=1700$ and $\beta=50$. The red circle is for $\ell_x=17.5$, the blue circle is for $\ell_x=20$ and the green circle is for $\ell_x=22.5$.}
\label{fig:floq_dom_dep}
\end{figure}
\begin{table}
\centering
\begin{tabular}{cccccccccccccccc}
$\alpha$   & & $Re$  & &  $\beta$ & & $\ell_x$  & & $\ell_y$ & & $N_{\text{el}}$           & & $St_{\text{BF}}$ & & $\mu$ & \\
$10^\circ$ & & $1700$& &  $50$    & & $17.5$ & & $7$   & & $92\times 10^3$    & & $1.3302$  & & $(-0.802,0)$ & \\
$10^\circ$ & & $1700$& &  $50$    & & $17.5$ & & $10$  & & $11.5 \times 10^4$ & & $1.3447$  & & $(-0.711,0)$ & \\
$10^\circ$ & & $1700$& &  $50$    & & $17.5$ & & $14$  & & $13.8 \times 10^4$ & & $1.3582$  & & $(-0.637,0)$ & \\
$10^\circ$ & & $1700$& &  $50$    & & $17.5$ & & $20$  & & $14.1 \times 10^4$ & & $1.3687$  & & $(-0.581,0)$ & \\
$10^\circ$ & & $1700$& &  $50$    & & $17.5$ & & $25$  & & $13.9 \times 10^4$ & & $1.3751$  & & $(-0.548,0)$ & \\
$10^\circ$ & & $1700$& &  $50$    & & $17.5$ & & $30$  & & $13.1 \times 10^4$ & & $1.3778$  & & $(-0.534,0)$ & \\
$10^\circ$ & & $1700$& &  $50$    & & $17.5$ & & $35$  & & $12.2 \times 10^4$ & & $1.3792$  & & $(-0.528,0)$ & \\
$10^\circ$ & & $1700$& &  $50$    & & $20  $ & & $25$  & & $15.2 \times 10^4$ & & $1.3714$  & & $(-0.567,0)$ & \\
$10^\circ$ & & $1700$& &  $50$    & & $22.5$ & & $25$  & & $15.9 \times 10^4$ & & $1.3702$  & & $(-0.572,0)$ & \\
\hline
$ 7^\circ$ & & $4500$& &  $20$    & & $17.5$ & & $14$  & & $13.8 \times 10^4$ & & $2.3764$  & & $(1.0557,0)$ & \\
$ 7^\circ$ & & $4500$& &  $20$    & & $20  $ & & $30$  & & $15.4 \times 10^4$ & & $2.3624$  & & $(1.0612,0)$ & \\
\hline
$-5^\circ$ & & $1450$& &  $0 $    & & $17.5$ & & $14$  & & $13.8 \times 10^4$ & & $1.0738$  & & $(0.7458,0.6514)$ & \\
$-5^\circ$ & & $1450$& &  $0 $    & & $20  $ & & $30$  & & $15.4 \times 10^4$ & & $1.0710$  & & $(0.7539,0.6529)$ & \\
\hline
$-5^\circ$ & & $1450$& &  $0 $    & & $17.5$ & & $14$  & & $13.8 \times 10^4$ & & $1.0738$  & & $(-0.9281,0.1171)$ & \\
$-5^\circ$ & & $1450$& &  $0 $    & & $20  $ & & $30$  & & $15.4 \times 10^4$ & & $1.0710$  & & $(-0.8972,0.1571)$ & \\
\end{tabular}
\caption{Dependence of the Floquet multiplier on the size of the computational domain for $\alpha=10^\circ$ (with $Re=1700$ and $\beta=50$), $\alpha=7^\circ$ (with $Re=4500$ and $\beta=20$) and $\alpha=-5^\circ$ (with $Re=1450$ and $\beta=0$). The baseline grid for $8^\circ \le \alpha \le 10^\circ$ has $\ell_y=30$, $\ell_x=20$ and $N_{\text{el}} = 15.4 \times 10^4$. The baseline grid for $-5^\circ \le \alpha \le 7^\circ$ has $\ell_x=14$, $\ell_x=17.5$ and $N_{\text{el}} = 13.8 \times 10^4$.}
\label{tab:floq_dom_dep}
\end{table}
We have investigated the dependence of the results on the grid size and grid resolution by repeating the complete Floquet analysis by varying the extent of the computational domain and the grid resolution. The convergence study has been performed for three different angles of attack, $\alpha=10^\circ$ and $\alpha=-5^\circ$ (the extrema of the range of $\alpha$ considered in this work) and $\alpha=7^\circ$. This allows us to consider the different growing modes detected in \S\ref{sec:sec-bif}, i.e. mode $sS$ for $\alpha=10^\circ$, mode $sA$ for $\alpha=7^\circ$ and modes $sQPb$ and $sS2$ for $\alpha=-5^\circ$.

For $\alpha=10^\circ$ the streamwise extent of the computational domain has been varied between $\ell_x=17.5$ and $\ell_x=22.5$, while the vertical extent between $\ell_y=7$ and $\ell_y=35$ (recall that all quantities are made dimensionless using the airfoil chord $c$ as length scale). For all the grids, the grid resolution close to the body is maintained approximately constant. In this case we have considered $Re=1700$ and $\beta=50$. When fixing $\ell_x=17.5$ an increase of the vertical extent of the computational domain from $\ell_y=7$ to $\ell_y=35$ leads to a variation of the base flow frequency of $|\Delta St_{\text{BF}}| \approx 3.5\%$, and to a variation of the the multiplier associated with mode $sS$ of $\left|\Delta \left|\mu_{sS}\right| \right| \approx 34.16\%$ (see table \ref{tab:floq_dom_dep} and figure \ref{fig:floq_dom_dep}). In contrast, when fixing the vertical extent of the grid to $\ell_y=25$ an increase of the streamwise extent from $\ell_x=17.5$ to $\ell_x=22.5$ leads to $|\Delta St_{\text{BF}}| \approx 0.35 \%$ and $\left|  \Delta |\mu_{sS} | \right| \approx 4\%$. Based on this convergence study, for $\alpha=10^\circ$, we have chosen $(\ell_x,\ell_y)=(20,30)$ with a number of elements $N_{\text{el}} \approx 15.4 \times 10^4$, after having verified that an increase of the grid resolution ($N_{\text{el}} \approx 18.5 \times 10^4$) leads to a negligible variation ($\lessapprox 1\%$) of the results. 

For the smaller angles of attack, instead, we have considered $Re=4500$ and $\beta=20$ (for $\alpha=7^\circ$), and $Re=1450$ and $\beta=0$ (for $\alpha=-5^\circ$). This enables us to investigate the sensitivity of modes $sA$, $sQPB$ and $sS2$ on the grid. For these $\alpha$ two grids have been considered. The baseline grid used for $\alpha=10^\circ$ (with $\ell_x=20$ and $\ell_y=30$) and a smaller grid with $(\ell_x,\ell_y)=(17.5,14)$; see table \ref{tab:floq_dom_dep}. For $\alpha=7^\circ$ we found a variation of $|\Delta St_{\text{BF}}| \approx 0.59\%$ and $ | \Delta |\mu_{sA}|| \approx 0.51\%$. For $\alpha=-5^\circ$, instead, we measure a variation of $|\Delta St_{\text{BF}}| \approx 0.26\%$, $|\Delta |\mu_{sS2}|| \approx 2.6\%$ and $|\Delta |\mu_{sQPb} | | \approx 0.71 \%$. Based on these results, and given that the nature of the different modes are well described by all the considered grids, for $-5^\circ \le \alpha \le 7^\circ$ we have adopted the smaller grid with $\ell_x=17.5$ and $\ell_y=14$, with $N_{\text{el}} = 13.8 \times 10^4$. We have verified the adequacy of the grid resolution using $\alpha=7^\circ$ and mode $sA$ as reference. We have increased the number of elements to $N_{\text{el}} = 15.8 \times 10^4$ and $N_{\text{el}} = 18.7 \times 10^4$ (mainly in the near wake and close to the body) and observed a marginal variation ($\lessapprox 1\%$) of the results. 

\bibliographystyle{jfm}

\end{document}